\documentclass[12pt, a4paper]{article}
\usepackage{graphicx}
\usepackage{amssymb}
\usepackage{amsmath}
\usepackage{bm}
\usepackage{color}
\usepackage{theorem}
\usepackage{subcaption}
\usepackage{listings}
\usepackage{colortbl}
\usepackage{tabularx}
\usepackage{longtable}
\usepackage[utf8]{inputenc}
\usepackage[T1]{fontenc}
\usepackage{lmodern}
\usepackage[top=30truemm,bottom=30truemm,left=25truemm,right=25truemm]{geometry}

\usepackage[sort&compress,numbers, merge]{natbib}

\definecolor{Orange}{cmyk}{0,0.61,0.87,0}
\definecolor{JungleGreen}{cmyk}{0.99,0,0.52,0}
\definecolor{OliveGreen}{cmyk}{0.64,0,0.95,0.40}
\definecolor{Brown}{cmyk}{0,0.81,1,0.60}
\definecolor{RoyalBlue}{cmyk}{0.71,0.53,0,0.12}
\definecolor{Gray}{cmyk}{0,0,0,0.40}
\definecolor{LightPink}{cmyk}{0.0,0.25,0,0}
\definecolor{LLightPink}{cmyk}{0.0,0.10,0,0}
\definecolor{LightBlue}{cmyk}{0.25,0,0,0}
\definecolor{LightGray}{cmyk}{0,0,0,0.2}

\allowdisplaybreaks[1]

\usepackage[colorlinks=true, linkcolor=OliveGreen, citecolor=RoyalBlue,
urlcolor=RoyalBlue]{hyperref}

\renewcommand{\thefootnote}{\fnsymbol{footnote}}

\begin{document}

\begin{titlepage}

  \begin{flushright}
    {\tt
    }
\end{flushright}

\vskip 1.35cm
\begin{center}

{\Large
{\bf
Neutron Star Heating in Dark Matter Models \\[4pt]
for the Muon $g-2$ Discrepancy 
}
}

\vskip 1.5cm

Koichi Hamaguchi$^{a,b}$\footnote{
\href{mailto:hama@hep-th.phys.s.u-tokyo.ac.jp}{\tt
 hama@hep-th.phys.s.u-tokyo.ac.jp}},
Natsumi Nagata$^a$\footnote{\href{mailto:natsumi@hep-th.phys.s.u-tokyo.ac.jp}{\tt natsumi@hep-th.phys.s.u-tokyo.ac.jp}}, 
and 
Maura E. Ramirez-Quezada$^{a}$\footnote{\href{mailto:me.quezada@hep-th.phys.s.u-tokyo.ac.jp}{\tt me.quezada@hep-th.phys.s.u-tokyo.ac.jp}}

\vskip 0.8cm

{\it $^a$Department of Physics, University of Tokyo, Bunkyo-ku, Tokyo
 113--0033, Japan} \\[2pt]
 {\it ${}^b$Kavli Institute for the Physics and Mathematics of the
  Universe (Kavli IPMU), University of Tokyo, Kashiwa 277--8583, Japan} \\[2pt]

\date{\today}

\vskip 1.5cm

\begin{abstract}

  The observed value of the muon magnetic dipole moment, which deviates from the Standard Model prediction by $4.2\sigma$, can be explained in models with weakly-interacting massive particles (WIMPs) coupled to muons. However, a considerable range of parameter space of such models will remain unexplored in the future LHC experiments and dark matter (DM) direct searches. In this work we discuss the temperature observation of neutron stars (NSs) as a promising way to probe such models given that WIMPs  are efficiently captured by NSs through DM-muon or spin-dependent DM-nucleon scattering. The captured WIMPs eventually annihilate in the star core and heat the NS. This effect can be observed in old NSs as it keeps the NS surface temperature at a few thousand K at most, which is much higher than the predicted values of the standard NS cooling theory for NSs older than $\sim 10^7$~years.  We consider two classes of representative models, where the DM couples or does not couple to the Higgs field at tree level, and show that the maximal DM heating is realized in both scenarios.

\end{abstract}

\end{center}
\end{titlepage}

\renewcommand{\thefootnote}{\arabic{footnote}}
\setcounter{footnote}{0}

\section{Introduction}

The latest measurement of the muon anomalous magnetic moment at Fermilab~\cite{Muong-2:2021ojo} confirmed the deviation from the Standard Model (SM) prediction~\cite{Aoyama:2020ynm,Davier:2017zfy,Keshavarzi:2018mgv,Colangelo:2018mtw,Hoferichter:2019mqg,Davier:2019can,Keshavarzi:2019abf,Keshavarzi:2019abf,Kurz:2014wya,Chakraborty:2017tqp,Borsanyi:2017zdw,Blum:2018mom,Giusti:2019xct,Shintani:2019wai,FermilabLattice:2019ugu,Gerardin:2019rua,Aubin:2019usy,Giusti:2019hkz,Masjuan:2017tvw,Colangelo:2017fiz,Hoferichter:2018kwz,Gerardin:2019vio,Bijnens:2019ghy,Colangelo:2019uex,Pauk:2014rta,Danilkin:2016hnh,Jegerlehner:2017gek,Knecht:2018sci,Eichmann:2019bqf,Roig:2019reh,Colangelo:2014qya,Blum:2019ugy,Masjuan:2017tvw,Colangelo:2017fiz,Hoferichter:2018kwz,Gerardin:2019vio,Bijnens:2019ghy,Colangelo:2019uex,Pauk:2014rta,Danilkin:2016hnh,Jegerlehner:2017gek,Knecht:2018sci,Eichmann:2019bqf,Roig:2019reh,Blum:2019ugy,Aoyama:2012wk,Aoyama:2019ryr,Czarnecki:2002nt,Gnendiger:2013pva,Davier:2017zfy,Keshavarzi:2018mgv,Colangelo:2018mtw,Hoferichter:2019mqg,Davier:2019can,Keshavarzi:2019abf,Kurz:2014wya,Melnikov:2003xd,Masjuan:2017tvw,Colangelo:2017fiz,Hoferichter:2018kwz,Gerardin:2019vio,Bijnens:2019ghy,Colangelo:2019uex,Pauk:2014rta,Danilkin:2016hnh,Jegerlehner:2017gek,Knecht:2018sci,Eichmann:2019bqf,Roig:2019reh,Blum:2019ugy,Colangelo:2014qya,Aoyama:2012wk,Aoyama:2019ryr,Czarnecki:2002nt,Gnendiger:2013pva,Davier:2017zfy,Keshavarzi:2018mgv,Colangelo:2018mtw,Hoferichter:2019mqg,Davier:2019can,Keshavarzi:2019abf,Kurz:2014wya,Masjuan:2017tvw,Colangelo:2017fiz,Hoferichter:2018kwz,Gerardin:2019vio,Bijnens:2019ghy,Colangelo:2019uex,Blum:2019ugy,Colangelo:2014qya} previously observed at the Brookhaven National Laboratory~\cite{Muong-2:2006rrc}. The combined value of these two measurements is found to be larger than the recommended value  in Ref.~\cite{Aoyama:2020ynm} by
\begin{equation}
  \Delta a_\mu = 251(59) \times 10^{-11}~,
  \label{eq:delamu_exp}
\end{equation}
whose significance is $4.2\sigma$.\footnote{Note that this significance is considerably reduced if we adopt the recent result from the QCD lattice simulation~\cite{Borsanyi:2020mff} for the estimate of the hadronic vacuum polarization contribution, instead of that obtained with the data-driven method~\cite{Aoyama:2020ynm}. The compatibility of this lattice calculation with existing experimental/lattice results is the subject of much debate for the moment (see, for instance, Refs.~\cite{Lehner:2020crt, Crivellin:2020zul,  Keshavarzi:2020bfy, deRafael:2020uif, Malaescu:2020zuc, DiLuzio:2021uty,Colangelo:2020lcg}); we therefore use Eq.~\eqref{eq:delamu_exp} as a benchmark value for the muon $g-2$ discrepancy in this paper. } This muon $g-2$ discrepancy has attracted much attention for decades as it could be a sign of physics beyond the SM. 

It is known that this discrepancy can be explained in models with weakly-interacting massive particles (WIMPs) coupling to muons (see, \textit{e.g.}, Refs.~\cite{Kanemitsu:2012dc, Kowalska:2017iqv, Calibbi:2018rzv, Kawamura:2020qxo, Horigome:2021qof, Arcadi:2021cwg, Bai:2021bau, Athron:2021iuf, Acuna:2021rbg, Ghosh:2022zef} for recent relevant studies). A well-known class of models that realize such a setup are supersymmetric (SUSY) extensions of the SM, where the interactions of muons with neutralinos, charginos, and sleptons give additional contributions to the muon $g-2$ at the loop level~\cite{Lopez:1993vi, Chattopadhyay:1995ae, Moroi:1995yh}. See Refs.~\cite{Endo:2021zal, Iwamoto:2021aaf, Gu:2021mjd, Yin:2021mls, Wang:2021bcx, Abdughani:2021pdc, Cao:2021tuh, Chakraborti:2021dli, Ibe:2021cvf, Cox:2021nbo, Heinemeyer:2021opc,Baum:2021qzx, Zhang:2021gun, Ahmed:2021htr, Aboubrahim:2021xfi, Chakraborti:2021bmv, Baer:2021aax, Altmannshofer:2021hfu, Aboubrahim:2021phn, Chakraborti:2021squ, Zheng:2021wnu, Jeong:2021qey, Li:2021pnt, Kim:2021suj, Ellis:2021zmg, Aboubrahim:2021ily, Nakai:2021mha, Li:2021cte, Lamborn:2021snt, Ellis:2021vpp, Chakraborti:2021mbr, Ali:2021kxa, Gomez:2022qrb, Chakraborti:2022vds, Agashe:2022uih, Endo:2022qnm, Chigusa:2022xpq,Cao:2022chy} for recent studies on the SUSY interpretations of the muon $g-2$ measurements. An attractive feature of this class of models is that they contain a promising candidate for dark matter (DM) in the Universe. Indeed, previous studies have revealed that it is possible to reproduce the observed value of the DM density~\cite{Planck:2018vyg}, while explaining the muon $g-2$ discrepancy and evading the current experimental limits.

However, if the couplings to muons of such new particles are sizable, their masses are predicted to be  $\sim 1\,\rm TeV$ in order to explain both, the muon $g-2$ discrepancy and the observed DM density. It is hard to discover such heavy colorless particles at the LHC experiments and very challenging to fully test such models  even in future experiments.
On the other hand, DM direct search experiments are capable of detecting DM around $\mathcal{O} (1)~\rm TeV$ mass. Nevertheless, the DM detection rate depends highly on the size of the DM-Higgs coupling, and considerable regions of parameter space are beyond the reach of  the next-generation DM direct detection experiments.

The temperature observation of neutron stars (NSs) offers a promising way to probe these scenarios by means of the DM accretion in NS core. DM particles are captured in NSs if they lose a considerable amount of their kinetic energy when scattering off stellar matter~\cite{Goldman:1989nd}. In this work we show that the WIMP DM particles in the models motivated by the muon $g-2$ discrepancy are efficiently captured by NSs through their interactions with nucleons and muons. The captured DM particles eventually annihilate inside the NSs, giving their energy to the NSs as heat. This heating effect modifies the NS temperature evolution and, in particular, keeps the  surface temperature at $\text{a few} \times 10^3$~K at late times~\cite{Kouvaris:2007ay, Bertone:2007ae, Kouvaris:2010vv, deLavallaz:2010wp}. This is in contrast to the prediction of the standard NS cooling theory~\cite{Yakovlev:1999sk, Yakovlev:2000jp, Yakovlev:2004iq, Page:2004fy, Page:2009fu, Potekhin:2015qsa}, where the surface temperature is predicted to be less than $10^3$~K for NSs older than $\sim 5 \times 10^6$~years. The surface temperature of $\sim \text{a few} \times 10^3$~K can be observed for nearby NSs with future infrared telescopes~\cite{Baryakhtar:2017dbj} such as the James Webb Space Telescope (JWST)~\cite{Gardner:2006ky}, making it possible to probe the DM heating effect. See Refs.~\cite{Bramante:2017xlb, Raj:2017wrv,  Chen:2018ohx, Bell:2018pkk, Garani:2018kkd, Camargo:2019wou, Bell:2019pyc, Hamaguchi:2019oev, Garani:2019fpa, Acevedo:2019agu, Joglekar:2019vzy, Keung:2020teb,  Yanagi:2020yvg, Joglekar:2020liw, Bell:2020jou, Bell:2020lmm, Anzuini:2021lnv, Zeng:2021moz, Bramante:2021dyx, Tinyakov:2021lnt,Maity:2021fxw, Fujiwara:2022uiq,Ilie:2020vec} for recent studies on the DM heating.

In this paper, we show that in the muon $g-2$ motivated scenarios for DM, the DM heating mechanism efficiently operates in NSs in the parameter regions out of reach in future DM direct detection experiments. We consider two representative models with a Majorana fermion DM. We will see that these two models have qualitatively different phenomenology. 
In the first case, model I, DM directly couples to the Higgs field. The tree-level DM-Higgs coupling yields a sizable DM-nucleon spin-independent (SI) scattering cross section, and therefore the constraints from DM direct detection experiments tend to be severe. In the second case, model II, there is no tree-level DM-Higgs coupling and hence  the DM-nucleon SI scattering cross section is highly suppressed.
We see that in both models the DM capture rate in NSs is large enough so that the DM heating operates efficiently. It is  found that the presence of muons in NSs plays an important role in the DM capture.

The paper is organized as follows. In Sec.~\ref{sec:Models}, we describe the two DM models discussed through this work. Then, we show the formulae for the calculation of the muon $g-2$ and the SI DM-nucleon scattering cross section in Sec.~\ref{sec:muongmi2} and Sec.~\ref{sec:direct_detection}, respectively. The discussions on the DM capture and heating in NSs are given in Sec.~\ref{sec:DMheating}. In Sec.~\ref{sec:results}, we present the predictions of the models for the muon $g-2$, the DM-nucleon scattering cross sections, and the DM-muon scattering cross sections, and discuss the prospects of the NS temperature observations to probe these models. Finally, Sec.~\ref{sec:conclusion} summarizes our conclusions and discussions.

\section{Models}
\label{sec:Models}
We consider DM models that can explain the muon $g-2$ discrepancy while evading the existing experimental limits. To assure the stability of the DM particle, we introduce a $\mathbb{Z}_2$ symmetry under which the DM is odd while all the SM fields are even. We further assume that the DM couples to both $\mu_L$ and $\mu_R$ to give an additional contribution to the muon magnetic moment at the one-loop level.\footnote{We could have assumed that the DM couples to either $\mu_L$ or $\mu_R$. However, it is quite difficult to explain the observed size of the muon $g-2$ discrepancy in this setup without conflicting with the experimental limits, since the new contribution to the muon magnetic moment is always suppressed by the muon mass  (see, for instance, Refs.~\cite{Kanemitsu:2012dc, Kowalska:2017iqv, Calibbi:2018rzv, Kawamura:2020qxo, Athron:2021iuf}). Therefore, we do not consider this possibility in the present work.} 
For such couplings to be present, we need to introduce two extra $\mathbb{Z}_2$-odd fields. 

There is a large (actually infinite) number of possibilities for this class of ``three-field extension models''. The purpose of the present work is not to thoroughly examine these possibilities, but to explore typical predictions in these type of models. Here, we consider two representative models, with qualitatively different phenomenology, for the case I (II) where the DM does (not) couple to the Higgs field at tree level. We present the models for each of these scenarios in Sec.~\ref{sec:model1} and Sec.~\ref{sec:model2}, respectively.

\subsection{Model I}
\label{sec:model1}

We first describe a model where the DM couples to the Higgs field at the renormalizable level. In this case, the DM is a mixed state of the two gauge eigenstates. Due to the presence of the direct coupling to the Higgs field, the DM-nucleon scattering cross section can be sizable. This is an advantageous feature for testability of this model in DM direct detection experiments.

\begin{table}[t!]
  \begin{center}
 \caption{New particles and their quantum numbers in Model I.   }
 \label{tab:model1}
 \vspace{5pt}
 \begin{tabular}{cccccc}
 \hline
 \hline
 Field & Spin & SU(3)$_C$ & SU(2)$_L$ & U(1)$_Y$ & $\mathbb{Z}_2$  \\
 \hline
 ${\chi}_S$ & $1/2$ & {\bf 1} & {\bf 1} & $0$ & $-$  \\
 ${\xi}_D$ & $1/2$ & {\bf 1} & {\bf 2} & $-1/2$ & $-$  \\
 ${\eta}_D$ & $1/2$ & {\bf 1} & {\bf 2} & $1/2$ & $-$  \\
 $\widetilde{L}$ & $0$ & {\bf 1} & {\bf 2} & $-{1}/{2}$ & $-$ \\
 \hline
 \hline
 \end{tabular}
  \end{center}
 \end{table}

We summarize the new ($\mathbb{Z}_2$-odd) fields in Model I and their quantum numbers in Table~\ref{tab:model1}, where the fermion fields are expressed in terms of two-component left-handed Weyl fields. 
There are two sets of fermion fields, a SU(2)$_L$ singlet fermion, $\chi_S$, with hypercharge $ Y = 0$, and a Dirac fermion ($\xi_D$ and $\eta_D$) of a SU(2)$_L$ doublet with $Y = -1/2$. The model also contains a SU(2)$_L$ doublet scalar, $\widetilde{L}$,  with $Y = -1/2$. This particle content is identical to the FLR1 model in Ref.~\cite{Calibbi:2018rzv}. In the framework of SUSY, $\chi_S$, $\xi_D$ and $\eta_D$, and $\widetilde{L}$ can be regarded as bino (or singlino in the next-to-minimal supersymmetric model), higgsinos, and the left-handed slepton, respectively.

The Lagrangian terms relevant for our discussions are 
\begin{equation}
  \mathcal{L}_{\mathrm{int}} = {\cal L}_{\rm mass} 
  + {\cal L}_{\mathrm{Yukawa}} 
  + {\cal L}_{\rm quart}  ~,
\end{equation}
with 
\begin{align}
  {\cal L}_{\rm mass} 
  &= 
  - 
  \left( \frac{1}{2} M_{F_S} \chi_S \chi_S  + M_{F_D} \xi_D \eta_D + \text{h.c.} \right)
  -  M_{\tilde{L}}^2 |\widetilde{L}|^2 
   ~, \label{eq:mass1} \\[3pt]
   {\cal L}_{\mathrm{Yukawa}} &= 
   - y_{1H} \chi_S (\xi_D \cdot H )
   - y_{2H} \chi_S \eta_D H^\dagger 
   - y_1 \chi_S L_\mu \widetilde{L}^\dagger 
   - y_2 \mu_R^c (\xi_D \cdot \widetilde{L} )
   +{\rm h.c.} ~, \label{eq:yukawa1} \\[3pt] 
   {\cal L}_{\rm quart} 
& = - \lambda_{L} |\widetilde{L}|^2 |H|^2
- \lambda^\prime_L \widetilde{L}^\dagger \tau_a \widetilde{L} 
H^\dagger \tau_a H
+\dots ~,
\label{eq:quart1} 
\end{align}
where $(A \cdot B) \equiv \epsilon_{\alpha\beta} A^\alpha B^\beta$ with $\epsilon_{\alpha\beta}$ the totally antisymmetric tensor ($\epsilon_{12} = -\epsilon_{21} = 1$).  $L_\mu$ and $\mu_R^c$ are the left- and right-handed second-generation lepton fields, respectively, $\tau^a$ ($a = 1,2,3$) are the Pauli matrices, and the dots in Eq.~\eqref{eq:quart1} indicate the self-coupling terms of the scalar fields (such as $|\widetilde{L}|^4$), which are irrelevant for the following discussions. The SU(2)$_L$ components of $\xi_D$, $\eta_D$, and $\widetilde{L}$ are 
\begin{equation}
  \xi_D =
  \begin{pmatrix}
   \xi_{D^0} \\ \xi_{D^-}
  \end{pmatrix}
  ~, \qquad 
  \eta_D =
  \begin{pmatrix}
   \eta_{D^+} \\ \eta_{D^0}
  \end{pmatrix}
  ~, \qquad 
  \widetilde{L} =
  \begin{pmatrix}
   \widetilde{\nu} \\ \widetilde{e}
  \end{pmatrix}
  ~.
\end{equation}

Generically speaking, the new fields can couple also to the first/third-generation lepton fields in a similar manner as in the third and fourth terms in Eq.~\eqref{eq:yukawa1}. We neglect such terms in the following analysis for simplicity. We also take all of the coefficients in Eq.~\eqref{eq:mass1} and Eq.~\eqref{eq:yukawa1} to be real to evade constraints from the measurements of electric dipole moments.

After the Higgs field acquires a vacuum expectation value (VEV),
\begin{equation}
  \langle H \rangle =\frac{1}{\sqrt{2}} 
 \begin{pmatrix}
  0 \\v
 \end{pmatrix}
 ~,
 \end{equation}
 with $v\simeq 246$~GeV, the mass terms for the new fields are given by 
 \begin{align}
  \mathcal{L}_{\mathrm{mass}} =  &- \frac{1}{2} \left( \chi_S, \xi_{D^0}, \eta_{D^0} \right)
   \mathcal{M}_\chi
  \begin{pmatrix}
    \chi_S \\ \xi_{D^0}\\ \eta_{D^0} 
  \end{pmatrix}
  - M_{F_D} \xi_{D^-} \eta_{D^+} + \mathrm{h.c.}  
  \nonumber \\ 
  &- M_{\tilde{e}}^2 |\widetilde{e}|^2 
  -M_{\tilde{\nu}}^2  \left|
 \widetilde{\nu}
 \right|^2
  ~,
\end{align}
with 
\begin{equation}
  \mathcal{M}_{\chi} = 
  \begin{pmatrix}
    M_{F_S} & \frac{y_{1H} v}{\sqrt{2}} & \frac{y_{2H} v}{\sqrt{2}}  \\ 
    \frac{y_{1H} v}{\sqrt{2}} & 0 & M_{F_D} \\ 
    \frac{y_{2H} v}{\sqrt{2}}  & M_{F_D} & 0
  \end{pmatrix}
  ~, 
  \quad 
  M_{\tilde{e}}^2  = M_{\tilde{L}}^2 + \frac{\lambda_L + \lambda^\prime_L}{2}v^2   ~, 
  \quad 
  M_{\tilde{\nu}}^2  = M_{\tilde{L}}^2 +\frac{\lambda_L-\lambda^\prime_L}{2}
 v^2 ~.
\end{equation}
As the mass matrix $M_\chi$ is a symmetric matrix, it is diagonalized by means of a unitary matrix $V_\chi$: 
\begin{equation}
  V_\chi^T \mathcal{M}_\chi V_\chi = \mathrm{diag} (M_{\chi_1}, M_{\chi_2}, M_{\chi_3}) ~, 
\end{equation}
with 
\begin{equation}
  \begin{pmatrix}
    \chi_S \\ \xi_{D^0}\\ \eta_{D^0} 
  \end{pmatrix}
  = V_\chi
  \begin{pmatrix}
    \chi_1 \\ \chi_2 \\ \chi_3
   \end{pmatrix}
  ~.
\end{equation}
We take $0<M_{\chi_1} \leq M_{\chi_2} \leq M_{\chi_3}$ without loss of generality. The lightest state $\chi_1$ with mass $M_{\chi_1}$ is the DM in this model. We summarize the relevant interaction terms expressed in the mass eigenbasis in Appendix~\ref{sec:masseigen1} for convenience.

\subsection{Model II}
\label{sec:model2}

Next, we consider a model where the DM does not couple to the Higgs field at tree level. It interacts with the SM particles only through the second-generation leptons. We thus expect the DM direct detection rate to be  highly suppressed in this case.

\begin{table}[t!]
  \begin{center}
 \caption{New particles and their quantum numbers in Model II.   }
 \label{tab:model2}
 \vspace{5pt}
 \begin{tabular}{cccccc}
 \hline
 \hline
 Field & Spin & SU(3)$_C$ & SU(2)$_L$ & U(1)$_Y$ & $\mathbb{Z}_2$  \\
 \hline
 ${\chi}_S$ & $1/2$ & {\bf 1} & {\bf 1} & $0$ & $-$  \\
 $\widetilde{L}$ & $0$ & {\bf 1} & {\bf 2} & $-{1}/{2}$ & $-$ \\
 $\widetilde{\bar{e}}$ & $0$ & {\bf 1} & {\bf 1} & $1$ & $-$ \\
 \hline
 \hline
 \end{tabular}
  \end{center}
 \end{table}

The particle content of the second model is shown in Table~\ref{tab:model2}. The SU(2)$_L$ singlet fermion $\chi_S$ and the SU(2)$_L$ doublet scalar $\widetilde{L}$ are the same as in Model~I. In addition to these two fields, there is a SU(2)$_L$ singlet scalar field, $\widetilde{\bar{e}}$,  with $Y = +1$, which has the same quantum numbers as the right-handed slepton in SUSY theories. The singlet fermion $\chi_S$ is the DM candidate in this model. This model is identical to one of the models discussed in Ref.~\cite{Kawamura:2020qxo}.

At the renormalizable level, the relevant Lagrangian terms are given by 
\begin{equation}
  \mathcal{L}_{\mathrm{int}} = {\cal L}_{\rm mass} 
  + {\cal L}_{\mathrm{Yukawa}} + {\cal L}_{\rm tri} 
  + {\cal L}_{\rm quart}  ~,
\end{equation}
with 
\begin{align}
  {\cal L}_{\rm mass} 
  &= 
  - 
  \left( \frac{1}{2} M_{F_S} \chi_S \chi_S  + \text{h.c.} \right)
  -  M_{\tilde{L}}^2 |\widetilde{L}|^2 
  -  M_{\tilde{\bar{e}}}^2 |\widetilde{\bar{e}}|^2 
   ~, \label{eq:mass2} \\[3pt]
   {\cal L}_{\mathrm{Yukawa}} &= -y_1\, \chi_S L_\mu \widetilde{L}^\dagger
   - y_2 \, \chi_S \mu^c_R \widetilde{\bar{e}}^\dagger 
   +{\rm h.c.} ~, \label{eq:yukawa2} \\[3pt] 
   {\cal L}_{\rm tri}
   & =
   -a_H \, \widetilde{\bar{e}} \widetilde{L} H^\dagger
   +{\rm h.c.} ~, \label{eq:tri2} \\[3pt]
   {\cal L}_{\rm quart} 
& = - \sum_{f= L, \bar{e}} \lambda_f |\widetilde{f}|^2 |H|^2
- \lambda^\prime_L \widetilde{L}^\dagger \tau_a \widetilde{L} 
H^\dagger \tau_a H
+\dots ~.
\label{eq:quart2}
\end{align}
We again neglect the couplings of the DM with the first/third-generation leptons and assume all of the couplings to be real. 

Below the electroweak symmetry breaking scale, the mass terms become 
\begin{align}
  {\cal L}_{\rm mass} 
  &= 
  - 
  \left( \frac{1}{2} M_{F_S} \chi_S \chi_S  + \text{h.c.} \right)
  -\left(\widetilde{e}^*,~
 \widetilde{\bar{e}}\right) {\cal M}_e^2 
 \begin{pmatrix}
  \widetilde{e} \\ \widetilde{\bar{e}}^*
 \end{pmatrix}
 -M_{\tilde{\nu}}^2  \left|
 \widetilde{\nu}
 \right|^2 ~,
 \end{align}
with 
\begin{align}
  {\cal M}_e^2 
&=
\begin{pmatrix}
  M_{\tilde{L}}^2 +\frac{\lambda_L + \lambda^\prime_L}{2}v^2 
& \frac{v}{\sqrt{2}} a_H \\
 \frac{v}{\sqrt{2}} a_H & M_{\tilde{\bar{e}}}^2
 +\frac{\lambda_{\bar{e}}}{2} v^2 
\end{pmatrix}
~,
\qquad 
M_{\tilde{\nu}}^2  = M_{\tilde{L}}^2 +\frac{\lambda_L-\lambda^\prime_L}{2}
 v^2 ~.
\end{align}
The mass matrix ${\cal M}_e^2$ is diagonalized with a unitary matrix $U_e$ as 
\begin{equation}
  U_e^\dagger {\cal M}_e^2 U_e^{} 
 =
 \mathrm{diag}\left( M_{e_1}^2 ,M_{e_2}^2  \right)
 ~,
 \end{equation}
 with the mass eigenstates given by
 \begin{equation}
  \begin{pmatrix}
    \widetilde{e} \\ \widetilde{\bar{e}}^*
  \end{pmatrix}
 = U_e 
 \begin{pmatrix}
  \widetilde{e}_1 \\ \widetilde{e}_2
 \end{pmatrix}
 ~.
 \end{equation}
We show the interaction terms in the mass eigenbasis in Appendix~\ref{sec:masseigen2}.

\section{Muon \texorpdfstring{$g-2$}{TEXT}}
\label{sec:muongmi2}

\subsection{Model I}

The anomalous magnetic dipole moment induced by the new particles in Model I is computed at one-loop level as follows:\footnote{We have adopted the same notation for the mass functions as in Ref.~\cite{Calibbi:2018rzv}. } 
\begin{align}
  \Delta a_\mu = &- \frac{m_\mu}{8\pi^2 M_{\tilde{e}}^2 } \sum_{i = 1,2,3} M_{\chi_i} \mathrm{Re} \left[ y_1 y_2 \left( V_\chi \right)_{1i} \left( V_\chi \right)_{2i} \right] f_{LR}^S \biggl(\frac{M_{\chi_i}^2}{M_{\tilde{e}}^2}\biggr) \nonumber \\[2pt] 
  &- \frac{m_\mu^2}{8\pi^2 M_{\tilde{e}}^2 } \sum_{i = 1,2,3}
  \left[ \left| y_1 \left( V_\chi \right)_{1i}\right|^2 + \left| y_2 \left( V_\chi \right)_{2i}\right|^2 \right]
  f_{LL}^S \biggl(\frac{M_{\chi_i}^2}{M_{\tilde{e}}^2}\biggr) \nonumber \\[2pt] 
  &+ \frac{m_\mu^2 |y_2|^2 }{8\pi^2 M_{\tilde{\nu}}^2 } 
  f_{LL}^F \biggl(\frac{M_{F_D}^2}{M_{\tilde{\nu}}^2}\biggr) ~, 
  \label{eq:delamu1}
\end{align}
where $m_\mu$ is the muon mass and 
\begin{align}
  f_{LR}^S (x) &= \frac{1 - x^2 + 2x \ln x}{2(1-x)^3} ~, \\ 
  f_{LL}^F (x) &= \frac{2+ 3x-6x^2+x^3 + 6 x \ln x}{12 (1-x)^4} ~, \\ 
  f_{LL}^S (x) &= \frac{1 - 6 x + 3 x^2 + 2 x^3 - 6 x^2 \ln x}{12 (1-x)^4} ~.
\end{align}
Note that the first term in Eq.~\eqref{eq:delamu1} generically dominates the second and third terms which  are additionally suppressed by the small muon mass.

\subsection{Model II}

For Model II, we have 
\begin{align}
  \Delta a_\mu = &- \frac{m_\mu M_{F_S}}{8\pi^2 } \sum_{i = 1,2}
  \frac{1}{M_{e_i}^2 } \mathrm{Re} \left[ y_1 y_2 \left( U_e \right)^*_{1i} \left( U_e \right)_{2i} \right]f_{LR}^S \biggl(\frac{M_{F_S}^2}{M_{e_i}^2}\biggr) \nonumber \\[2pt] 
  &- \frac{m_\mu^2}{8\pi^2 } \sum_{i = 1,2}
  \frac{1}{M_{e_i}^2 }\left[ \left| y_1 \left( U_e \right)^*_{1i}\right|^2 + \left| y_2 \left( U_e\right)_{2i}\right|^2 \right]
  f_{LL}^S \biggl(\frac{M_{F_S}^2}{M_{{e}_i}^2}\biggr)~,
\end{align}
where the first term tends to dominate the second term as in the previous case.

\section{DM direct detection}
\label{sec:direct_detection}

In both Models I \& II, the DM can scatter off nuclei on the Earth via interactions with SM particles. Such scattering events can be probed in DM direct search experiments. However, the detection rate of DM is found to be quite different between these two models. In Model I,  DM particles interact with nucleons through the tree-level exchange of the Higgs boson, yielding a relatively large DM-nucleon scattering cross section. In Model II, the  DM-nucleon scattering is induced at  one-loop level since  there is no tree-level coupling between  DM and the Higgs field. Consequently, the scattering cross section  is  highly suppressed and beyond the reach of DM direct detection experiments.

First, we compute the SI DM-nucleon scattering cross section in Model I. We focus on SI interactions since the experimental limits are  much stronger than for the spin-dependent (SD) case.  The SI DM-nucleon interaction in this model is induced by the tree-level Higgs-boson exchange\footnote{Since the DM candidate  in this model has SU(2)$_L$ doublet components $\xi_D$ and $\eta_D$, the SI DM-nucleon scattering can also occur through the exchange of electroweak gauge bosons at the loop level. However, the contribution of such processes is found to be negligibly small for an SU(2)$_L$ doublet DM particle~\cite{Hisano:2011cs, Hisano:2015rsa}, and therefore we can safely ignore this contribution in the present case. } and described by the effective interaction of the form 
\begin{equation}
  \mathcal{L}_N = f_N \overline{\psi_1^0} \psi^0_1 \bar{N}N ~,
\end{equation}
where $\psi^0_1$ is the DM field represented in terms of a four-component Majorana fermion (see Eq.~\eqref{eq:dmmaj}), $N = p, n$ stands for nucleons, and $f_N$ is the effective coupling computed as~\cite{Shifman:1978zn} 
\begin{equation}
  f_N = m_N \biggl[\sum_{q = u, d,s} f_q f^{(N)}_{T_q} + \frac{2}{27} \sum_{Q = c, b, t} f_Q f^{(N)}_{T_G}\biggr] ~,
  \label{eq:fn}
\end{equation}
where $m_N$ is the nucleon mass and $f_q$ is the DM-quark coupling 
\begin{equation}
  f_q = 
  \frac{1}{2\sqrt{2} v m_{h}^2} \left[ \left( C_{\chi h L} \right)_{11} + \left( C_{\chi h R} \right)_{11}  \right] ~,
\end{equation}
with $m_h$ the Higgs-boson mass and $(C_{\chi h L/R} )_{ij}$ given in Eq.~\eqref{eq:cchih}.  $f^{(N)}_{T_q} \equiv \langle N | m_q \bar{q}q |N\rangle /m_N$ are the nucleon matrix elements for the quark operators, for which we use the values obtained in a recent compilation~\cite{Ellis:2018dmb}: $f^p_{T_u} = 0.018(5)$, $f^n_{T_u} = 0.013(3)$, $f^p_{T_d} = 0.027(7)$, $f^n_{T_d} = 0.040(10)$, $f^p_{T_s} = f^n_{T_s} = 0.037(17)$. $f_{T_G}^{(N)}$ corresponds to the gluon contribution to the nucleon mass given by $f_{T_G}^{(N)} \equiv 1 - \sum_{q = u,d,s} f^{(N)}_{T_q} $. The SI DM-nucleon scattering cross section is then given by 
\begin{equation}
  \sigma_{\mathrm{SI}}^{(N)} = \frac{4}{\pi} \biggl(\frac{m_N M_{\mathrm{DM}}}{m_N + M_{\mathrm{DM}}}\biggr)^2 f_N^2\,,
  \label{eq:sigsiN}
\end{equation}
with $M_{\mathrm{DM}} \equiv M_{\chi_1}$.

Finally, in Model II there is no tree-level coupling between the DM and the Higgs field. This is induced at the one-loop level and thus suppressed by a factor of $\sim y_i^2/(4\pi)^2$. For the parameter choice adopted in the following analysis, this factor is $\sim 10^{-3}$, and hence the resultant scattering cross section is suppressed by a factor of $\sim 10^{-6}$. This leads to values of $\sigma_{\mathrm{SI}}^{(N)}$ out of reach of future DM direct detection experiments.

\section{DM capture in NSs}
\label{sec:DMheating}

NSs are promising cosmic laboratories to probe scenarios such as the ones discussed in this work via DM accretion in their cores.
These compact stars create strong gravitational potentials, attracting DM particles. These DM particles become mildly relativistic when they approach the NS. When DM particles traverse the NS, they might lose considerable amount of their kinetic energy by scattering off the stellar material~\cite{Goldman:1989nd}. In fact, the DM particle  loses its whole initial kinetic energy after a single scattering if its  mass is $\lesssim 10^3~\rm TeV$. 
As we see in Sec.~\ref{sec:results}, the favored range of the DM mass in our models is $\sim 1~\rm TeV$, and therefore we can safely assume that DM  particles  are captured by a NS after a single scattering inside the NS. 

The regime in which we compute the DM capture rate depends on the DM scattering cross section.  For instance, if the DM scattering cross section is above a threshold cross section, $\sigma_{\rm th}$,  every DM particle traversing the NS is captured, \textit{i.e.}, the ``maximum capture probability'' is realized. This threshold cross section has been evaluated in the literature and depends on the NS mass, the NS equation of state, and the particle species scattered by the DM.

In the models presented in this work,  DM particles predominantly interact with nucleons and muons. The threshold cross sections for the interaction with such  particles are estimated to be $\sigma^N_{\mathrm{th}} \simeq [1.7 \times 10^{-45},1.4 \times 10^{-44}]~\mathrm{cm}^2$ and $\sigma^\mu_{\mathrm{th}} \simeq  8 \times 10^{-44}~\mathrm{cm}^2$ for the case of interactions with neutrons~\cite{Bell:2020jou,Anzuini:2021lnv} and muons~\cite{Bell:2020lmm} respectively. In this work,  we regard these results  as representative values. We will see in Sec.~\ref{sec:results} that in the favored parameter regions of our models, the DM-neutron or DM-muon scattering cross section is much larger than these threshold values. In this case, the DM capture rate is given by that in the geometric limit~\cite{Goldman:1989nd, Kouvaris:2007ay, Bell:2018pkk}
\begin{equation}
   C_{G}=\frac{\pi R_{\rm NS}^2(1-B(R_{\rm NS}))}{v_{\rm NS} B(R_{\rm NS})} \cdot \frac{\rho_{\mathrm{DM}}}{M_{\mathrm{DM}}} \cdot \mathrm{Erf}\left(\sqrt{\frac{3}{2}}\frac{v_{\rm NS}}{v_d}\right) ~,
   \label{eq:cg}
\end{equation}
where $R_{\mathrm{NS}}$ is the NS radius, $v_{\rm NS}$ is the NS speed (with respect to our Galaxy), $v_d$ is the DM velocity dispersion, $\rho_{\mathrm{DM}}= 0.42 ~\mathrm{GeV}\cdot \mathrm{cm}^{-3}$ is the DM local energy density~\cite{Pato:2015dua}, and $B(R_{\mathrm{NS}})$ is the time component of the metric on the NS surface, which is given by 
\begin{equation}
  B(R_{\mathrm{NS}})=1-\frac{2GM_{\mathrm{NS}}}{R_{\mathrm{NS}}}~, 
  \label{eq:brns}
\end{equation}
where $G$ is the gravitational constant and $M_{\mathrm{NS}}$ is the NS mass. Notice that this capture rate is independent of the DM interactions and determined solely by the NS properties and the DM distribution. 

On the other hand, if the DM scattering cross section is below  the threshold cross section, the capture rate is suppressed accordingly, with a suppression factor approximately given by $\sim \sigma/\sigma_{\mathrm{th}}$. Finally, when  the DM capture is of the order of the threshold cross section, the NS opacity should be taken into account in the capture rate computation. See, \textit{e.g.}, Refs.~\cite{Bell:2020lmm,Bell:2020jou,Bell:2020obw} for more detailed and complete treatments for the cases where the NS cannot be regarded as an optically-thick object.

\subsection{DM heating in NSs}
DM particles captured by the NS eventually annihilate in the NS core, giving their energy (including their rest energy) to the NS as heat~\cite{Kouvaris:2007ay}. For old NSs of our interest ($\sim 10^7$~years), the DM annihilation-capture equilibrium has already been achieved, and thus the DM annihilation rate is equal to $C_G/2$ in Eq.~\eqref{eq:cg}. The heating luminosity, observed at the distance, due to the presence of DM in the NS, is estimated to be
\begin{equation}
  L_H^{\infty} =  B(R_{\mathrm{NS}}) C_G M_{\mathrm{DM}} \left[ \chi + (\gamma -1) \right] ~,
  \label{eq:lh}
\end{equation}
where $\chi$ is the fraction of the annihilation energy transferred to heat and $\gamma = B(R_{\mathrm{NS}}) ^{-1/2}$ is the Lorentz factor of the incoming DM particle. The $(\gamma-1)$ factor represents the contribution of the DM kinetic energy~\cite{Baryakhtar:2017dbj}. Note that this heating luminosity does not depend explicit on the DM mass $M_{\rm DM}$, since $C_G \propto M_{\mathrm{DM}}^{-1}$ as shown in Eq.~\eqref{eq:cg}.

In the standard NS cooling theory,\footnote{The standard NS cooling theory is known to be consistent with the temperature observations of young and middle-age NSs~\cite{Potekhin:2020ttj}. For the latest data of NS temperatures, see Ref.~\cite{tempdata}.} a NS cools mainly through the photon emission at late times. The photon luminosity is given by $L_\gamma = 4\pi R_{\mathrm{NS}}^2 \sigma_{\mathrm{SB}} T_s^4$, where $\sigma_{\mathrm{SB}}$ is the Stefan-Boltzmann constant and $T_s$ is the NS surface temperature. In the presence of a heating source, the photon emission luminosity eventually balances with the heating luminosity,
\begin{equation}
    L_H^\infty = L_\gamma^\infty \equiv B(R_{\mathrm{NS}}) L_\gamma~.
\end{equation} 
From this equation, we obtain the late-time surface temperature, which is typically $\simeq \text{a few} \times 10^3$~K. In the case of the standard NS cooling without heating sources, NSs cool down to $\lesssim 10^{3}$~K after $\sim 5 \times 10^6$~years. Hence, an observation of an old NS with $T_s \simeq \text{a few} \times 10^3$~K will be a strong hit for the operation of  DM heating. As discussed in Ref.~\cite{Baryakhtar:2017dbj}, future infrared telescopes, such as the JWST~\cite{Gardner:2006ky}, are expected to be sensitive to this size of temperature for nearby NSs. However, we note that this prospect is based on the standard NS cooling theory, but its applicability to old NSs has not been established yet. 

Recent observations found that there are several isolated old NSs that have temperatures significantly higher than the prediction in the standard NS cooling theory as well as that with the DM heating~\cite{Kargaltsev:2003eb, Mignani:2008jr, Durant:2011je,  Rangelov:2016syg,  Pavlov:2017eeu, Abramkin:2021fzy} (see Ref.~\cite{Yanagi:2019vrr} for a list of such observations). This implies the presence of extra heating sources other than the DM heating~\cite{Gonzalez:2010ta}, such as the effect of non-equilibrium beta processes (dubbed as the rotochemical heating)~\cite{Reisenegger:1994be, 1992A&A...262..131H, 1993A&A...271..187G, Fernandez:2005cg, Villain:2005ns, Petrovich:2009yh, Pi:2009eq, Gonzalez-Jimenez:2014iia, Yanagi:2019vrr}, the vortex creep heating~\cite{1984ApJ...276..325A, 1989ApJ...346..808S, 1991ApJ...381L..47V, 1993ApJ...408..186U, VanRiper:1994vp, Larson:1998it}, and the rotationally-induced deep crustal heating \cite{Gusakov:2015kaa}. Such  additional heating sources could hide the DM heating effect, depending on the heating mechanism and NS properties.  In the case of rotochemical heating, for instance, the DM heating effect is always concealed for millisecond pulsars but can still be observed in ordinary pulsars with an initial period $\gtrsim 10$~ms~\cite{Hamaguchi:2019oev}. The detailed assessment of the observational probability of the DM heating is beyond the scope of the present work, and we simply assume that it is possible to detect the signature of the DM heating in nearby NSs in future observations.

\subsection{DM-nucleon scattering}
\label{eq:dm-nucleon_scattering}

As discussed in Sec.~\ref{sec:direct_detection}, the SI DM-nucleon scattering occurs at the tree and loop levels in Model I and II, respectively. Additionally, in NSs, the SD scattering is also relevant. In Model I, such scattering processes are induced by the tree-level exchange of the $Z$ boson. The resultant scattering cross section is,\footnote{Here, we show the scattering cross section in the non-relativistic limit, which is valid for the DM-nucleon scattering on the Earth but can be modified by an $\mathcal{O} (1)$ factor for that on NSs due to a large momentum transfer~\cite{Anzuini:2021lnv}. We take account of this potential modification as theoretical uncertainty when we compare the prediction for the SD DM-nucleon scattering cross section with the threshold cross section in Sec.~\ref{sec:results}.}   
\begin{equation}
  \sigma_{\mathrm{SD}}^{(N)} = \frac{12}{\pi} \biggl(\frac{m_N M_{\mathrm{DM}}}{m_N + M_{\mathrm{DM}}}\biggr)^2 a_N^2\,,
  \label{eq:sigsdN}
\end{equation}
where $a_N$ denotes the SD DM-nucleon coupling and is given by 
\begin{equation}
  a_N  = \sum_{q = u,d,s} d_q \Delta q_N ~.
\end{equation}
where $\Delta q_N$ are the spin fractions: $2 s_\mu \Delta q_N \equiv \langle N | \bar{q} \gamma_\mu \gamma_5 q | N \rangle$, with $s_\mu$ the spin four-vector of the nucleon. For $\Delta q_N$, we use the values obtained by QCD lattice simulations: $\Delta u_p = 0.862(17)$, $\Delta u_n = - 0.424(16)$, $\Delta d_p = - 0.424(16)$, $\Delta d_n = 0.862(17)$, $\Delta s_p = \Delta s_n = -0.0458(73)$~\cite{Alexandrou:2019brg}.
In terms of the SD DM-quark couplings, the interaction is of the form
\begin{equation}
  \mathcal{L}_q^{(\mathrm{SD})} = d_q \overline{\psi_1^0} \gamma_\mu \gamma_5 \psi^0_1 \, \bar{q} \gamma^\mu \gamma_5 q ~. 
  \label{eq:lqsd}
\end{equation}
 The coupling $d_q$ in Eq.~\eqref{eq:lqsd} is calculated in Model I as 
\begin{align}
  d_q = - \frac{1}{2v^2} \left[ \left|\left(V_\chi\right)_{21}\right|^2 - \left|\left(V_\chi\right)_{31}\right|^2 \right] T_q^3 ~,
\end{align}
with $T_q^3 = 1/2$ for $q = u$ and $-1/2$ for $q = d,s$.

In Model II, the SD scattering is induced at  one-loop level and, as we discuss in Sec.~\ref{sec:direct_detection}, its cross section is suppressed by a factor of $\sim 10^{-6}$ compared to the tree-level induced one.  As we see below, the cross section of the latter is $\mathcal{O}(10^{-42})~\mathrm{cm}^2$ in the parameter regions where the muon $g-2$ discrepancy can be explained. Therefore it is unlikely that the loop-induced SD scattering cross section exceeds the threshold cross section. For this reason, we do not evaluate the SD scattering cross section for Model II.

\subsection{DM--\texorpdfstring{$\mu$}{TEXT} scattering}
\label{sec:dm-muon_scattering}

We now compute the DM-muon scattering cross sections. The differential scattering cross section is 
\begin{equation}
  \frac{d \sigma_{\chi \mu}}{dt} = \frac{1}{16 \pi \lambda (s, M_\mathrm{DM}^2, m_\mu^2)} \cdot \frac{1}{4} \sum_{\mathrm{spins}} |\mathcal{A}|^2
  ~,
  \label{eq:dmmudiffcross}
\end{equation}
where $t \equiv (p_\chi - p_\chi^\prime)^2$  and  $s \equiv (p_\chi + p_\mu)^2$; with $p_\chi$, $p_\chi^\prime$ and $p_\mu$ the incoming and outgoing DM four-momenta, and  the initial-state muon four-momentum, respectively. $\lambda (x, y, z) \equiv x^2 + y^2 + z^2 - 2 xy - 2 yz -2 zx$ is the K\"{a}ll\'{e}n function while $\mathcal{A}$ is the invariant scattering amplitude. The kinematically allowed range of the $t$ variable is 
\begin{equation}
  - \frac{\lambda (s, M_\mathrm{DM}^2, m_\mu^2)}{s} \leq t \leq 0 ~.
  \label{eq:trange}
\end{equation}

We compute the scattering cross section~\eqref{eq:dmmudiffcross} using the following approximation. First, we replace the $s$ variable with the averaged value with respect to the momentum direction: 
\begin{equation}
  s \to \bar{s} \equiv M_{\mathrm{DM}}^2 + 2 E_\chi E_\mu + m_\mu^2 ~,
\end{equation}
where $E_\chi = p_\chi^0$ and $E_\mu = p_\mu^0$. The DM energy is approximated by $E_\chi = M_{\mathrm{DM}}/\sqrt{B (R_{\mathrm{NS}})}$. The muon energy depends on the NS equation of state and thus is highly uncertain.  In the following analysis, we vary $E_\mu$ in the range $[m_\mu, m_\mu/\sqrt{B (R_{\mathrm{NS}})}]$, and regard the resultant change in $\sigma_{\chi \mu}$ as the theoretical error in our estimation. For more precise treatment, see Ref.~\cite{Bell:2020lmm}. 

Notice that $E_\mu \ll E_{\chi}$ 
in the present setup, and thus we can expand all of the quantities in terms of $E_\mu$. For example, $s \simeq M_{\mathrm{DM}}^2$ and $\bar{s} - M_{\mathrm{DM}}^2 \simeq  2 E_\chi E_\mu$. We also find 
\begin{align}
  \lambda (\bar{s}, M_{\mathrm{DM}}^2, m_\mu^2) = \left[ \bar{s} - \left( M_{\mathrm{DM}}^2 + m_\mu^2 \right) \right]^2 - 4 M_{\mathrm{DM}}^2  m_\mu^2 
  \simeq  \left( \bar{s} - M_{\mathrm{DM}}^2 \right)^2 - 4 M_{\mathrm{DM}}^2  m_\mu^2 ~,
\end{align} 
is an $\mathcal{O}(E_\chi^2 E_\mu^2)$ quantity. It then follows, from Eq.~\eqref{eq:trange}, that $t$ is  $\mathcal{O} (E_\mu^2)$. 

The leading-order contribution to $|\mathcal{A}|^2$ is $\mathcal{O} (E_\mu^2/E_\chi^2)$. Hence, it is sufficient to keep the terms up to the first order in the $t$ expansion: 
\begin{equation}
  \frac{1}{4} \sum_{\mathrm{spins}} |\mathcal{A}|^2 \simeq \alpha_0 + \alpha_1 \, (- t)~,
  \label{eq:ampexp}
\end{equation}
where $\alpha_0$ and $\alpha_1$ are independent of $t$. The expressions of $\alpha_0$ and $\alpha_1$ are quite lengthy, so we show them in Appendix~\ref{sec:amplitude}. By integrating Eq.~\eqref{eq:dmmudiffcross} with respect to the $t$ variable over the range in Eq.~\eqref{eq:trange}, we obtain
\begin{align}
  \sigma_{\chi \mu} &= \frac{1}{16 \pi \bar{s}} \left[ \alpha_0 + \frac{\lambda (\bar{s}, M_\mathrm{DM}^2, m_\mu^2)}{2 \bar{s}} \alpha_1 \right]  \nonumber \\ 
  & \simeq   \frac{1}{16 \pi M_{\mathrm{DM}}^2} \left[ \alpha_0 + \frac{2 \left( E_\chi^2 E_\mu^2 - M_{\mathrm{DM} }^2 m_\mu^2 \right)}{ M_{\mathrm{DM}}^2} \alpha_1 \right] \nonumber \\ 
  & \simeq   \frac{1}{16 \pi M_{\mathrm{DM}}^2} \left[ \alpha_0 + \frac{2 \left(  E_\mu^2 - B(R_{\mathrm{NS}}) \, m_\mu^2 \right)}{ B(R_{\mathrm{NS}}) } \alpha_1 \right] ~.
\end{align}

\section{Results}
\label{sec:results}
Having computed the SI (SD) DM-nucleon and DM muon cross sections, now we show the predictions for both, Model I and Model II in the present section.

It is in principle possible to fix a parameter of the models by requiring the thermal relic abundance of the DM to agree with the observed value of the DM density~\cite{Planck:2018vyg}. However, in the following analysis, we do not strictly impose this condition and just mention ball-park values of the DM mass that can satisfy this condition. Note that if the DM relic abundance is lower than the observed DM density,\footnote{Even in this case, the DM candidates in our models can still occupy the entire DM density if they are produced non-thermally as well. On the other hand, if the thermal relic abundance of the DM is larger than the observed value, we need to assume a non-trivial cosmological history to reduce the abundance, such as the late-time entropy production. }  the limits from DM direct detection experiments are relaxed and the heating luminosity in Eq.~\eqref{eq:lh} decreases accordingly.

\subsection{Model I (singlet-like)}

\begin{figure}
  \centering
  \subcaptionbox{\label{fig:mdm_Damu}
  DM mass dependence 
  }
  {\includegraphics[width=0.48\textwidth]{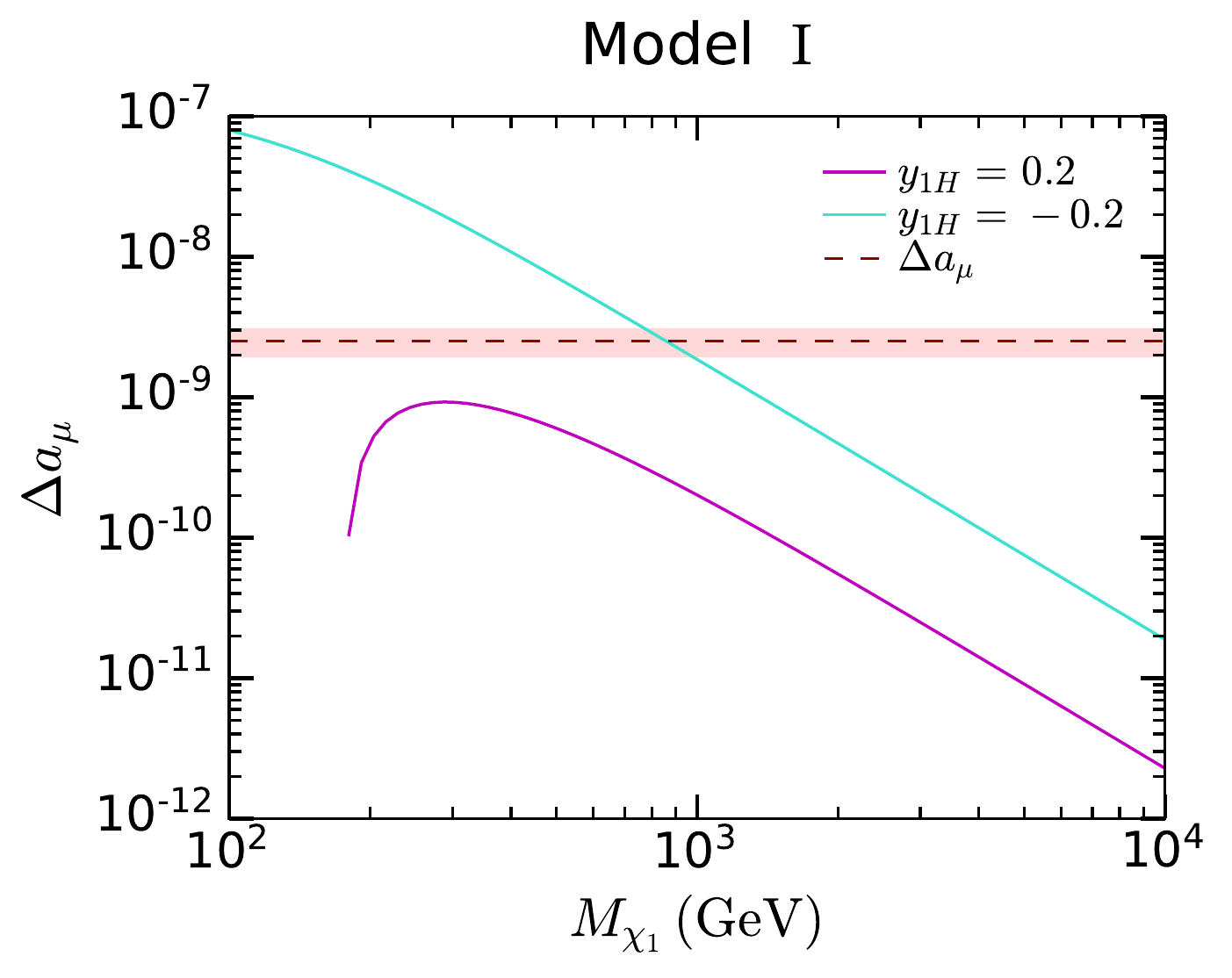}}
  \subcaptionbox{\label{fig:y1H_Da_mu}
  $y_{1H}$ dependence 
  }
  { 
  \includegraphics[width=0.48\textwidth]{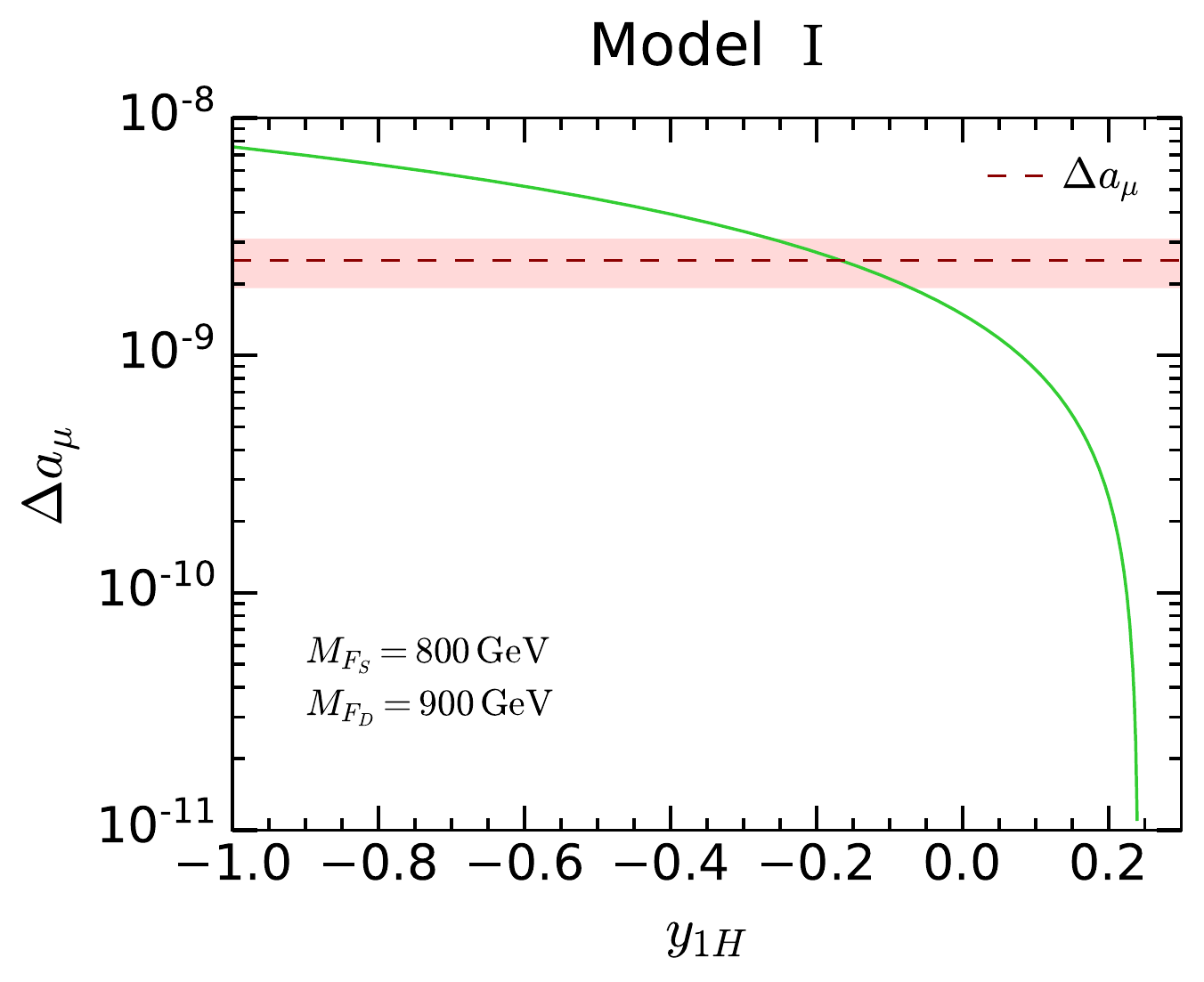}}
  \caption{
    In the left panel we show $\Delta a_\mu$ as a function of DM mass $M_{\chi_1}$ for two benchmark parameter points $y_{1H} = 0.2$ and $y_{1H} = -0.2$, with $M_{F_D}/M_{F_S} = 1.1$ and $M_{\tilde{L}}/M_{F_S} = 1.2$. In the right panel, we show $\Delta a_\mu$ as a function of  $y_{1H}$, fixing $M_{F_S} = 800~\mathrm{GeV}$, $ M_{F_D} = 900~\mathrm{GeV}$, and $  M_{\tilde{L}} = 1000\,\rm GeV$. The rest of the parameters in the both panels are set to be $y_1=y_2= \lambda_L = \lambda_L'=0.5$ and $ y_{2H} = 0.3$. The horizontal dashed line indicates the measured value of $\Delta a_\mu$, with its error indicated by the red band. 
  }
  \label{fig:Da_mu_singlet}
  \end{figure}
In this section we compute the DM interaction cross sections in Model I. In Fig.~\ref{fig:mdm_Damu}, we show $\Delta a_\mu$ as a function of the DM mass $M_{\chi_1}$ for $y_{1H} = 0.2$ and $-0.2$ in the magenta and cyan solid lines, respectively. The other parameters are fixed to be $y_{2H} = 0.3$, $y_1=y_2= \lambda_L = \lambda_L'=0.5$, $M_{F_D}/M_{F_S} = 1.1$, and $M_{\tilde{L}}/M_{F_S} = 1.2$. With this parameter choice, we have a singlet-like DM candidate since $M_{F_S} < M_{F_D}$. 
The horizontal dashed line indicates the measured value of $\Delta a_\mu$ in Eq.~\eqref{eq:delamu_exp}, with its error indicated by the red band. We see that for $y_{1H} = -0.2$ 
this model can explain the observed discrepancy in the muon $g-2$ if the DM mass is $\simeq 800$~GeV. It is also shown in Ref.~\cite{Calibbi:2018rzv} that the thermal relic abundance of $\chi_1$ can coincide with the observed DM density with this size of DM mass. Since the new particles in this model are not charged under SU(3)$_C$, they safely evade the current LHC limits for their masses $\gtrsim 400$~GeV~\cite{ATLAS:2019lff}. For the case where $y_{1H} = 0.2$, 
the predicted values of $\Delta a_\mu$ are always below the experimental value. To examine the dependence on $y_{1H}$, in Fig.~\ref{fig:y1H_Da_mu}, we plot $\Delta a_\mu$ as a function of $y_{1H}$. For this plot we have fixed   the parameters to $M_{F_S} = 800~\mathrm{GeV}$, $ M_{F_D} = 900~\mathrm{GeV}$, $  M_{\tilde{L}} = 1000~\mathrm{GeV}$, $y_1=y_2= \lambda_L = \lambda_L'=0.5$, and $ y_{2H} = 0.3$. This figure shows that $\Delta a_\mu$ decreases as $y_{1H}$ increases, and becomes negative for $y_{1H} \gtrsim 0.25$. 

\begin{figure}
  \centering
  \subcaptionbox{\label{fig:mdm_xsec_SI}
  DM mass dependence 
  }
  {\includegraphics[width=0.48\textwidth]{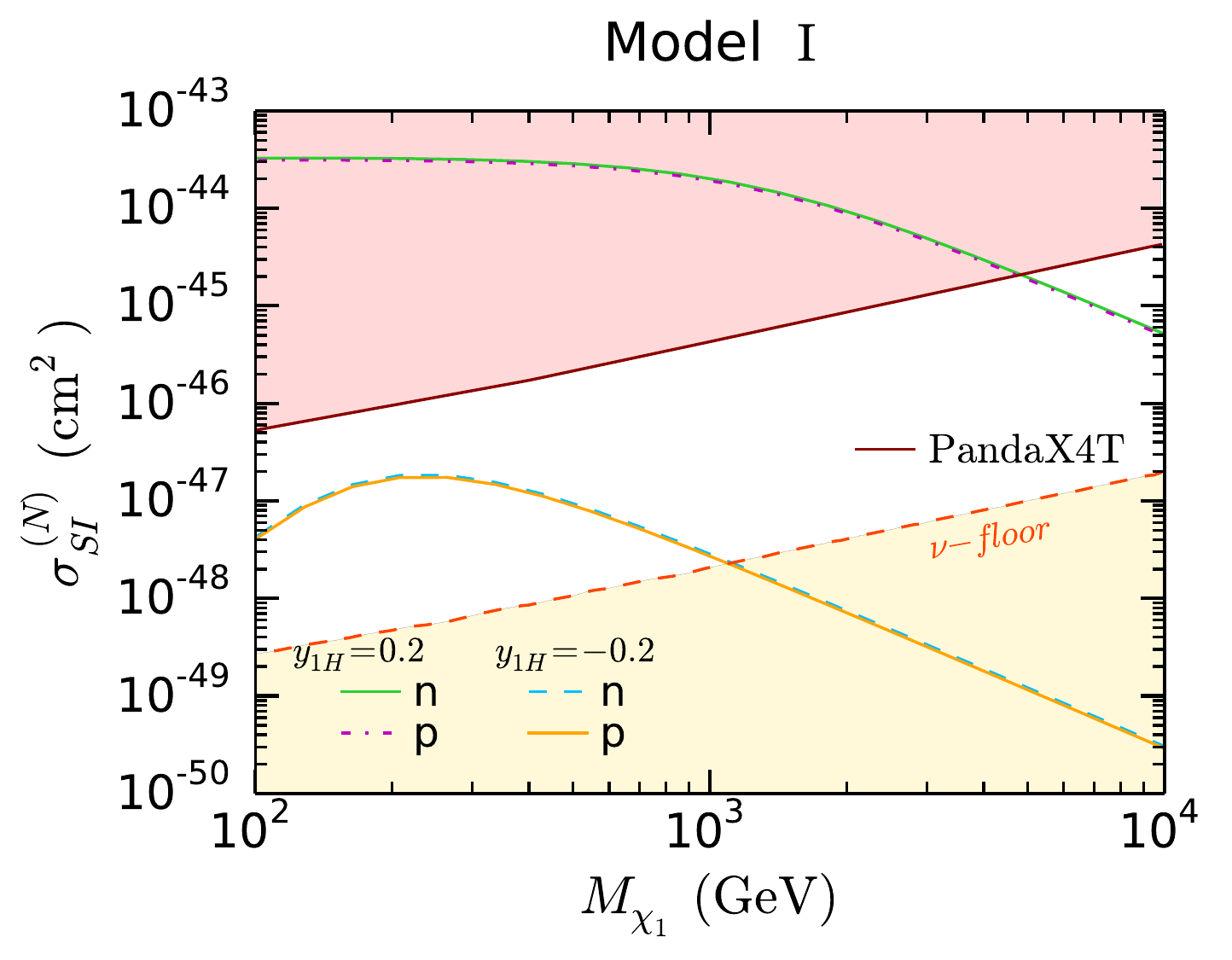}}
  \subcaptionbox{\label{fig:y1H_xsec_SI}
  $y_{1H}$ dependence 
  }
  { 
  \includegraphics[width=0.48\textwidth]{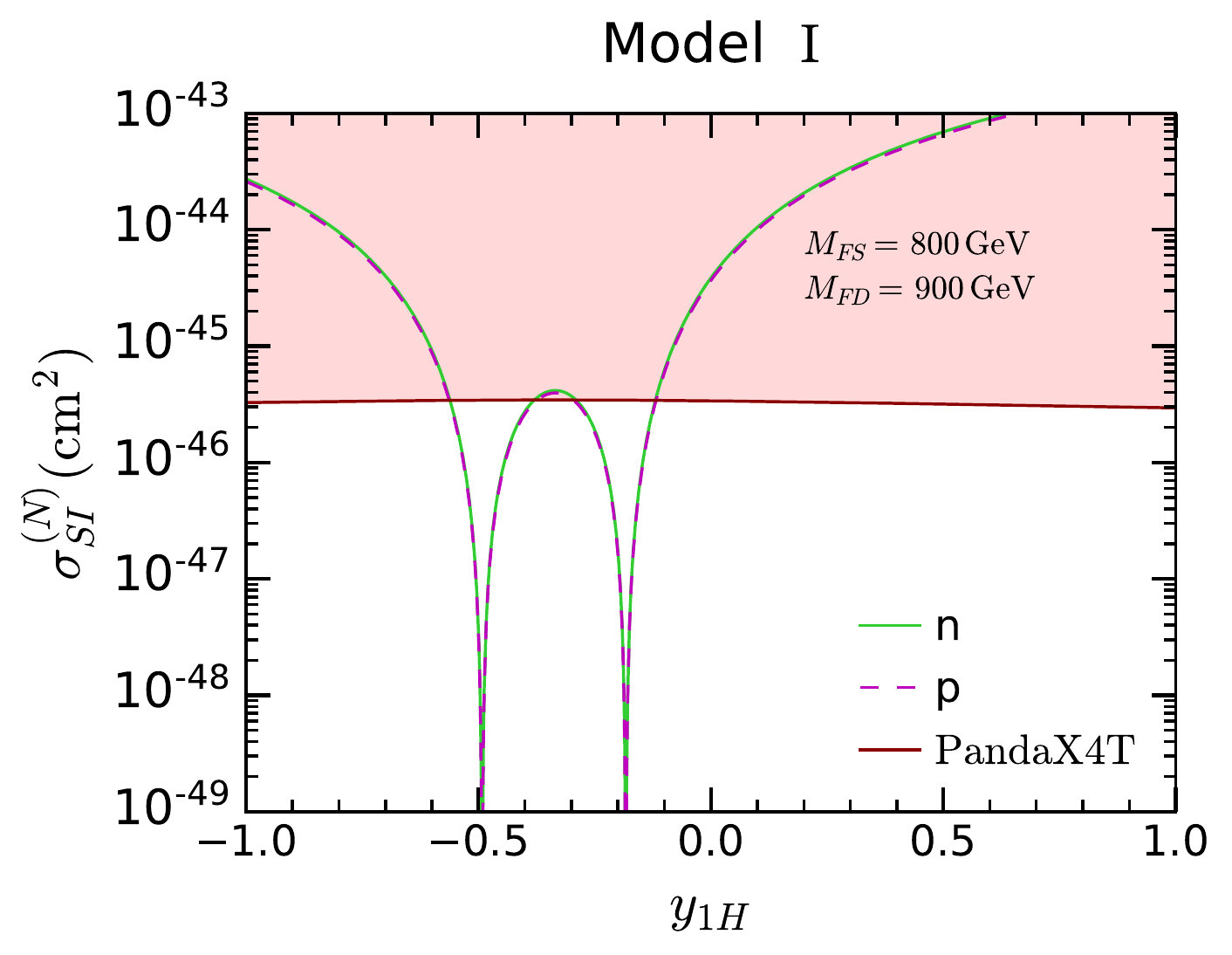}}
  \caption{
    The DM-nucleon SI scattering cross sections as functions of (a) the DM mass $M_{\chi_1}$ and (b) $y_{1H}$. The rest of the parameters are fixed as per in Fig.~\ref{fig:Da_mu_singlet}. The red shaded area is excluded by the PandaX4T experiment~\cite{PandaX-4T:2021bab} and the yellow shaded region corresponds to the neutrino floor~\cite{Billard:2021uyg}. 
  }
  \label{fig:DM_nucleon_SI}
\end{figure}

Fig.~\ref{fig:DM_nucleon_SI} shows the DM-nucleon SI scattering cross sections as functions of the DM mass $M_{\chi_1}$ (Fig.~\ref{fig:mdm_xsec_SI}) and $y_{1H}$ (Fig.~\ref{fig:y1H_xsec_SI}). The rest of the parameters are fixed as per in Fig.~\ref{fig:Da_mu_singlet}. The red shaded area is excluded by the PandaX4T experiment~\cite{PandaX-4T:2021bab} and the yellow shaded region is out of reach of the present detection strategy due to the neutrino background---the so-called neutrino floor~\cite{Billard:2021uyg}. From these plots, we see that for $y_{1H} = +0.2$, the SI scattering cross section is too large to evade the current experimental limit for $M_{\chi_1} \simeq 1~\rm TeV$. For $y_{1H} = -0.2$, on the other hand, the SI cross section is much smaller than the current limit, almost on the border of the neutrino floor for $M_{\chi_1} \simeq 1\,\rm TeV$. This indicates that it is difficult to test this case in the next-generation DM direct detection experiments. 

\begin{figure}
  \centering
  \subcaptionbox{\label{fig:mdm_xsec_SD}
  DM mass dependence 
  }
  {\includegraphics[width=0.48\textwidth]{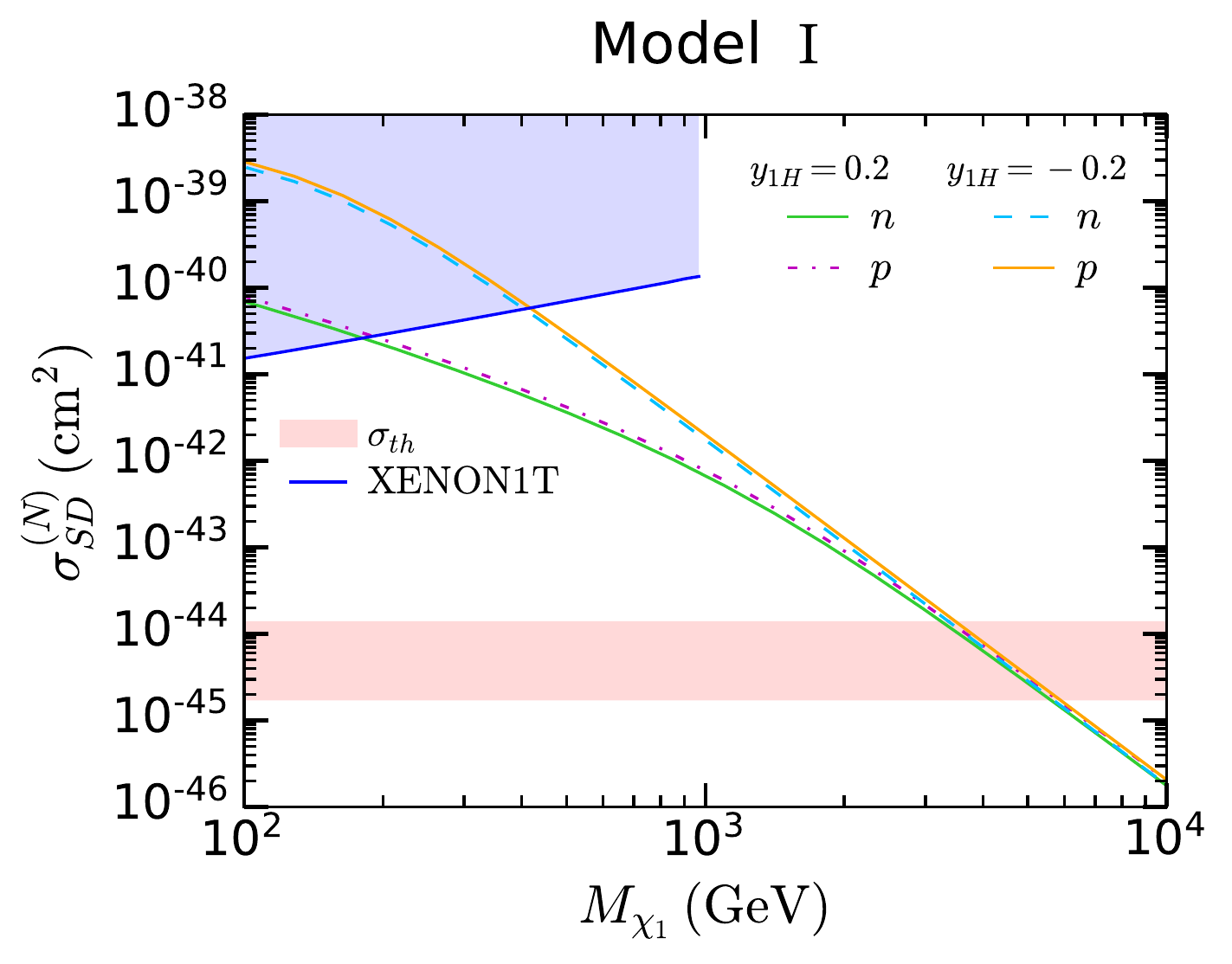}}
  \subcaptionbox{\label{fig:y1H_xsec_SD}
  $y_{1H}$ dependence 
  }
  { 
  \includegraphics[width=0.48\textwidth]{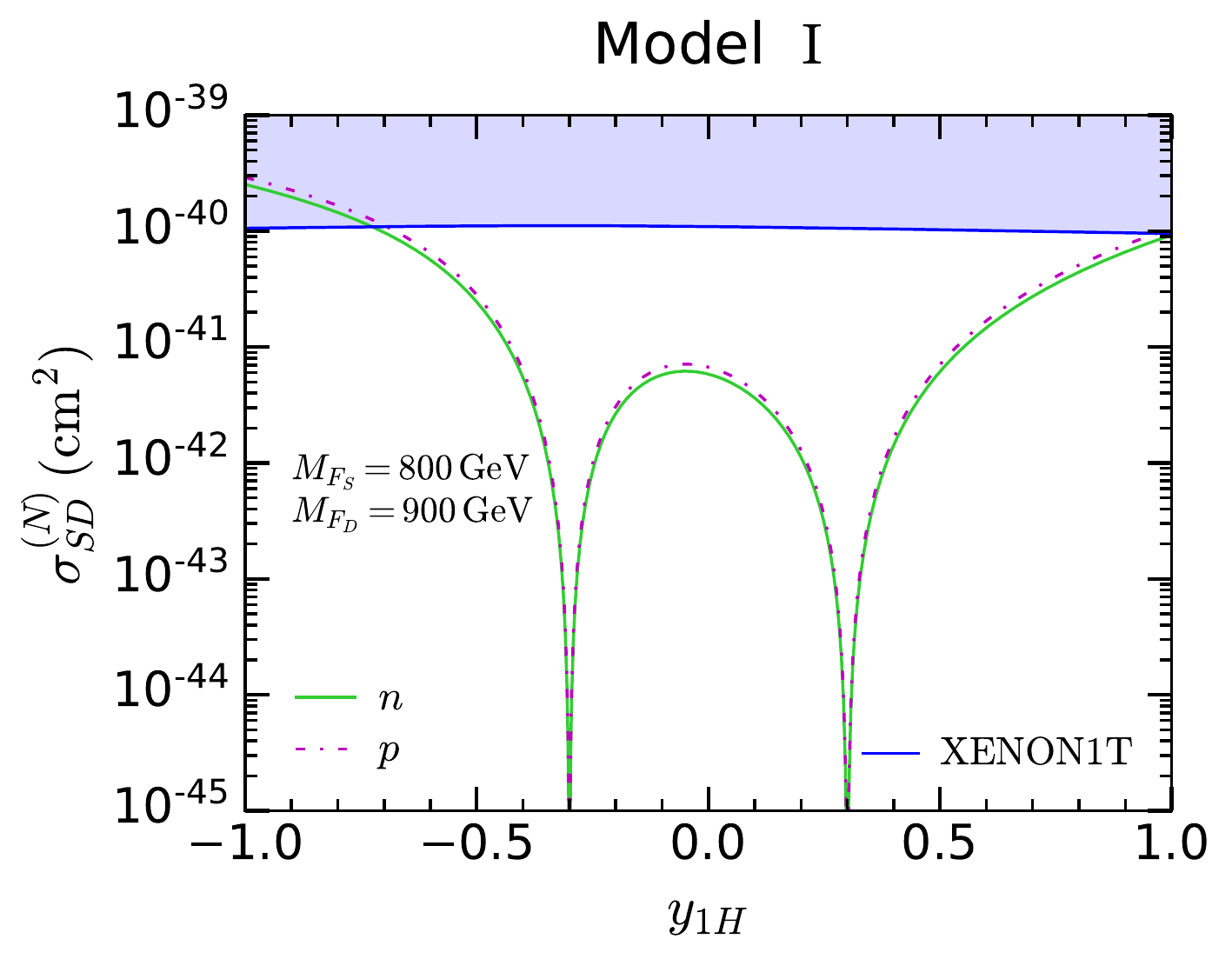}}
  \caption{The DM-nucleon SD scattering cross sections as functions of (a) the DM mass $M_{\chi_1}$ and (b) $y_{1H}$. The rest of the parameters are fixed as per in Fig.~\ref{fig:Da_mu_singlet}. The blue shaded area is excluded by the XENON1T experiment~\cite{XENON:2019rxp}. The horizontal red band represents the threshold cross section for neutron obtained in Refs.~\cite{Bell:2020jou,Anzuini:2021lnv}, $\sigma_{\mathrm{th}} \simeq [1.7 \times 10^{-45},1.4 \times 10^{-44}] \,~\mathrm{cm}^2$. }
  \label{fig:DM_nucleon_SD}
\end{figure}

In Fig.~\ref{fig:DM_nucleon_SD}, the DM-nucleon SD scattering cross sections are shown as functions of the DM mass $M_{\chi_1}$ (Fig.~\ref{fig:mdm_xsec_SD}) and $y_{1H}$ (Fig.~\ref{fig:y1H_xsec_SD}) with the same parameter choice as in Fig.~\ref{fig:Da_mu_singlet}. The blue shaded region is excluded by XENON1T~\cite{XENON:2019rxp} for neutron.\footnote{For proton, PICO-60~\cite{PICO:2019vsc} gives the most stringent limit, which is slightly weaker than the XENON1T bound on the DM-neutron SD scattering cross section.} We also show the threshold cross section for DM interactions with neutron obtained in Refs.~\cite{Bell:2020jou,Anzuini:2021lnv}, $\sigma_{\mathrm{th}} \simeq [1.7 \times 10^{-45},1.4 \times 10^{-44}] \,~\mathrm{cm}^2$, in the horizontal red band. As seen from these plots, the SD scattering cross sections are predicted to be smaller than the XENON1T limit~\cite{XENON:2019rxp} in the parameter regions where the muon $g-2$ discrepancy can be explained.  These cross sections are much larger than the threshold cross section $\sigma_{\mathrm{th}}$, and therefore the singlet-like DM candidate in Model I is efficiently captured by NSs.

\begin{figure}
  \centering
  \subcaptionbox{\label{fig:mdm_xsec_DM-muon}
  DM mass dependence 
  }
  {\includegraphics[width=0.48\textwidth]{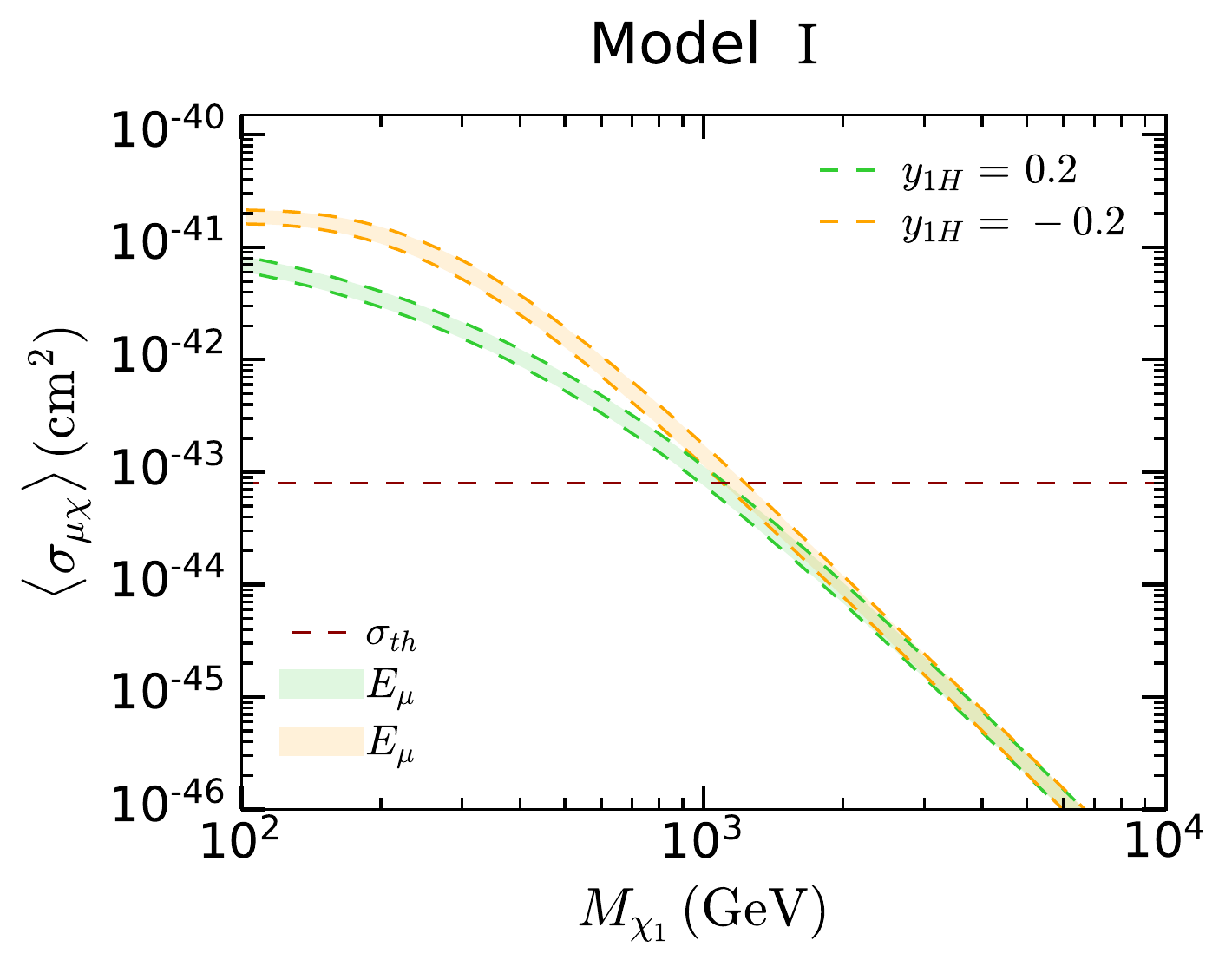}}
  \subcaptionbox{\label{fig:y1H_xsec_DM-muon}
  $y_{1H}$ dependence 
  }
  { 
  \includegraphics[width=0.48\textwidth]{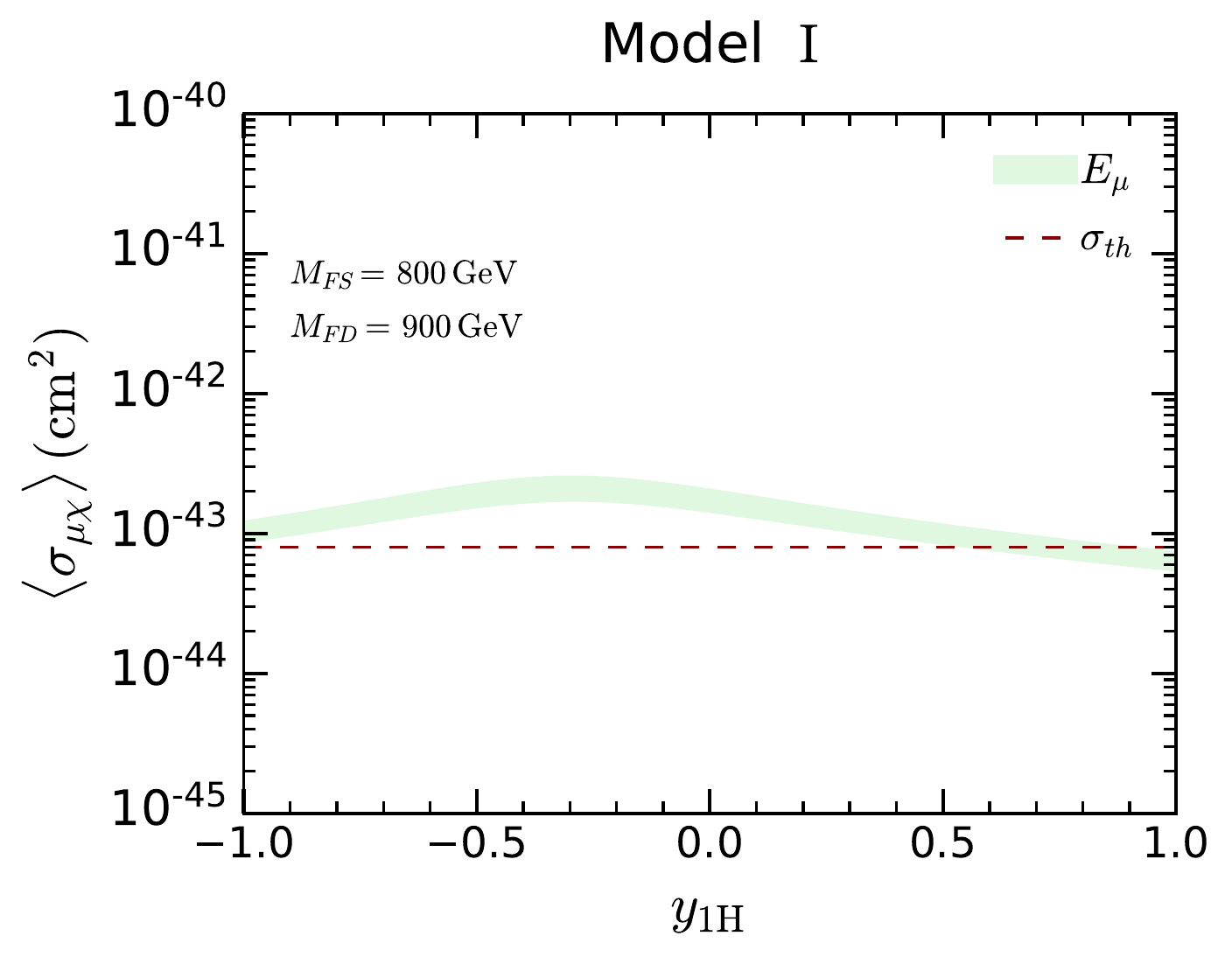}}
  \caption{The DM-muon scattering cross sections as functions of (a) the DM mass $M_{\chi_1}$ and (b) $y_{1H}$. The rest of the parameters are fixed as per in Fig.~\ref{fig:Da_mu_singlet}. $E_\mu$ is varied in the range $[m_\mu, m_\mu/\sqrt{B (R_{\mathrm{NS}})}]$ for $M_{\mathrm{NS}} = 1.5~M_{\odot}$ and $R_{\mathrm{NS}} = 12.593$~km and the resultant change in the cross section is indicated by the band. The horizontal red dashed line shows the threshold cross section for muon, $\sigma_{\mathrm{th}} \simeq  8 \times 10^{-44}~\mathrm{cm}^2$~\cite{Bell:2020lmm}.  }
  \label{fig:xsec_DM-muon}
\end{figure}

We also compute the DM-muon scattering cross sections adopting the same parameters as in Fig.~\ref{fig:Da_mu_singlet}. In this work, we fix the NS mass and radius to be $M_{\mathrm{NS}} = 1.5~M_{\odot}$ and $R_{\mathrm{NS}} = 12.593$~km, respectively, which give $B (R_{\mathrm{NS}}) = 0.648$ in Eq.~\eqref{eq:brns}.  The DM-muon scattering cross section is shown in Fig.~\ref{fig:xsec_DM-muon}. As we noted in Sec.~\ref{sec:dm-muon_scattering}, we vary $E_\mu$ in the range $[m_\mu, m_\mu/\sqrt{B (R_{\mathrm{NS}})}]$, and the resultant change in the cross section is indicated by the band. We find that the cross section is larger for  larger $E_\mu$.   The horizontal red dashed line shows the threshold cross section for muon, $\sigma_{\mathrm{th}} \simeq  8 \times 10^{-44}~\mathrm{cm}^2$~\cite{Bell:2020lmm}. We see that the DM-muon scattering cross section also exceeds the threshold cross section in the parameter range motivated by the muon $g-2$ anomaly. Finally, we  see  in Fig.~\ref{fig:y1H_xsec_DM-muon} that the DM-muon scattering cross section rarely depends on the coupling $y_{1H}$, in contrast to the DM-nucleon scattering.

\subsection{Model I (doublet-like)}

\begin{figure}
  \centering
  \subcaptionbox{\label{fig:mdm_Damu_2}
  DM mass dependence 
  }
  {\includegraphics[width=0.48\textwidth]{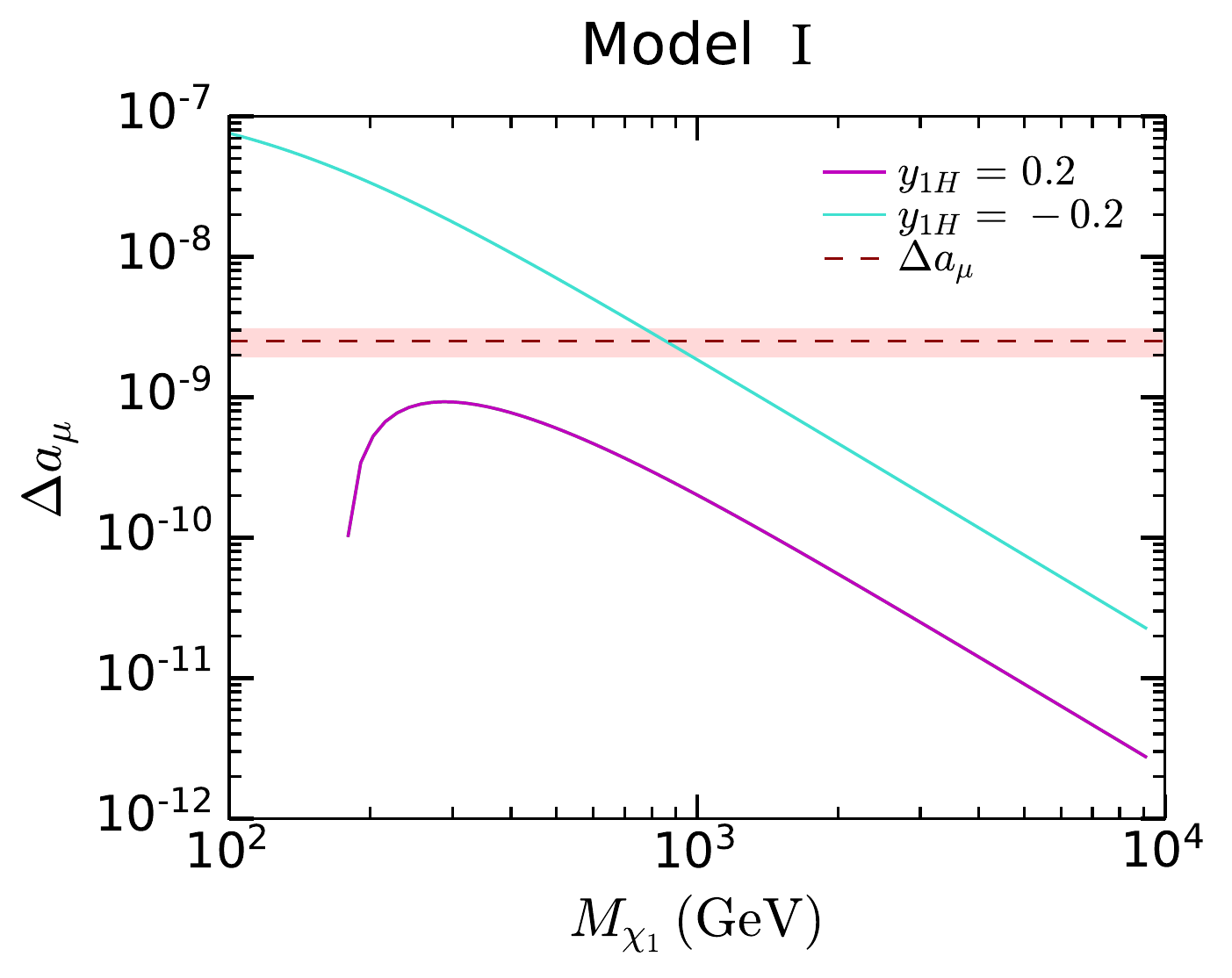}}
  \subcaptionbox{\label{fig:y1H_Da_mu_2}
  $y_{1H}$ dependence 
  }
  { 
  \includegraphics[width=0.48\textwidth]{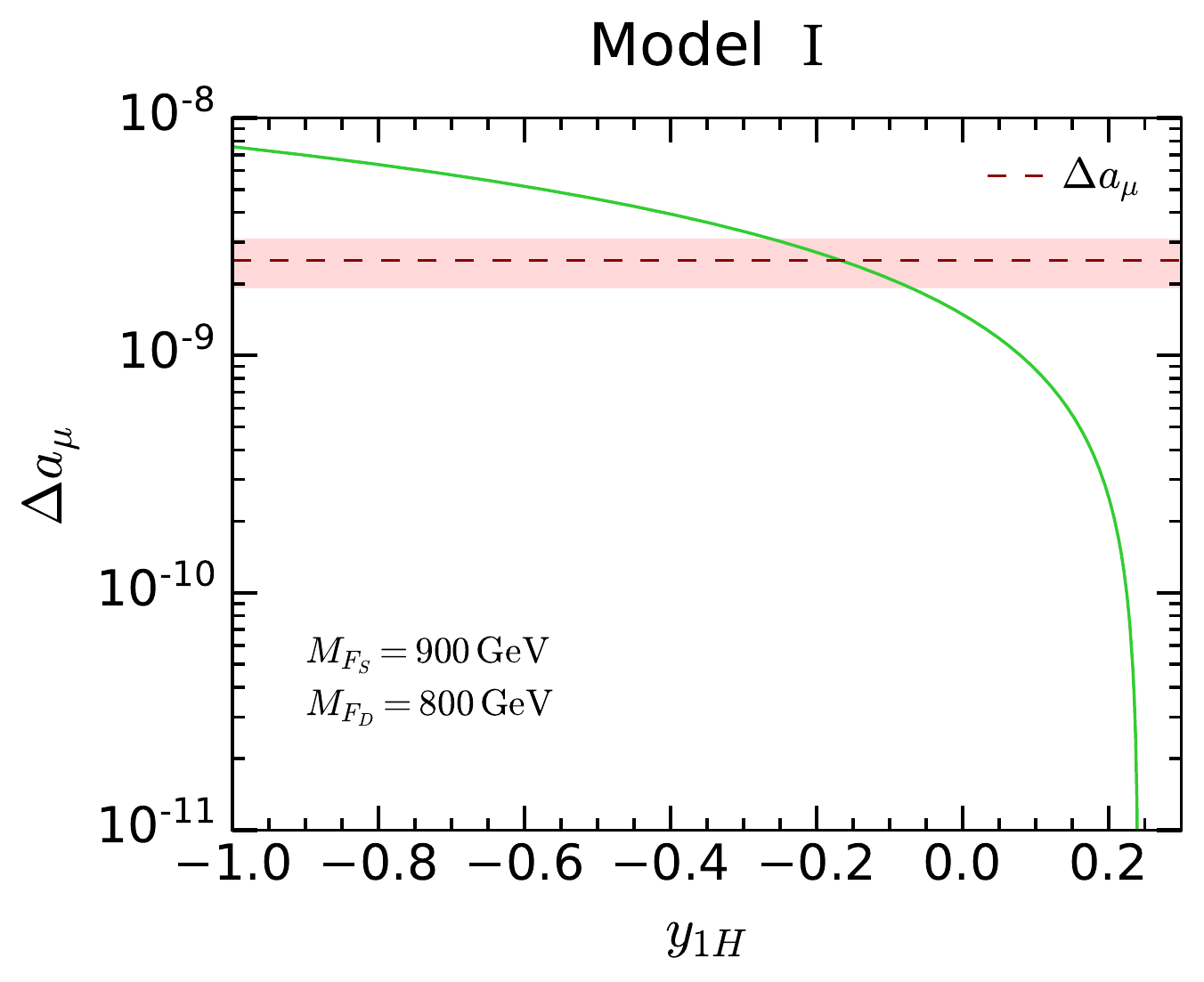}}
  \caption{
    $\Delta a_\mu$ as a function of (a) the DM mass $M_{\chi_1}$ for $y_{1H} = \pm 0.2$, $M_{F_S}/M_{F_D} = 1.1$, and $M_{\tilde{L}}/M_{F_D} = 1.2$; (b) $y_{1H}$ for $M_{F_D} = 800~\mathrm{GeV}$, $ M_{F_S} = 900~\mathrm{GeV}$, and $  M_{\tilde{L}} = 1000\,\rm GeV$. The rest of the parameters are set to be $y_1=y_2= \lambda_L = \lambda_L'=0.5$ and $ y_{2H} = 0.3$. The horizontal dashed line indicates the measured value of $\Delta a_\mu$, with its error indicated by the red band. 
  }
  \label{fig:Da_mu_doublet}
  \end{figure}

Next, we consider the case where $M_{F_D} < M_{F_S}$, \textit{i.e.}, the DM is doublet-like. In Fig.~\ref{fig:mdm_Damu_2}, we show $\Delta a_\mu$ as a function of the DM mass $M_{\chi_1}$ for $y_{1H} = 0.2$ and $-0.2$ in the magenta and cyan solid lines, respectively, with $M_{F_S}/M_{F_D} = 1.1$ and $M_{\tilde{L}}/M_{F_D} = 1.2$. In Fig.~\ref{fig:y1H_Da_mu_2}, we plot $\Delta a_\mu$ as a function of $y_{1H}$ for $M_{F_D} = 800~\mathrm{GeV}$, $ M_{F_S} = 900~\mathrm{GeV}$, and $  M_{\tilde{L}} = 1000~\mathrm{GeV}$. The rest of the parameters in these figures are fixed to be $y_1=y_2= \lambda_L = \lambda_L'=0.5$ and $ y_{2H} = 0.3$. The horizontal dashed line indicates the measured value of $\Delta a_\mu$ in Eq.~\eqref{eq:delamu_exp}, with its error indicated by the red band. The behavior of $\Delta a_\mu$ is similar to that in Fig.~\ref{fig:Da_mu_singlet} and a negative value of $y_{1H}$ and $\sim 1\,\rm TeV$ DM mass can explain the observed discrepancy in the muon $g-2$.

\begin{figure}
  \centering
  \subcaptionbox{\label{fig:mdm_xsec_SI_2}
  DM mass dependence 
  }
  {\includegraphics[width=0.48\textwidth]{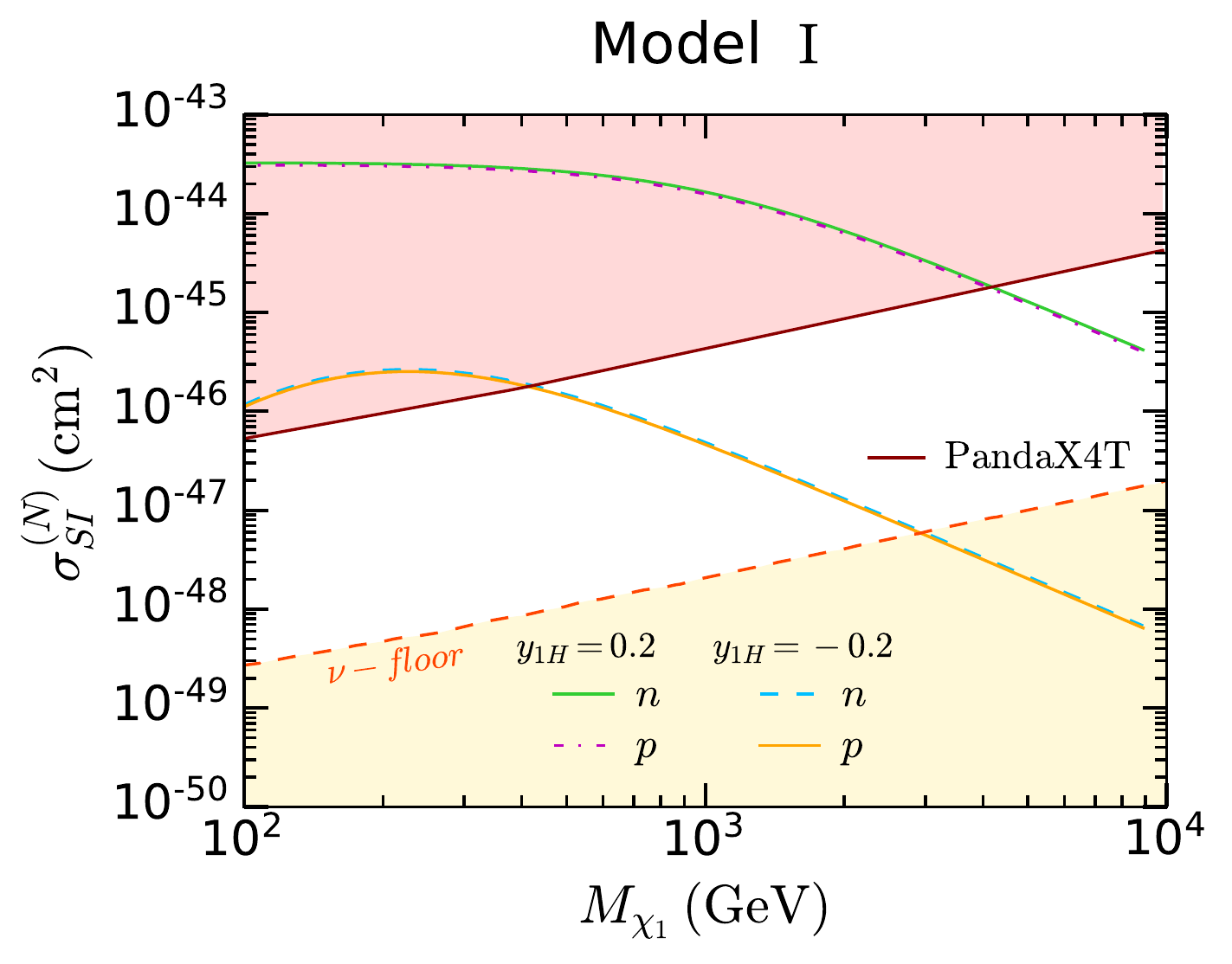}}
  \subcaptionbox{\label{fig:y1H_xsec_SI_2}
  $y_{1H}$ dependence 
  }
  { 
  \includegraphics[width=0.48\textwidth]{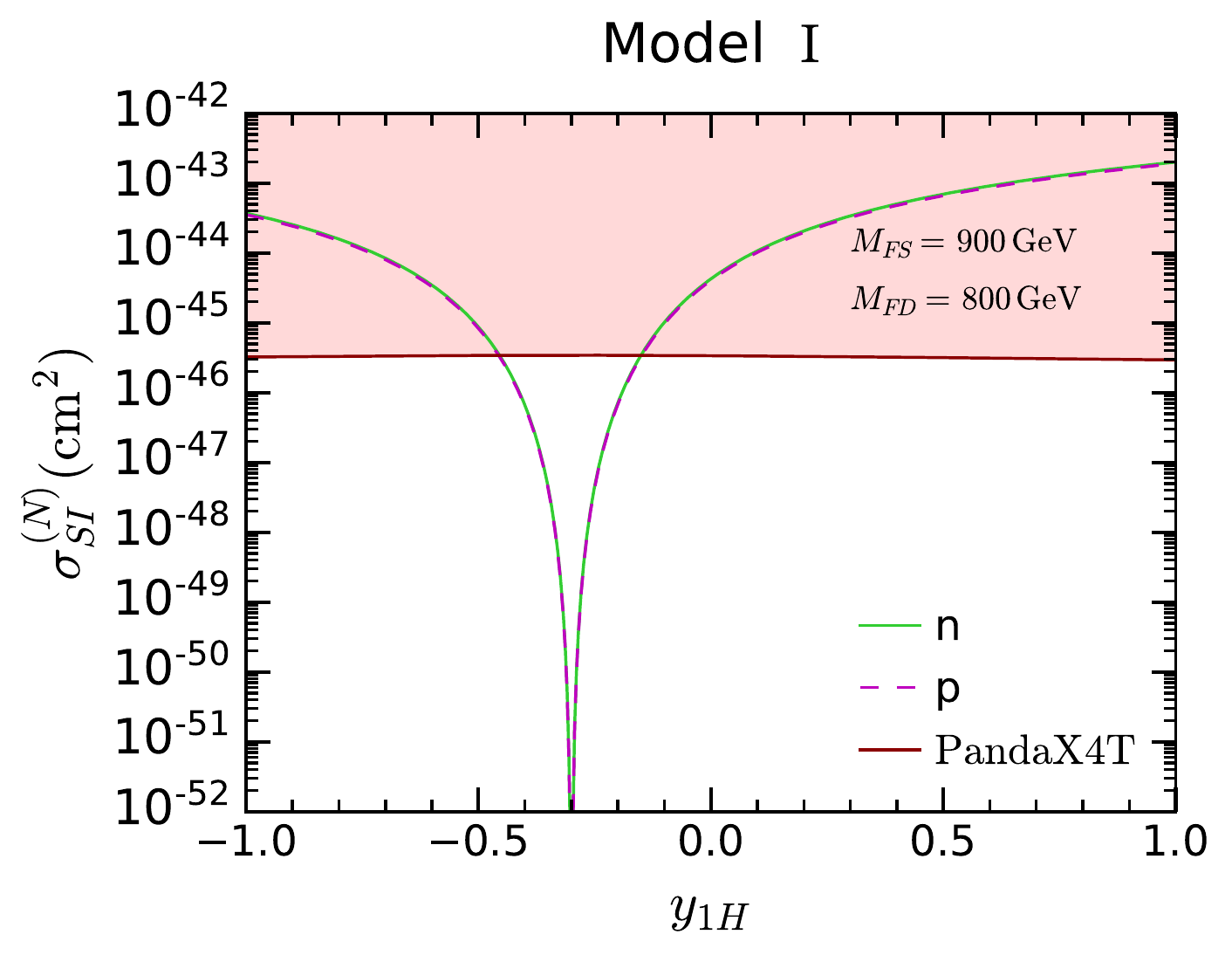}}
  \caption{
    The DM-nucleon SI scattering cross sections as functions of (a) the DM mass $M_{\chi_1}$ and (b) $y_{1H}$. The rest of the parameters are fixed as in Fig.~\ref{fig:Da_mu_doublet}. The red shaded area is excluded by the PandaX4T experiment~\cite{PandaX-4T:2021bab} and the yellow shaded region corresponds to the neutrino floor~\cite{Billard:2021uyg}. }
  
  \label{fig:DM_nucleon_SI_2}
\end{figure}

The DM-nucleon SI scattering cross sections are shown as functions of the DM mass $M_{\chi_1}$ and $y_{1H}$ in Fig.~\ref{fig:mdm_xsec_SI_2}  and Fig.~\ref{fig:y1H_xsec_SI_2}, respectively. The rest of the parameters are fixed as in Fig.~\ref{fig:Da_mu_doublet}. The red shaded region represents the PandaX4T bound~\cite{PandaX-4T:2021bab} and the yellow shaded region corresponds to the neutrino floor~\cite{Billard:2021uyg}. We see that the behavior of the SI scattering cross sections for $y_{1H} = + 0.2$ is similar to that in Fig.~\ref{fig:mdm_xsec_SI}. For $y_{1H} = -0.2$, on the other hand, the SI cross sections are predicted to be larger than those for the singlet-like case, and within reach of future DM direct detection experiments for the DM mass $\sim 1\,\rm TeV$.  

\begin{figure}
  \centering
  \subcaptionbox{\label{fig:mdm_xsec_SD_2}
  DM mass dependence 
  }
  {\includegraphics[width=0.48\textwidth]{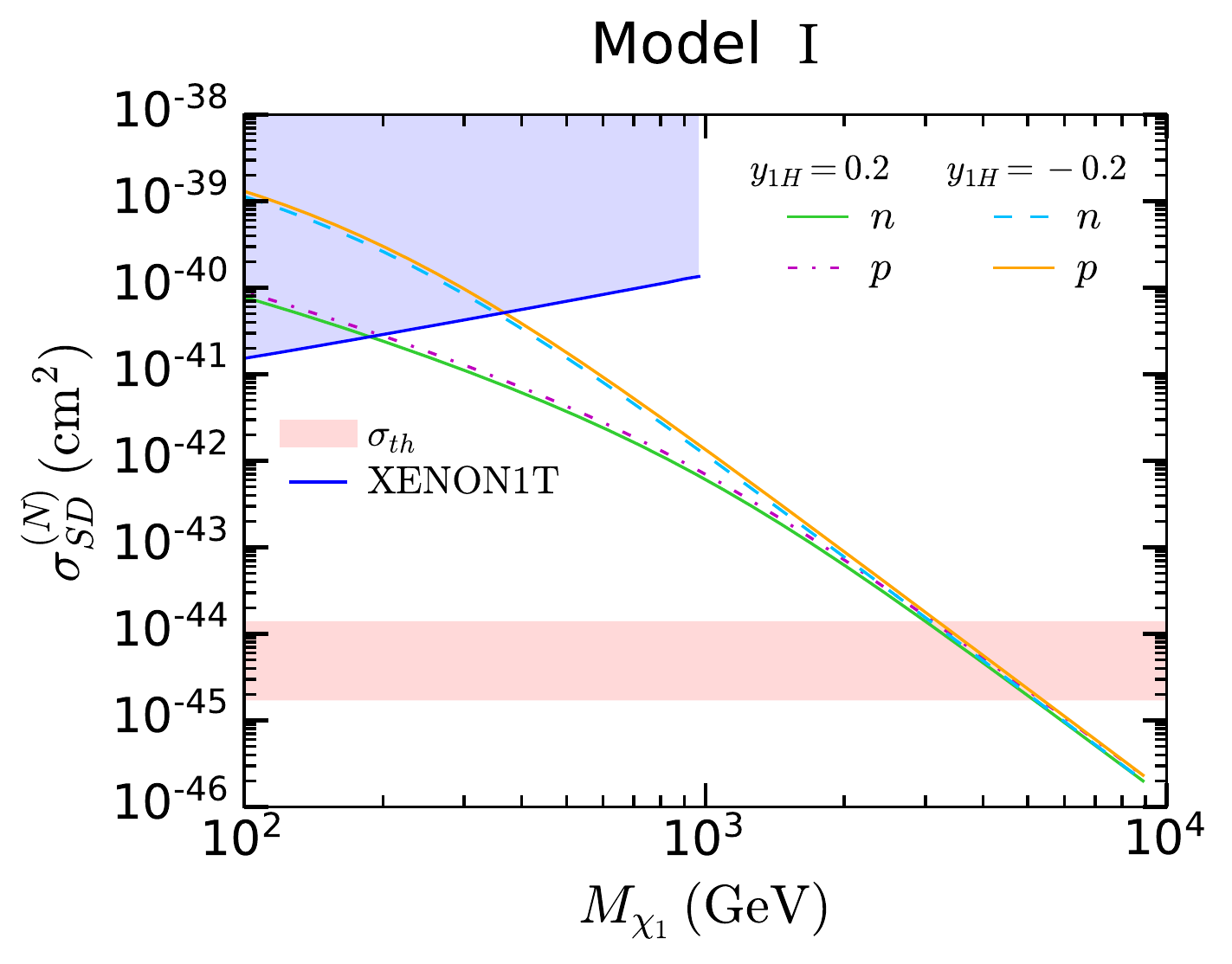}}
  \subcaptionbox{\label{fig:y1H_xsec_SD_2}
  $y_{1H}$ dependence 
  }
  { 
  \includegraphics[width=0.48\textwidth]{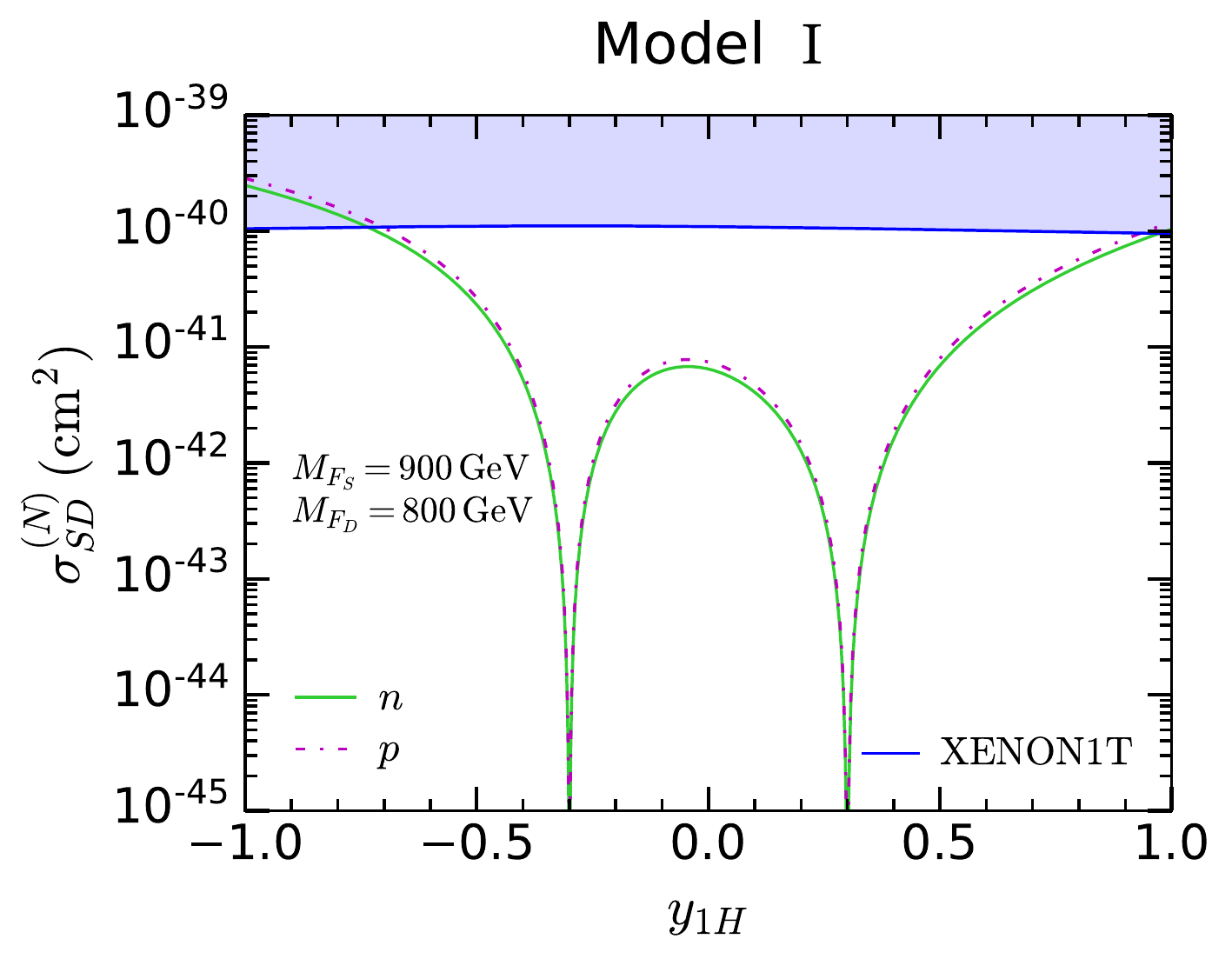}}
  \caption{The DM-nucleon SD scattering cross sections as functions of (a) the DM mass $M_{\chi_1}$ and (b) $y_{1H}$. The rest of the parameters are fixed as in Fig.~\ref{fig:Da_mu_doublet}. The blue shaded area is excluded by the XENON1T experiment~\cite{XENON:2019rxp}. The horizontal red band represents the threshold cross section for neutron obtained in Refs.~\cite{Bell:2020jou,Anzuini:2021lnv}. }
  \label{fig:DM_nucleon_SD_2}
\end{figure}

In Fig.~\ref{fig:DM_nucleon_SD_2}, the DM-nucleon SD scattering cross sections are shown as functions of the DM mass $M_{\chi_1}$ (Fig.~\ref{fig:mdm_xsec_SD_2}) and $y_{1H}$ (Fig.~\ref{fig:y1H_xsec_SD_2}) with the same parameter choice as in Fig.~\ref{fig:Da_mu_doublet}. The blue shaded region is excluded by XENON1T~\cite{XENON:2019rxp} for neutron and the horizontal red band represents the threshold cross section for neutron obtained in Refs.~\cite{Bell:2020jou,Anzuini:2021lnv}. We see that the behavior of the SD cross sections is similar to that in Fig.~\ref{fig:DM_nucleon_SD_2} and the XENON1T limit is evaded in the $g-2$ favored parameter regions. The DM is efficiently captured by NSs for $M_{\chi_1} \lesssim 3\,\rm TeV$.

\begin{figure}
  \centering
  \subcaptionbox{\label{fig:mdm_xsec_DM-muon_doublet}
  DM mass dependence 
  }
  {\includegraphics[width=0.48\textwidth]{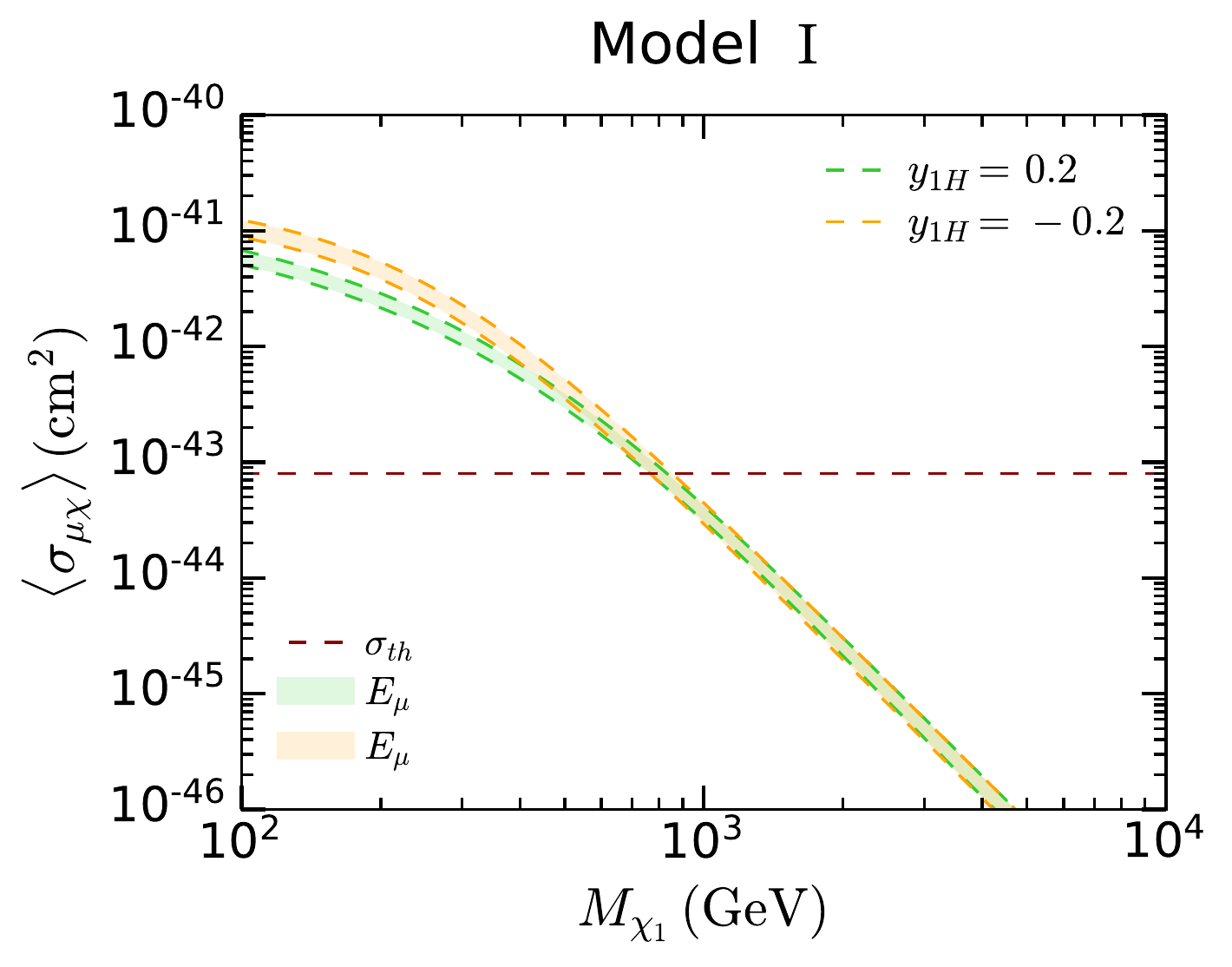}}
  \subcaptionbox{\label{fig:y1H_xsec_DM-muon_doublet}
  $y_{1H}$ dependence 
  }
  { 
  \includegraphics[width=0.48\textwidth]{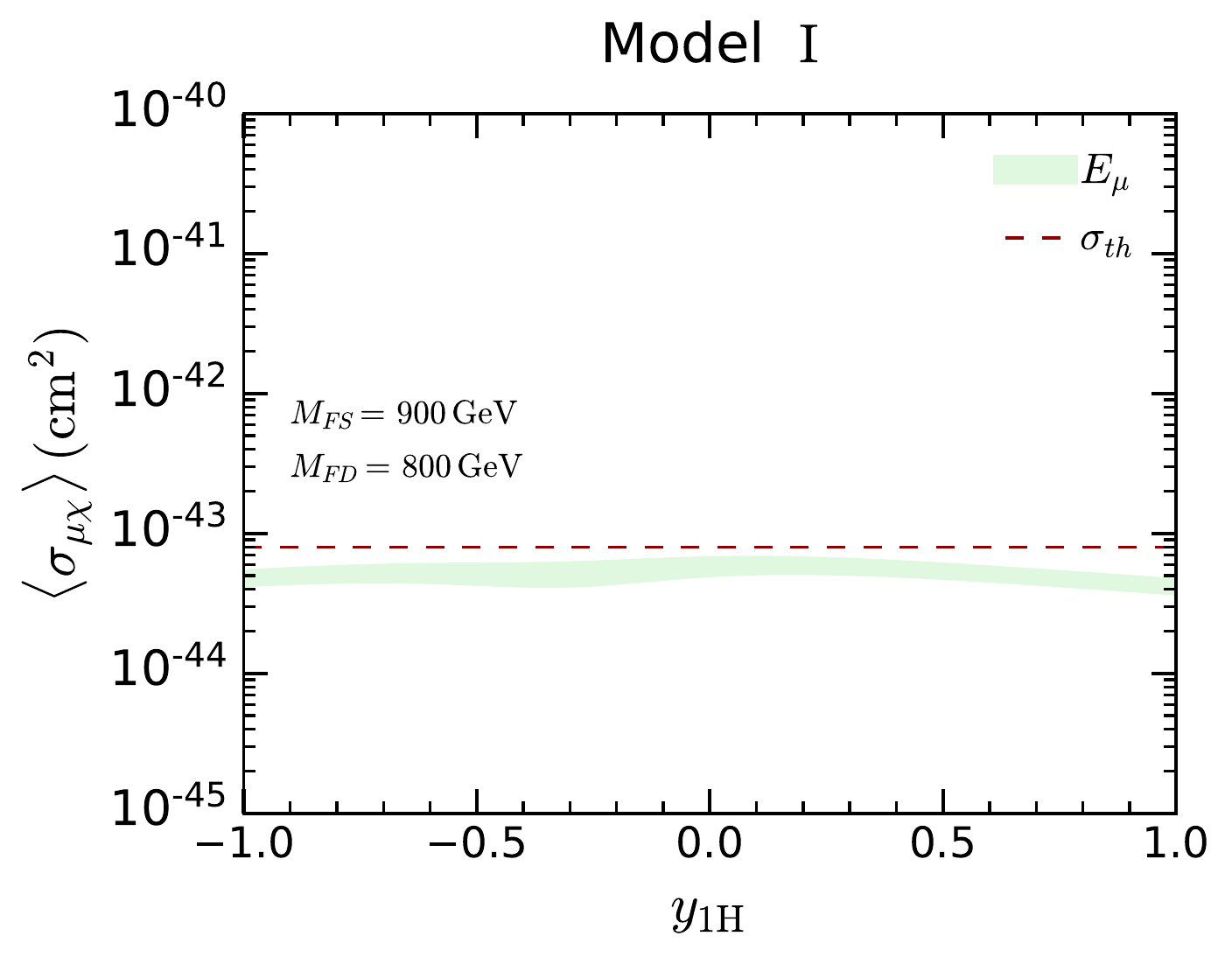}}
  \caption{The DM-muon scattering cross sections as functions of (a) the DM mass $M_{\chi_1}$ and (b) $y_{1H}$. The rest of the parameters are fixed as per in Fig.~\ref{fig:Da_mu_doublet}. $E_\mu$ is varied in the range $[m_\mu, m_\mu/\sqrt{B (R_{\mathrm{NS}})}]$ for $M_{\mathrm{NS}} = 1.5~M_{\odot}$ and $R_{\mathrm{NS}} = 12.593$~km and the resultant change in the cross section is indicated by the band. The horizontal red dashed line shows the threshold cross section for muon, $\sigma_{\mathrm{th}} \simeq  8 \times 10^{-44}~\mathrm{cm}^2$~\cite{Bell:2020lmm}.  }
  \label{fig:xsec_DM-muon_doublet}
\end{figure}

Figure~\ref{fig:xsec_DM-muon_doublet} shows the DM-muon scattering cross sections, where the parameters are fixed as in Fig.~\ref{fig:Da_mu_doublet}. We again vary $E_\mu$ in the range $[m_\mu, m_\mu/\sqrt{B (R_{\mathrm{NS}})}]$, for the same NS configuration, which is indicated by the band. The horizontal red dashed line shows the threshold cross section for muon, $\sigma_{\mathrm{th}} \simeq  8 \times 10^{-44}~\mathrm{cm}^2$~\cite{Bell:2020lmm}. The DM-muon scattering cross section is predicted to be smaller than that in the singlet-like DM case and nearly equal to the threshold cross section in the $g-2$ favored parameter range. 

\subsection{Model II}

\begin{figure}
  \centering
  \subcaptionbox{\label{fig:mdm_Damu_model2}
  $\Delta a_\mu$
  }
  {\includegraphics[width=0.48\textwidth]{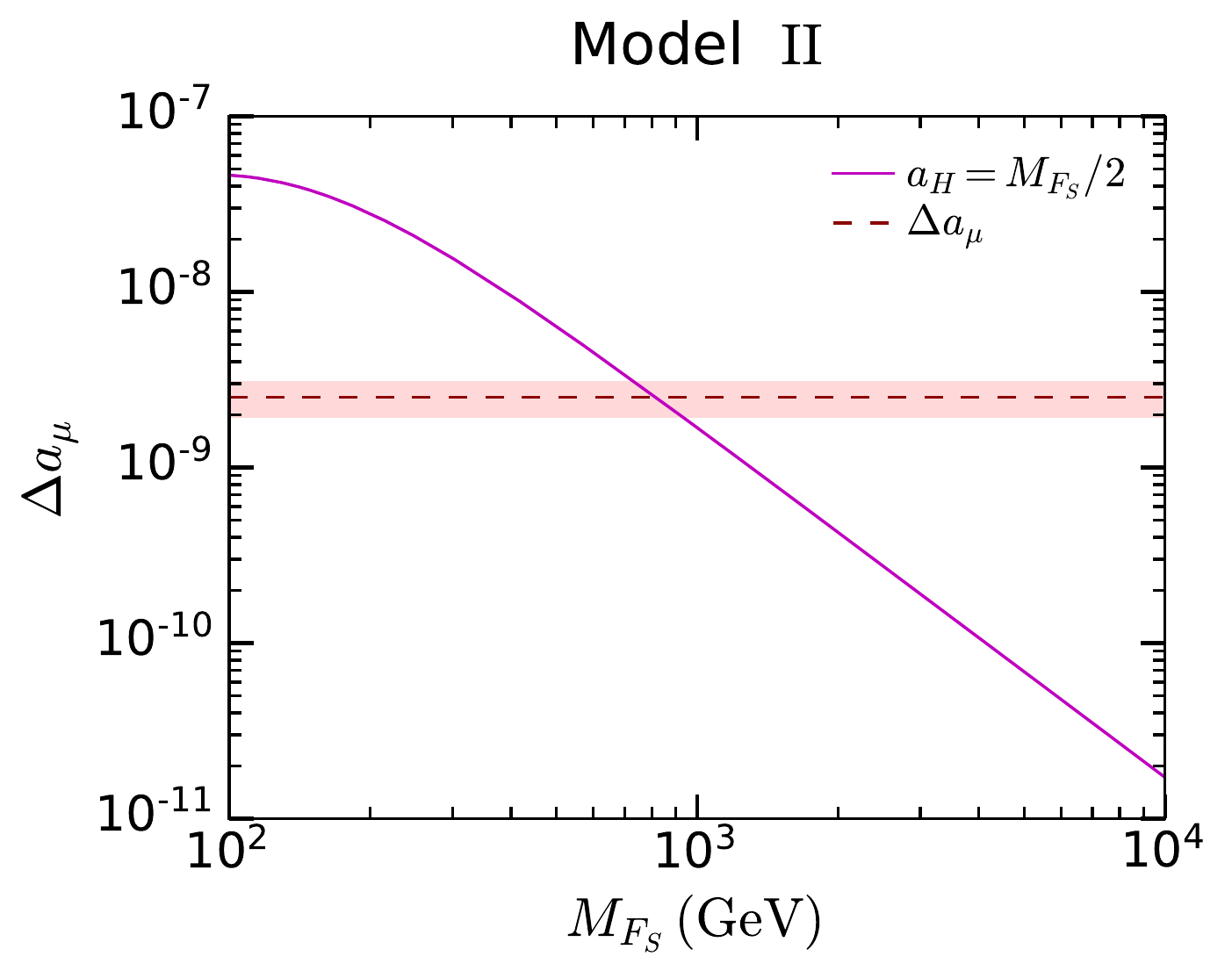}}
  \subcaptionbox{\label{fig:mdm_xsec_DM-muon_model2}
  DM-muon scattering cross section 
  }
  { 
  \includegraphics[width=0.48\textwidth]{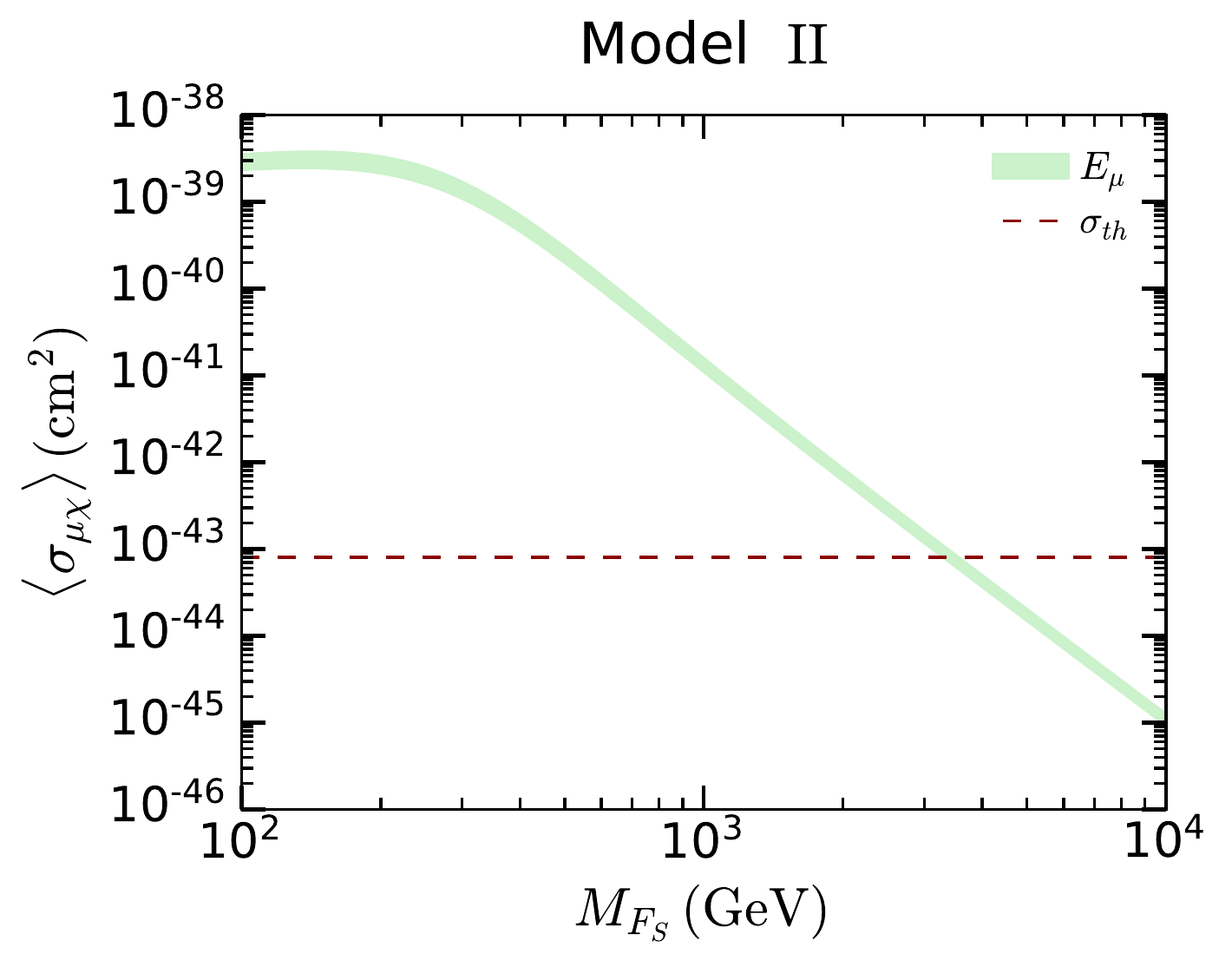}}
  \caption{
    (a) $\Delta a_\mu$ as a function of the DM mass $M_{F_S}$. The horizontal dashed line indicates the measured value of $\Delta a_\mu$, with its error indicated by the red band.  (b) DM-muon scattering cross section as a function of the DM mass $M_{F_S}$. $E_\mu$ is varied in the range $[m_\mu, m_\mu/\sqrt{B (R_{\mathrm{NS}})}]$ for $M_{\mathrm{NS}} = 1.5~M_{\odot}$ and $R_{\mathrm{NS}} = 12.593$~km and the resultant change in the cross section is indicated by the band. The horizontal red dashed line shows the threshold cross section for muon, $\sigma_{\mathrm{th}} \simeq  8 \times 10^{-44}~\mathrm{cm}^2$~\cite{Bell:2020lmm}. We set $M_{\tilde{\bar{e}}}/M_{F_S} = 1.1$, $M_{\tilde{L}}/M_{F_S} = 1.2$, $a_H = M_{F_S}/2$, and $y_1=y_2= \lambda_L = \lambda_{\bar{e}} = \lambda_L'=0.5$. 
  }
  \label{fig:model2}
  \end{figure}

Finally, we study Model II. As discussed in Sec.~\ref{sec:direct_detection} and Sec.~\ref{eq:dm-nucleon_scattering}, the DM-nucleon scattering is induced at the loop level in this case. As a result, the SI scattering cross section is too small to be probed in future DM direct searches and the SD scattering cross section lies below the threshold cross section in the parameter regions of our interest. We thus consider only $\Delta a_\mu$ and the DM-muon scattering cross section for Model II, which are shown as functions of the DM mass in Fig.~\ref{fig:mdm_Damu_model2} and Fig.~\ref{fig:mdm_xsec_DM-muon_model2}, respectively. We set $M_{\tilde{\bar{e}}}/M_{F_S} = 1.1$, $M_{\tilde{L}}/M_{F_S} = 1.2$, $a_H = M_{F_S}/2$, and $y_1=y_2= \lambda_L = \lambda_{\bar{e}} = \lambda_L'=0.5$ in both of the plots. The  observed deviation in the muon $g-2$ can be explained for the DM mass $\simeq 800~\rm GeV$ with this parameter choice. As shown in Ref.~\cite{Kawamura:2020qxo}, the observed DM density can be explained with this size of the DM mass, without conflicting with the LHC limits. For this DM mass, the DM-muon scattering cross section is much larger than the threshold cross section, as illustrated in Fig.~\ref{fig:mdm_xsec_DM-muon_model2}. Consequently, in Model II, the DM efficiently accumulates in NSs through the DM-muon scattering, not via the DM-nucleon scattering as in Model I. Given that it is very difficult to probe this DM candidate in DM direct detection experiments and the LHC experiments, the DM search using the NS temperature observation plays an important role in testing this scenario in the future.

\section{Conclusion and discussion}
\label{sec:conclusion}

We have studied two representative DM models, Model I and II, where WIMP DM particles have renormalizable couplings to muons. In both of these models, the experimental value of the muon $g-2$ can be explained with a DM mass of $\sim 1~\mathrm{TeV}$. Such a heavy DM particle, as well as heavier colorless states in the models, is beyond the reach of the LHC experiments. In Model I, the SI DM-nucleon scattering cross section is predicted to be generically rather large, but in the muon $g-2$ favored parameter regions, it is found to be much smaller than the current experimental limit and may be probed in future DM direct detection experiments. The SD DM-nucleon scattering cross section is larger than the threshold cross section. On the other hand, in Model II, both the SI and SD DM-nucleon scattering cross sections are highly suppressed, and it is hard to probe this model in DM direct detection experiments. However, we find that even in this case the DM particles efficiently accumulate in NSs since the DM-muon scattering cross section is sufficiently larger than the threshold cross section. As a result, in both of these models, the DM capture in NSs is effective and the DM heating operates maximally. Our study thus indicates that the temperature observation of old NSs provides a promising way of testing the WIMP DM models for the muon $g-2$ discrepancy. 

Although we have considered only simple setups in this paper, we expect that the same conclusion holds in a variety of more realistic WIMP DM models. For example, in the framework of SUSY, the bino-slepton system provides a setup similar to Model II. As demonstrated in Ref.~\cite{Endo:2021zal}, this system can explain the muon $g-2$ discrepancy and the observed DM density with a $\sim 100$~GeV bino DM and sleptons. The DM-muon coupling in this case is given by the hypercharge gauge coupling, $g' \simeq 0.36$, which is smaller than $y_1 = y_2 = 0.5$ taken in Fig.~\ref{fig:model2}. However, the bino DM mass is smaller than the favored value of $M_{F_S}$ in Fig.~\ref{fig:model2} by a factor of $\sim 8$, and therefore the DM-muon scattering cross section is much larger than that in Model II. This simple estimate shows that we can test the SUSY explanations of the muon $g-2$ discrepancy by means of the search of the DM heating in NSs. Such a model-specific study is beyond the scope of this paper, and we defer it to another occasion.

\section*{Acknowledgments}

MRQ would like to thank Shyam Balaji for useful discussion.
This work is supported in part by the Grant-in-Aid for Innovative Areas (No.19H05810 [KH], No.19H05802 [KH], No.18H05542 [NN]), Scientific Research B (No.20H01897 [KH, NN, and MRQ]), and Young Scientists (No.21K13916 [NN]).

\newpage
\section*{Appendix}
\appendix

\section{Interactions in the mass eigenbasis}
\label{sec:masseigen}
\renewcommand{\theequation}{A.\arabic{equation}}
\setcounter{equation}{0}

In this Appendix, we summarize the interaction terms expressed in terms of mass eigenfields.

\subsection{Model I}
\label{sec:masseigen1}

\subsubsection{Gauge interactions}

The interactions of the new particles with a photon are 
\begin{align}
  \mathcal{L}_{A} &=  e \overline{\psi^-} \gamma^\mu \psi^- A_\mu + e \widetilde{e}^* i \overleftrightarrow{\partial^\mu} \widetilde{e} A_\mu  ~, 
\end{align}
where $e > 0$ is the electric charge of positron, $A \overleftrightarrow{\partial^\mu} B \equiv A (\partial^\mu B) - (\partial^\mu A)B$, and we introduce the four-component Dirac notation,
\begin{equation}
  \psi^- \equiv 
  \begin{pmatrix}
    \xi_{D^-} \\ \eta^\dagger_{D^+}
  \end{pmatrix}
  ~.
\end{equation}
For the couplings with a $Z$ boson, only the dark matter-$Z$ coupling is relevant for our discussion: 
\begin{align}
  \mathcal{L}_{{\mathrm{DM-}}Z} = - \frac{g_Z}{4} \left[ \left|\left(V_\chi\right)_{21}\right|^2 - \left|\left(V_\chi\right)_{31}\right|^2 \right] \overline{\psi^0_1} \gamma^\mu \gamma_5 \psi^0_1 Z_\mu ~,
\end{align}
where $g_Z \equiv \sqrt{g^{\prime 2} + g^2}$ with $g'$ and $g$ the U(1)$_Y$ and SU(2)$_L$ gauge coupling constants, respectively, and $\psi^0_i$ $(i = 1,2,3)$ are four-component Majorana fermions defined by 
\begin{equation}
  \psi^0_i \equiv 
  \begin{pmatrix}
    \chi_i \\ \chi_i^\dagger
  \end{pmatrix}
  ~.
  \label{eq:dmmaj}
\end{equation}

\subsubsection{Yukawa interactions}

In the unitary gauge 
\begin{equation}
 H =\frac{1}{\sqrt{2}} 
 \begin{pmatrix}
  0 \\v+h
 \end{pmatrix}
 ~,
 \label{eq:unit_gauge}
\end{equation}
the Yukawa interactions are written as 
\begin{align}
  \mathcal{L}_{\mathrm{Yukawa}} = &- \frac{h}{\sqrt{2}} \overline{\psi^0_i} \left[ \left( C_{\chi h L} \right)_{ij} P_L + \left( C_{\chi h R} \right)_{ij} P_R  \right] \psi_j^0  
  \nonumber\\[2pt] 
  &- 
  \left\{ \overline{\psi^0_i}  \left[ y_1 \left( V_\chi \right)_{1i}  P_L + y_2^* \left( V_\chi \right)^*_{2i} P_R   \right] \mu \, \widetilde{e}^* + \mathrm{h.c.} \right\}  
  \nonumber\\[2pt] 
  &- \left[y_1 \left( V_\chi \right)_{1i}  \overline{\psi^0_i} P_L \nu \widetilde{\nu}^*
  - y_2 \bar{\mu} P_L \psi^- \widetilde{\nu}
  + \mathrm{h.c.} \right] ~,
\end{align}
with $P_{L/R} \equiv (1\mp \gamma_5)/2$ and 
\begin{align}
  \left( C_{\chi hL} \right)_{ij}  &\equiv \left( V_\chi \right)_{1i} \left[ y_{1H} \left( V_\chi \right)_{2j} + y_{2H} \left( V_\chi \right)_{3j}\right] ~, \qquad 
  \left( C_{\chi hR} \right)_{ij}  \equiv \left( C_{\chi hL} \right)_{ji}^*  ~.
  \label{eq:cchih}
\end{align}

\subsection{Model II}
\label{sec:masseigen2}

\subsubsection{Gauge interactions}

For interactions with a photon, we have 
\begin{align}
  \mathcal{L}_{A} &=   e \sum_{i = 1,2} \widetilde{e}_i^* i \overleftrightarrow{\partial^\mu} \widetilde{e}_i  A_\mu ~, 
\end{align}
while for those with a $Z$-boson, 
\begin{align}
  \mathcal{L}_{Z} &= \frac{g_Z}{2} \widetilde{\nu}^* i \overleftrightarrow{\partial^\mu} \widetilde{\nu} Z_\mu 
  + g_Z \left[ - \frac{1}{2} \left( U_e^* \right)_{1i} \left( U_e \right)_{1j} + \sin^2 \theta_W \delta_{ij} \right] \widetilde{e}_i^* i \overleftrightarrow{\partial^\mu} \widetilde{e}_j  Z_\mu 
\end{align}
where $\theta_W$ is the weak mixing angle.

\subsubsection{Yukawa interactions}

The Yukawa interactions in Model II are 
\begin{align}
  \mathcal{L}_{\mathrm{Yukawa}} = 
  &- 
  \left\{ \overline{\psi^0}  \left[ y_1 \left( U_e \right)_{1i}^*  P_L + y_2^* \left( U_e \right)^*_{2i} P_R   \right] \mu \, \widetilde{e}_i^* + \mathrm{h.c.} \right\}  
  \nonumber\\[2pt] 
  &- \left[y_1  \overline{\psi^0} P_L \nu \widetilde{\nu}^*
  + \mathrm{h.c.} \right] ~,
\end{align}
where $\psi^0$ is a four-component Majorana fermion defined by 
\begin{equation}
  \psi^0 \equiv 
  \begin{pmatrix}
    \chi_S \\ \chi_S^\dagger
  \end{pmatrix}
  ~.
\end{equation}

\subsubsection{Scalar trilinear interaction}

In the unitary gauge~\eqref{eq:unit_gauge}, the scalar trilinear coupling  is obtained from Eq.~\eqref{eq:tri2} and Eq.~\eqref{eq:quart2} as 
\begin{align}
  \mathcal{L}_{\mathrm{tri}} &= -  \left( C_{h\widetilde{e} } \right)_{ij} 
  h \widetilde{e}_i^* \widetilde{e}_j - v (\lambda_L - \lambda_L') h \left|\widetilde{\nu} \right|^2 ~,
\end{align}
where 
\begin{align}
  \left( C_{h\widetilde{e} } \right)_{ij} &\equiv \frac{1}{\sqrt{2}}
  \left[ a_H \left( U_e \right)^*_{2i} \left( U_e \right)_{1j} + a_H^* \left( U_e \right)^*_{1i} \left( U_e \right)_{2j} \right]\nonumber \\ 
  &+ v \left[ (\lambda_L + \lambda_L') \left( U_e \right)_{1i}^* \left( U_e \right)_{1j} + \lambda_{\bar{e}} \left( U_e \right)_{2i}^* \left( U_e \right)_{2j}  \right] ~.
  \label{eq:chetil}
\end{align}

\section{Amplitudes for the DM--\texorpdfstring{$\mu$}{TEXT} scattering }
\label{sec:amplitude}
\renewcommand{\theequation}{B.\arabic{equation}}
\setcounter{equation}{0}

In this section, we show the expressions for the scattering amplitudes in Model I and II in Sec.~\ref{sec:scatmod1} and Sec.~\ref{sec:scatmod2}, respectively. We expand the invariant scattering amplitude as in Eq.~\eqref{eq:ampexp}: 
\begin{equation}
  \frac{1}{4} \sum_{\mathrm{spins}} |\mathcal{A}|^2 \simeq \alpha_0 + \alpha_1 \, (- t )~,
\end{equation}
and give the expressions of $\alpha_0$ and $\alpha_1$. We take account of the hierarchy, $ \bar{s} \simeq M_{\mathrm{DM}}^2 \gg \bar{s}- M_{\mathrm{DM}}^2 \simeq 2 E_\chi E_\mu \gg |t|, E_\mu^2$, and keep only the leading order terms, which turn out to be $\mathcal{O} (E_\mu^2/E_\chi^2)$.

\subsection{Model I}
\label{sec:scatmod1}

\begin{align}
  \alpha_0 = \frac{1}{2 \left( M_{\tilde{e}}^2 - M_{\chi_1}^2 \right)^2} 
  &\Bigl[
    \left( s - M_{\chi_1}^2 \right)^2
    \left\{ \left| y_1 (V_\chi)_{11}\right|^2 + \left| y_2 (V_\chi)_{21}\right|^2 \right\}^2
    \nonumber \\ 
    &+ 16 m_\mu^2 M_{\chi_1}^2 
    \left\{ \mathrm{Re} \left[ y_1 y_2 (V_\chi)_{11} (V_\chi)_{21} \right]\right\}^2  \nonumber \\ 
    &+ 2 m_\mu^2 M_{\chi_1}^2
    \left\{ \left| y_1 (V_\chi)_{11}\right|^4 + \left| y_2 (V_\chi)_{21}\right|^4 \right\} \nonumber \\ 
    &- 2  \left| y_1 (V_\chi)_{11}\right|^2  \left| y_2 (V_\chi)_{21}\right|^2 \left\{
      (s - M_{\chi_1}^2)^2 - 4 m_\mu^2 M_{\chi_1}^2
    \right\} \nonumber \\ 
    &-4m_\mu^2 M_{\chi_1}^2 \mathrm{Re} \left\{
    y_1^2 y_2^2 (V_\chi)_{11}^2 (V_\chi)_{21}^2 
    \right\}
  \Bigr] ~.
\end{align}
\begin{align}
  \alpha_1 &=\frac{M_{\chi_1}^2 \left\{ \left| y_1 (V_\chi)_{11}\right|^2 + \left| y_2 (V_\chi)_{21}\right|^2 \right\}^2 }{2 \left( M_{\tilde{e}}^2 - M_{\chi_1}^2 \right)^2} 
  ~.
\end{align}

\subsection{Model II}
\label{sec:scatmod2}
\begin{align}
    \alpha_0=\sum_{i = 1,2}\frac{1}{(M_{F_S}^2-M_{e_i}^2)(M_{F_S}^2-M_{e_j}^2)}&\times\nonumber\\
    \times\Bigg[\frac{1}{2}(M_{F_S}^2-s)^2&\big[y_1^4|(U_e)_{1i}|^2|(U_e)_{1j}|^2+y_2^4|(U_e)_{2i}|^2|(U_e)_{2j}|^2\big]\nonumber\\
    +m_\mu^2 M_{F_S}^2\Big[ 3&y_1^2y_2^2|(U_e)_{1i}|^2|(U_e)_{2j}|^2+3y_1^2y_2^2|(U_e)_{1j}|^2|(U_e)_{2i}|^2\nonumber\\[0.1cm]
    &+y_1^4|(U_e)_{1i}|^2|(U_e)_{1j}|^2+y_2^4|(U_e)_{2i}|^2|(U_e)_{2j}|^2\nonumber\\[0.2cm]
     &+2y_1^2y_2^2\mathrm{Re}[(U_e)_{1i}(U_e^\ast)_{1j}(U_e^\ast)_{2i}(U_e)_{2j}]\nonumber\\
       &+2y_1^2y_2^2\mathrm{Re}[(U_e)_{1i}(U_e)_{1j}(U_e^\ast)_{2i}(U_e^\ast)_{2j}]\Big]\Bigg]
\end{align}
\begin{align}
    \alpha_1=&\sum_{i = 1,2}\frac{M_{F_S}^2}{2(M_{F_S}^2-M_{e_i}^2)(M_{F_S}^2-M_{e_j}^2)}\times\nonumber\\&\times\Bigg[y_1^4|(U_e)_{1i}|^2|(U_e)_{1j}|^2+y_2^4|(U_e)_{2i}|^2|(U_e)_{2j}|^2+2\mathrm{Re}[(U_e)_{1i}(U_e^\ast)_{1j}(U_e^\ast)_{2i}(U_e)_{2j}]\Bigg]
  \end{align}  


\bibliographystyle{utphysmod}
\bibliography{ref}

\providecommand{\href}[2]{#2}\begingroup\raggedright\begin{thebibliography}{100}

\bibitem{Muong-2:2021ojo}
{\bfseries Muon g-2} Collaboration, {\em {Measurement of the Positive Muon
  Anomalous Magnetic Moment to 0.46 ppm}},
  \href{https://dx.doi.org/10.1103/PhysRevLett.126.141801}{Phys.\  Rev.\
  Lett.\  {\bfseries 126} (2021) 141801} {\ttfamily
  [\href{https://arxiv.org/abs/2104.03281}{arXiv:2104.03281}]}.

\bibitem{Aoyama:2020ynm}
T.~Aoyama {\em et~al.}, {\em {The anomalous magnetic moment of the muon in the
  Standard Model}},
  \href{https://dx.doi.org/10.1016/j.physrep.2020.07.006}{Phys.\  Rept.\
  {\bfseries 887} (2020) 1--166} {\ttfamily
  [\href{https://arxiv.org/abs/2006.04822}{arXiv:2006.04822}]}.

\bibitem{Davier:2017zfy}
M.~Davier, A.~Hoecker, B.~Malaescu, and Z.~Zhang, {\em {Reevaluation of the
  hadronic vacuum polarisation contributions to the Standard Model predictions
  of the muon $g-2$ and ${\alpha (m_Z^2)}$ using newest hadronic cross-section
  data}}, \href{https://dx.doi.org/10.1140/epjc/s10052-017-5161-6}{Eur.\
  Phys.\  J.\  {\bfseries C77} (2017) 827}
{\ttfamily [\href{https://arxiv.org/abs/1706.09436}{arXiv:1706.09436}]}.

\bibitem{Keshavarzi:2018mgv}
A.~Keshavarzi, D.~Nomura, and T.~Teubner, {\em {Muon $g-2$ and $\alpha(M_Z^2)$:
  a new data-based analysis}},
  \href{https://dx.doi.org/10.1103/PhysRevD.97.114025}{Phys.\  Rev.\
  {\bfseries D97} (2018) 114025}
{\ttfamily [\href{https://arxiv.org/abs/1802.02995}{arXiv:1802.02995}]}.

\bibitem{Colangelo:2018mtw}
G.~Colangelo, M.~Hoferichter, and P.~Stoffer, {\em {Two-pion contribution to
  hadronic vacuum polarization}},
  \href{https://dx.doi.org/10.1007/JHEP02(2019)006}{JHEP {\bfseries 02} (2019)
  006}
{\ttfamily [\href{https://arxiv.org/abs/1810.00007}{arXiv:1810.00007}]}.

\bibitem{Hoferichter:2019mqg}
M.~Hoferichter, B.-L.~Hoid, and B.~Kubis, {\em {Three-pion contribution to
  hadronic vacuum polarization}},
  \href{https://dx.doi.org/10.1007/JHEP08(2019)137}{JHEP {\bfseries 08} (2019)
  137}
{\ttfamily [\href{https://arxiv.org/abs/1907.01556}{arXiv:1907.01556}]}.

\bibitem{Davier:2019can}
M.~Davier, A.~Hoecker, B.~Malaescu, and Z.~Zhang, {\em {A new evaluation of the
  hadronic vacuum polarisation contributions to the muon anomalous magnetic
  moment and to $\mathbf{\boldsymbol\alpha(m_Z^2)}$}},
  \href{https://dx.doi.org/10.1140/epjc/s10052-020-7792-2}{Eur.\  Phys.\  J.\
  {\bfseries C80} (2020) 241} {\ttfamily
  [\href{https://arxiv.org/abs/1908.00921}{arXiv:1908.00921}]}.
[Erratum: Eur. Phys. J. {\bf C80}, 410 (2020)].

\bibitem{Keshavarzi:2019abf}
A.~Keshavarzi, D.~Nomura, and T.~Teubner, {\em {The $g-2$ of charged leptons,
  $\alpha(M_Z^2)$ and the hyperfine splitting of muonium}},
  \href{https://dx.doi.org/10.1103/PhysRevD.101.014029}{Phys.\  Rev.\
  {\bfseries D101} (2020) 014029}
{\ttfamily [\href{https://arxiv.org/abs/1911.00367}{arXiv:1911.00367}]}.

\bibitem{Kurz:2014wya}
A.~Kurz, T.~Liu, P.~Marquard, and M.~Steinhauser, {\em {Hadronic contribution
  to the muon anomalous magnetic moment to next-to-next-to-leading order}},
  \href{https://dx.doi.org/10.1016/j.physletb.2014.05.043}{Phys.\  Lett.\
  {\bfseries B734} (2014) 144--147}
{\ttfamily [\href{https://arxiv.org/abs/1403.6400}{arXiv:1403.6400}]}.

\bibitem{Chakraborty:2017tqp}
{\bfseries Fermilab Lattice, LATTICE-HPQCD, MILC} Collaboration, {\em
  {Strong-Isospin-Breaking Correction to the Muon Anomalous Magnetic Moment
  from Lattice QCD at the Physical Point}},
  \href{https://dx.doi.org/10.1103/PhysRevLett.120.152001}{Phys.\  Rev.\
  Lett.\  {\bfseries 120} (2018) 152001}
{\ttfamily [\href{https://arxiv.org/abs/1710.11212}{arXiv:1710.11212}]}.

\bibitem{Borsanyi:2017zdw}
{\bfseries Budapest-Marseille-Wuppertal} Collaboration, {\em {Hadronic vacuum
  polarization contribution to the anomalous magnetic moments of leptons from
  first principles}},
  \href{https://dx.doi.org/10.1103/PhysRevLett.121.022002}{Phys.\  Rev.\
  Lett.\  {\bfseries 121} (2018) 022002}
{\ttfamily [\href{https://arxiv.org/abs/1711.04980}{arXiv:1711.04980}]}.

\bibitem{Blum:2018mom}
{\bfseries RBC, UKQCD} Collaboration, {\em {Calculation of the hadronic vacuum
  polarization contribution to the muon anomalous magnetic moment}},
  \href{https://dx.doi.org/10.1103/PhysRevLett.121.022003}{Phys.\  Rev.\
  Lett.\  {\bfseries 121} (2018) 022003}
{\ttfamily [\href{https://arxiv.org/abs/1801.07224}{arXiv:1801.07224}]}.

\bibitem{Giusti:2019xct}
{\bfseries ETM} Collaboration, {\em {Electromagnetic and strong
  isospin-breaking corrections to the muon $g - 2$ from Lattice QCD+QED}},
  \href{https://dx.doi.org/10.1103/PhysRevD.99.114502}{Phys.\  Rev.\
  {\bfseries D99} (2019) 114502}
{\ttfamily [\href{https://arxiv.org/abs/1901.10462}{arXiv:1901.10462}]}.

\bibitem{Shintani:2019wai}
E.~Shintani and Y.~Kuramashi, {\em {Study of systematic uncertainties in
  hadronic vacuum polarization contribution to muon $g-2$ with 2+1 flavor
  lattice QCD}}, \href{https://dx.doi.org/10.1103/PhysRevD.100.034517}{Phys.\
  Rev.\  {\bfseries D100} (2019) 034517}
{\ttfamily [\href{https://arxiv.org/abs/1902.00885}{arXiv:1902.00885}]}.

\bibitem{FermilabLattice:2019ugu}
{\bfseries Fermilab Lattice, LATTICE-HPQCD, MILC} Collaboration, {\em
  {Hadronic-vacuum-polarization contribution to the muon's anomalous magnetic
  moment from four-flavor lattice QCD}},
  \href{https://dx.doi.org/10.1103/PhysRevD.101.034512}{Phys.\  Rev.\
  {\bfseries D101} (2020) 034512}
{\ttfamily [\href{https://arxiv.org/abs/1902.04223}{arXiv:1902.04223}]}.

\bibitem{Gerardin:2019rua}
A.~G\'erardin, {\em et al.}, {\em {The leading hadronic contribution to
  $(g-2)_\mu$ from lattice QCD with $N_{\rm f}=2+1$ flavours of O($a$) improved
  Wilson quarks}}, \href{https://dx.doi.org/10.1103/PhysRevD.100.014510}{Phys.\
   Rev.\  {\bfseries D100} (2019) 014510}
{\ttfamily [\href{https://arxiv.org/abs/1904.03120}{arXiv:1904.03120}]}.

\bibitem{Aubin:2019usy}
C.~Aubin, {\em et al.}, {\em {Light quark vacuum polarization at the physical
  point and contribution to the muon $g-2$}},
  \href{https://dx.doi.org/10.1103/PhysRevD.101.014503}{Phys.\  Rev.\
  {\bfseries D101} (2020) 014503}
{\ttfamily [\href{https://arxiv.org/abs/1905.09307}{arXiv:1905.09307}]}.

\bibitem{Giusti:2019hkz}
D.~Giusti and S.~Simula, {\em {Lepton anomalous magnetic moments in Lattice
  QCD+QED}}, \href{https://dx.doi.org/10.22323/1.363.0104}{PoS {\bfseries
  LATTICE2019} (2019) 104}
{\ttfamily [\href{https://arxiv.org/abs/1910.03874}{arXiv:1910.03874}]}.

\bibitem{Masjuan:2017tvw}
P.~Masjuan and P.~S{\'a}nchez-Puertas, {\em {Pseudoscalar-pole contribution to
  the $(g_{\mu}-2)$: a rational approach}},
  \href{https://dx.doi.org/10.1103/PhysRevD.95.054026}{Phys.\  Rev.\
  {\bfseries D95} (2017) 054026}
{\ttfamily [\href{https://arxiv.org/abs/1701.05829}{arXiv:1701.05829}]}.

\bibitem{Colangelo:2017fiz}
G.~Colangelo, M.~Hoferichter, M.~Procura, and P.~Stoffer, {\em {Dispersion
  relation for hadronic light-by-light scattering: two-pion contributions}},
  \href{https://dx.doi.org/10.1007/JHEP04(2017)161}{JHEP {\bfseries 04} (2017)
  161}
{\ttfamily [\href{https://arxiv.org/abs/1702.07347}{arXiv:1702.07347}]}.

\bibitem{Hoferichter:2018kwz}
M.~Hoferichter, B.-L.~Hoid, B.~Kubis, S.~Leupold, and S.~P.~Schneider, {\em
  {Dispersion relation for hadronic light-by-light scattering: pion pole}},
  \href{https://dx.doi.org/10.1007/JHEP10(2018)141}{JHEP {\bfseries 10} (2018)
  141}
{\ttfamily [\href{https://arxiv.org/abs/1808.04823}{arXiv:1808.04823}]}.

\bibitem{Gerardin:2019vio}
A.~G{\'e}rardin, H.~B.~Meyer, and A.~Nyffeler, {\em {Lattice calculation of the
  pion transition form factor with $N_f=2+1$ Wilson quarks}},
  \href{https://dx.doi.org/10.1103/PhysRevD.100.034520}{Phys.\  Rev.\
  {\bfseries D100} (2019) 034520}
{\ttfamily [\href{https://arxiv.org/abs/1903.09471}{arXiv:1903.09471}]}.

\bibitem{Bijnens:2019ghy}
J.~Bijnens, N.~Hermansson-Truedsson, and A.~Rodr{\'i}guez-S{\'a}nchez, {\em
  {Short-distance constraints for the HLbL contribution to the muon anomalous
  magnetic moment}},
  \href{https://dx.doi.org/10.1016/j.physletb.2019.134994}{Phys.\  Lett.\
  {\bfseries B798} (2019) 134994}
{\ttfamily [\href{https://arxiv.org/abs/1908.03331}{arXiv:1908.03331}]}.

\bibitem{Colangelo:2019uex}
G.~Colangelo, F.~Hagelstein, M.~Hoferichter, L.~Laub, and P.~Stoffer, {\em
  {Longitudinal short-distance constraints for the hadronic light-by-light
  contribution to $(g-2)_\mu$ with large-$N_c$ Regge models}},
  \href{https://dx.doi.org/10.1007/JHEP03(2020)101}{JHEP {\bfseries 03} (2020)
  101}
{\ttfamily [\href{https://arxiv.org/abs/1910.13432}{arXiv:1910.13432}]}.

\bibitem{Pauk:2014rta}
V.~Pauk and M.~Vanderhaeghen, {\em {Single meson contributions to the muon`s
  anomalous magnetic moment}},
  \href{https://dx.doi.org/10.1140/epjc/s10052-014-3008-y}{Eur.\  Phys.\  J.\
  {\bfseries C74} (2014) 3008}
{\ttfamily [\href{https://arxiv.org/abs/1401.0832}{arXiv:1401.0832}]}.

\bibitem{Danilkin:2016hnh}
I.~Danilkin and M.~Vanderhaeghen, {\em {Light-by-light scattering sum rules in
  light of new data}},
  \href{https://dx.doi.org/10.1103/PhysRevD.95.014019}{Phys.\  Rev.\
  {\bfseries D95} (2017) 014019}
{\ttfamily [\href{https://arxiv.org/abs/1611.04646}{arXiv:1611.04646}]}.

\bibitem{Jegerlehner:2017gek}
F.~Jegerlehner, {\em {The Anomalous Magnetic Moment of the Muon}},
\href{https://dx.doi.org/10.1007/978-3-319-63577-4}{Springer Tracts Mod.\
  Phys.\  {\bfseries 274} (2017) 1--693}.

\bibitem{Knecht:2018sci}
M.~Knecht, S.~Narison, A.~Rabemananjara, and D.~Rabetiarivony, {\em {Scalar
  meson contributions to $a_\mu$ from hadronic light-by-light scattering}},
  \href{https://dx.doi.org/10.1016/j.physletb.2018.10.048}{Phys.\  Lett.\
  {\bfseries B787} (2018) 111--123}
{\ttfamily [\href{https://arxiv.org/abs/1808.03848}{arXiv:1808.03848}]}.

\bibitem{Eichmann:2019bqf}
G.~Eichmann, C.~S.~Fischer, and R.~Williams, {\em {Kaon-box contribution to the
  anomalous magnetic moment of the muon}},
  \href{https://dx.doi.org/10.1103/PhysRevD.101.054015}{Phys.\  Rev.\
  {\bfseries D101} (2020) 054015}
{\ttfamily [\href{https://arxiv.org/abs/1910.06795}{arXiv:1910.06795}]}.

\bibitem{Roig:2019reh}
P.~Roig and P.~S{\'a}nchez-Puertas, {\em {Axial-vector exchange contribution to
  the hadronic light-by-light piece of the muon anomalous magnetic moment}},
  \href{https://dx.doi.org/10.1103/PhysRevD.101.074019}{Phys.\  Rev.\
  {\bfseries D101} (2020) 074019}
{\ttfamily [\href{https://arxiv.org/abs/1910.02881}{arXiv:1910.02881}]}.

\bibitem{Colangelo:2014qya}
G.~Colangelo, M.~Hoferichter, A.~Nyffeler, M.~Passera, and P.~Stoffer, {\em
  {Remarks on higher-order hadronic corrections to the muon $g-2$}},
  \href{https://dx.doi.org/10.1016/j.physletb.2014.06.012}{Phys.\  Lett.\
  {\bfseries B735} (2014) 90--91}
{\ttfamily [\href{https://arxiv.org/abs/1403.7512}{arXiv:1403.7512}]}.

\bibitem{Blum:2019ugy}
T.~Blum, {\em et al.}, {\em {The hadronic light-by-light scattering
  contribution to the muon anomalous magnetic moment from lattice QCD}},
  \href{https://dx.doi.org/10.1103/PhysRevLett.124.132002}{Phys.\  Rev.\
  Lett.\  {\bfseries 124} (2020) 132002}
{\ttfamily [\href{https://arxiv.org/abs/1911.08123}{arXiv:1911.08123}]}.

\bibitem{Aoyama:2012wk}
T.~Aoyama, M.~Hayakawa, T.~Kinoshita, and M.~Nio, {\em {Complete Tenth-Order
  QED Contribution to the Muon $g-2$}},
  \href{https://dx.doi.org/10.1103/PhysRevLett.109.111808}{Phys.\  Rev.\
  Lett.\  {\bfseries 109} (2012) 111808}
{\ttfamily [\href{https://arxiv.org/abs/1205.5370}{arXiv:1205.5370}]}.

\bibitem{Aoyama:2019ryr}
T.~Aoyama, T.~Kinoshita, and M.~Nio, {\em {Theory of the Anomalous Magnetic
  Moment of the Electron}},
  \href{https://dx.doi.org/10.3390/atoms7010028}{Atoms {\bfseries 7} (2019)
  28}.

\bibitem{Czarnecki:2002nt}
A.~Czarnecki, W.~J.~Marciano, and A.~Vainshtein, {\em {Refinements in
  electroweak contributions to the muon anomalous magnetic moment}},
  \href{https://dx.doi.org/10.1103/PhysRevD.67.073006}{Phys.\  Rev.\
  {\bfseries D67} (2003) 073006} {\ttfamily
  [\href{https://arxiv.org/abs/hep-ph/0212229}{hep-ph/0212229}]}.
[Erratum: Phys. Rev. {\bf D73}, 119901 (2006)].

\bibitem{Gnendiger:2013pva}
C.~Gnendiger, D.~St{\"o}ckinger, and H.~St{\"o}ckinger-Kim, {\em {The
  electroweak contributions to $(g-2)_\mu$ after the Higgs boson mass
  measurement}}, \href{https://dx.doi.org/10.1103/PhysRevD.88.053005}{Phys.\
  Rev.\  {\bfseries D88} (2013) 053005}
{\ttfamily [\href{https://arxiv.org/abs/1306.5546}{arXiv:1306.5546}]}.

\bibitem{Melnikov:2003xd}
K.~Melnikov and A.~Vainshtein, {\em {Hadronic light-by-light scattering
  contribution to the muon anomalous magnetic moment revisited}},
  \href{https://dx.doi.org/10.1103/PhysRevD.70.113006}{Phys.\  Rev.\
  {\bfseries D70} (2004) 113006}
{\ttfamily [\href{https://arxiv.org/abs/hep-ph/0312226}{hep-ph/0312226}]}.

\bibitem{Muong-2:2006rrc}
{\bfseries Muon g-2} Collaboration, {\em {Final Report of the Muon E821
  Anomalous Magnetic Moment Measurement at BNL}},
  \href{https://dx.doi.org/10.1103/PhysRevD.73.072003}{Phys.\  Rev.\  D
  {\bfseries 73} (2006) 072003} {\ttfamily
  [\href{https://arxiv.org/abs/hep-ex/0602035}{hep-ex/0602035}]}.

\bibitem{Borsanyi:2020mff}
S.~Borsanyi {\em et~al.}, {\em {Leading hadronic contribution to the muon
  magnetic moment from lattice QCD}},
  \href{https://dx.doi.org/10.1038/s41586-021-03418-1}{Nature {\bfseries 593}
  (2021) 51--55} {\ttfamily
  [\href{https://arxiv.org/abs/2002.12347}{arXiv:2002.12347}]}.

\bibitem{Lehner:2020crt}
C.~Lehner and A.~S.~Meyer, {\em {Consistency of hadronic vacuum polarization
  between lattice QCD and the R-ratio}},
  \href{https://dx.doi.org/10.1103/PhysRevD.101.074515}{Phys.\  Rev.\  D
  {\bfseries 101} (2020) 074515} {\ttfamily
  [\href{https://arxiv.org/abs/2003.04177}{arXiv:2003.04177}]}.

\bibitem{Crivellin:2020zul}
A.~Crivellin, M.~Hoferichter, C.~A.~Manzari, and M.~Montull, {\em {Hadronic
  Vacuum Polarization: $(g-2)_\mu$ versus Global Electroweak Fits}},
  \href{https://dx.doi.org/10.1103/PhysRevLett.125.091801}{Phys.\  Rev.\
  Lett.\  {\bfseries 125} (2020) 091801} {\ttfamily
  [\href{https://arxiv.org/abs/2003.04886}{arXiv:2003.04886}]}.

\bibitem{Keshavarzi:2020bfy}
A.~Keshavarzi, W.~J.~Marciano, M.~Passera, and A.~Sirlin, {\em {Muon $g-2$ and
  $\Delta \alpha$ connection}},
  \href{https://dx.doi.org/10.1103/PhysRevD.102.033002}{Phys.\  Rev.\  D
  {\bfseries 102} (2020) 033002} {\ttfamily
  [\href{https://arxiv.org/abs/2006.12666}{arXiv:2006.12666}]}.

\bibitem{deRafael:2020uif}
E.~de~Rafael, {\em {Constraints between $\Delta\alpha_{\rm had}(M_Z^2)$ and
  $(g_{\mu}-2)_{\rm HVP}$}},
  \href{https://dx.doi.org/10.1103/PhysRevD.102.056025}{Phys.\  Rev.\  D
  {\bfseries 102} (2020) 056025} {\ttfamily
  [\href{https://arxiv.org/abs/2006.13880}{arXiv:2006.13880}]}.

\bibitem{Malaescu:2020zuc}
B.~Malaescu and M.~Schott, {\em {Impact of correlations between $a_{\mu }$ and
  $\alpha _\text {QED}$ on the EW fit}},
  \href{https://dx.doi.org/10.1140/epjc/s10052-021-08848-9}{Eur.\  Phys.\  J.\
  C {\bfseries 81} (2021) 46} {\ttfamily
  [\href{https://arxiv.org/abs/2008.08107}{arXiv:2008.08107}]}.

\bibitem{DiLuzio:2021uty}
L.~Di~Luzio, A.~Masiero, P.~Paradisi, and M.~Passera, {\em {New physics behind
  the new muon $g$-2 puzzle?}}, {\ttfamily
  \href{https://arxiv.org/abs/2112.08312}{arXiv:2112.08312}} (2021).

\bibitem{Colangelo:2020lcg}
G.~Colangelo, M.~Hoferichter, and P.~Stoffer, {\em {Constraints on the two-pion
  contribution to hadronic vacuum polarization}},
  \href{https://dx.doi.org/10.1016/j.physletb.2021.136073}{Phys.\  Lett.\  B
  {\bfseries 814} (2021) 136073} {\ttfamily
  [\href{https://arxiv.org/abs/2010.07943}{arXiv:2010.07943}]}.

\bibitem{Kanemitsu:2012dc}
S.~Kanemitsu and K.~Tobe, {\em {New physics for muon anomalous magnetic moment
  and its electroweak precision analysis}},
  \href{https://dx.doi.org/10.1103/PhysRevD.86.095025}{Phys.\  Rev.\  D
  {\bfseries 86} (2012) 095025} {\ttfamily
  [\href{https://arxiv.org/abs/1207.1313}{arXiv:1207.1313}]}.

\bibitem{Kowalska:2017iqv}
K.~Kowalska and E.~M.~Sessolo, {\em {Expectations for the muon g-2 in
  simplified models with dark matter}},
  \href{https://dx.doi.org/10.1007/JHEP09(2017)112}{JHEP {\bfseries 09} (2017)
  112} {\ttfamily [\href{https://arxiv.org/abs/1707.00753}{arXiv:1707.00753}]}.

\bibitem{Calibbi:2018rzv}
L.~Calibbi, R.~Ziegler, and J.~Zupan, {\em {Minimal models for dark matter and
  the muon g$-$2 anomaly}},
  \href{https://dx.doi.org/10.1007/JHEP07(2018)046}{JHEP {\bfseries 07} (2018)
  046} {\ttfamily [\href{https://arxiv.org/abs/1804.00009}{arXiv:1804.00009}]}.

\bibitem{Kawamura:2020qxo}
J.~Kawamura, S.~Okawa, and Y.~Omura, {\em {Current status and muon $g-2$
  explanation of lepton portal dark matter}},
  \href{https://dx.doi.org/10.1007/JHEP08(2020)042}{JHEP {\bfseries 08} (2020)
  042} {\ttfamily [\href{https://arxiv.org/abs/2002.12534}{arXiv:2002.12534}]}.

\bibitem{Horigome:2021qof}
S.-I.~Horigome, T.~Katayose, S.~Matsumoto, and I.~Saha, {\em {Leptophilic
  fermion WIMP: Role of future lepton colliders}},
  \href{https://dx.doi.org/10.1103/PhysRevD.104.055001}{Phys.\  Rev.\  D
  {\bfseries 104} (2021) 055001} {\ttfamily
  [\href{https://arxiv.org/abs/2102.08645}{arXiv:2102.08645}]}.

\bibitem{Arcadi:2021cwg}
G.~Arcadi, L.~Calibbi, M.~Fedele, and F.~Mescia, {\em {Muon $g-2$ and
  $B$-anomalies from Dark Matter}},
  \href{https://dx.doi.org/10.1103/PhysRevLett.127.061802}{Phys.\  Rev.\
  Lett.\  {\bfseries 127} (2021) 061802} {\ttfamily
  [\href{https://arxiv.org/abs/2104.03228}{arXiv:2104.03228}]}.

\bibitem{Bai:2021bau}
Y.~Bai and J.~Berger, {\em {Muon $g-2$ in Lepton Portal Dark Matter}},
  {\ttfamily \href{https://arxiv.org/abs/2104.03301}{arXiv:2104.03301}} (2021).

\bibitem{Athron:2021iuf}
P.~Athron, {\em et al.}, {\em {New physics explanations of $a_\mu$ in light of
  the FNAL muon $g-2$ measurement}},
  \href{https://dx.doi.org/10.1007/JHEP09(2021)080}{JHEP {\bfseries 09} (2021)
  080} {\ttfamily [\href{https://arxiv.org/abs/2104.03691}{arXiv:2104.03691}]}.

\bibitem{Acuna:2021rbg}
J.~T.~Acu\~na, P.~Stengel, and P.~Ullio, {\em {A Minimal Dark Matter Model for
  Muon g-2 with Scalar Lepton Partners up to the TeV Scale}}, {\ttfamily
  \href{https://arxiv.org/abs/2112.08992}{arXiv:2112.08992}} (2021).

\bibitem{Ghosh:2022zef}
T.~Ghosh, C.~Kelso, J.~Kumar, P.~Sandick, and P.~Stengel in {\em {2022 Snowmass
  Summer Study}}.
\newblock 2022.
\newblock {\ttfamily
  \href{https://arxiv.org/abs/2203.08107}{arXiv:2203.08107}}.

\bibitem{Lopez:1993vi}
J.~L.~Lopez, D.~V.~Nanopoulos, and X.~Wang, {\em {Large (g-2)-mu in SU(5) x
  U(1) supergravity models}},
  \href{https://dx.doi.org/10.1103/PhysRevD.49.366}{Phys.\  Rev.\  D {\bfseries
  49} (1994) 366--372} {\ttfamily
  [\href{https://arxiv.org/abs/hep-ph/9308336}{hep-ph/9308336}]}.

\bibitem{Chattopadhyay:1995ae}
U.~Chattopadhyay and P.~Nath, {\em {Probing supergravity grand unification in
  the Brookhaven g-2 experiment}},
  \href{https://dx.doi.org/10.1103/PhysRevD.53.1648}{Phys.\  Rev.\  D
  {\bfseries 53} (1996) 1648--1657} {\ttfamily
  [\href{https://arxiv.org/abs/hep-ph/9507386}{hep-ph/9507386}]}.

\bibitem{Moroi:1995yh}
T.~Moroi, {\em {The Muon anomalous magnetic dipole moment in the minimal
  supersymmetric standard model}},
  \href{https://dx.doi.org/10.1103/PhysRevD.53.6565}{Phys.\  Rev.\  D
  {\bfseries 53} (1996) 6565--6575} {\ttfamily
  [\href{https://arxiv.org/abs/hep-ph/9512396}{hep-ph/9512396}]}. [Erratum:
  Phys.Rev.D 56, 4424 (1997)].

\bibitem{Endo:2021zal}
M.~Endo, K.~Hamaguchi, S.~Iwamoto, and T.~Kitahara, {\em {Supersymmetric
  interpretation of the muon $g-2$ anomaly}},
  \href{https://dx.doi.org/10.1007/JHEP07(2021)075}{JHEP {\bfseries 07} (2021)
  075} {\ttfamily [\href{https://arxiv.org/abs/2104.03217}{arXiv:2104.03217}]}.

\bibitem{Iwamoto:2021aaf}
S.~Iwamoto, T.~T.~Yanagida, and N.~Yokozaki, {\em {Wino-Higgsino dark matter in
  MSSM from the $g-2$ anomaly}},
  \href{https://dx.doi.org/10.1016/j.physletb.2021.136768}{Phys.\  Lett.\  B
  {\bfseries 823} (2021) 136768} {\ttfamily
  [\href{https://arxiv.org/abs/2104.03223}{arXiv:2104.03223}]}.

\bibitem{Gu:2021mjd}
Y.~Gu, N.~Liu, L.~Su, and D.~Wang, {\em {Heavy bino and slepton for muon $g-2$
  anomaly}}, \href{https://dx.doi.org/10.1016/j.nuclphysb.2021.115481}{Nucl.\
  Phys.\  B {\bfseries 969} (2021) 115481} {\ttfamily
  [\href{https://arxiv.org/abs/2104.03239}{arXiv:2104.03239}]}.

\bibitem{Yin:2021mls}
W.~Yin, {\em {Muon $g-2$ anomaly in anomaly mediation}},
  \href{https://dx.doi.org/10.1007/JHEP06(2021)029}{JHEP {\bfseries 06} (2021)
  029} {\ttfamily [\href{https://arxiv.org/abs/2104.03259}{arXiv:2104.03259}]}.

\bibitem{Wang:2021bcx}
F.~Wang, L.~Wu, Y.~Xiao, J.~M.~Yang, and Y.~Zhang, {\em {GUT-scale constrained
  SUSY in light of new muon $g-2$ measurement}},
  \href{https://dx.doi.org/10.1016/j.nuclphysb.2021.115486}{Nucl.\  Phys.\  B
  {\bfseries 970} (2021) 115486} {\ttfamily
  [\href{https://arxiv.org/abs/2104.03262}{arXiv:2104.03262}]}.

\bibitem{Abdughani:2021pdc}
M.~Abdughani, {\em et al.}, {\em {A common origin of muon $g-2$ anomaly, Galaxy
  Center GeV excess and AMS-02 anti-proton excess in the NMSSM}},
  \href{https://dx.doi.org/10.1016/j.scib.2021.07.029}{Sci.\  Bull.\
  {\bfseries 66} (2021) 1545} {\ttfamily
  [\href{https://arxiv.org/abs/2104.03274}{arXiv:2104.03274}]}.

\bibitem{Cao:2021tuh}
J.~Cao, J.~Lian, Y.~Pan, D.~Zhang, and P.~Zhu, {\em {Improved (g - 2)$_{\mu}$
  measurement and singlino dark matter in $\mu$-term extended
  $\mathbb{Z}_{3}$-NMSSM}},
  \href{https://dx.doi.org/10.1007/JHEP09(2021)175}{JHEP {\bfseries 09} (2021)
  175} {\ttfamily [\href{https://arxiv.org/abs/2104.03284}{arXiv:2104.03284}]}.

\bibitem{Chakraborti:2021dli}
M.~Chakraborti, S.~Heinemeyer, and I.~Saha, {\em {The new
  \textquotedblleft{}MUON G-2\textquotedblright{} result and supersymmetry}},
  \href{https://dx.doi.org/10.1140/epjc/s10052-021-09900-4}{Eur.\  Phys.\  J.\
  C {\bfseries 81} (2021) 1114} {\ttfamily
  [\href{https://arxiv.org/abs/2104.03287}{arXiv:2104.03287}]}.

\bibitem{Ibe:2021cvf}
M.~Ibe, S.~Kobayashi, Y.~Nakayama, and S.~Shirai, {\em {Muon $g-2$ in Gauge
  Mediation without SUSY CP Problem}}, {\ttfamily
  \href{https://arxiv.org/abs/2104.03289}{arXiv:2104.03289}} (2021).

\bibitem{Cox:2021nbo}
P.~Cox, C.~Han, and T.~T.~Yanagida, {\em {Muon g-2 and coannihilating dark
  matter in the minimal supersymmetric standard model}},
  \href{https://dx.doi.org/10.1103/PhysRevD.104.075035}{Phys.\  Rev.\  D
  {\bfseries 104} (2021) 075035} {\ttfamily
  [\href{https://arxiv.org/abs/2104.03290}{arXiv:2104.03290}]}.

\bibitem{Heinemeyer:2021opc}
S.~Heinemeyer, {\em et al.}, {\em {The new $(g-2)_\mu $ result and the $\mu \nu
  $SSM}}, \href{https://dx.doi.org/10.1140/epjc/s10052-021-09601-y}{Eur.\
  Phys.\  J.\  C {\bfseries 81} (2021) 802} {\ttfamily
  [\href{https://arxiv.org/abs/2104.03294}{arXiv:2104.03294}]}.

\bibitem{Baum:2021qzx}
S.~Baum, M.~Carena, N.~R.~Shah, and C.~E.~M.~Wagner, {\em {The tiny (g-2) muon
  wobble from small-$\mu$ supersymmetry}},
  \href{https://dx.doi.org/10.1007/JHEP01(2022)025}{JHEP {\bfseries 01} (2022)
  025} {\ttfamily [\href{https://arxiv.org/abs/2104.03302}{arXiv:2104.03302}]}.

\bibitem{Zhang:2021gun}
H.-B.~Zhang, C.-X.~Liu, J.-L.~Yang, and T.-F.~Feng, {\em {Muon anomalous
  magnetic dipole moment in the $\mu\nu$SSM}}, {\ttfamily
  \href{https://arxiv.org/abs/2104.03489}{arXiv:2104.03489}} (2021).

\bibitem{Ahmed:2021htr}
W.~Ahmed, {\em et al.}, {\em {The natural explanation of the muon anomalous
  magnetic moment via the electroweak supersymmetry from the GmSUGRA in the
  MSSM}}, \href{https://dx.doi.org/10.1016/j.physletb.2022.136879}{Phys.\
  Lett.\  B {\bfseries 827} (2022) 136879} {\ttfamily
  [\href{https://arxiv.org/abs/2104.03491}{arXiv:2104.03491}]}.

\bibitem{Aboubrahim:2021xfi}
A.~Aboubrahim, M.~Klasen, and P.~Nath, {\em {What the Fermilab muon $g-$2
  experiment tells us about discovering supersymmetry at high luminosity and
  high energy upgrades to the LHC}},
  \href{https://dx.doi.org/10.1103/PhysRevD.104.035039}{Phys.\  Rev.\  D
  {\bfseries 104} (2021) 035039} {\ttfamily
  [\href{https://arxiv.org/abs/2104.03839}{arXiv:2104.03839}]}.

\bibitem{Chakraborti:2021bmv}
M.~Chakraborti, L.~Roszkowski, and S.~Trojanowski, {\em {GUT-constrained
  supersymmetry and dark matter in light of the new $(g-2)_\mu$
  determination}}, \href{https://dx.doi.org/10.1007/JHEP05(2021)252}{JHEP
  {\bfseries 05} (2021) 252} {\ttfamily
  [\href{https://arxiv.org/abs/2104.04458}{arXiv:2104.04458}]}.

\bibitem{Baer:2021aax}
H.~Baer, V.~Barger, and H.~Serce, {\em {Anomalous muon magnetic moment,
  supersymmetry, naturalness, LHC search limits and the landscape}},
  \href{https://dx.doi.org/10.1016/j.physletb.2021.136480}{Phys.\  Lett.\  B
  {\bfseries 820} (2021) 136480} {\ttfamily
  [\href{https://arxiv.org/abs/2104.07597}{arXiv:2104.07597}]}.

\bibitem{Altmannshofer:2021hfu}
W.~Altmannshofer, S.~A.~Gadam, S.~Gori, and N.~Hamer, {\em {Explaining (g $-$
  2)$_{\mu}$ with multi-TeV sleptons}},
  \href{https://dx.doi.org/10.1007/JHEP07(2021)118}{JHEP {\bfseries 07} (2021)
  118} {\ttfamily [\href{https://arxiv.org/abs/2104.08293}{arXiv:2104.08293}]}.

\bibitem{Aboubrahim:2021phn}
A.~Aboubrahim, P.~Nath, and R.~M.~Syed, {\em {Yukawa coupling unification in an
  SO(10) model consistent with Fermilab (g $-$ 2)$_{\mu}$ result}},
  \href{https://dx.doi.org/10.1007/JHEP06(2021)002}{JHEP {\bfseries 06} (2021)
  002} {\ttfamily [\href{https://arxiv.org/abs/2104.10114}{arXiv:2104.10114}]}.

\bibitem{Chakraborti:2021squ}
M.~Chakraborti, S.~Heinemeyer, and I.~Saha in {\em {International Workshop on
  Future Linear Colliders}}.
\newblock 2021.
\newblock {\ttfamily
  \href{https://arxiv.org/abs/2105.06408}{arXiv:2105.06408}}.

\bibitem{Zheng:2021wnu}
M.-D.~Zheng and H.-H.~Zhang, {\em {Studying the $b\rightarrow s \ell^+\ell^-$
  anomalies and $(g-2)_{\mu}$ in $R$-parity violating MSSM framework with the
  inverse seesaw mechanism}},
  \href{https://dx.doi.org/10.1103/PhysRevD.104.115023}{Phys.\  Rev.\  D
  {\bfseries 104} (2021) 115023} {\ttfamily
  [\href{https://arxiv.org/abs/2105.06954}{arXiv:2105.06954}]}.

\bibitem{Jeong:2021qey}
K.~S.~Jeong, J.~Kawamura, and C.~B.~Park, {\em {Mixed modulus and anomaly
  mediation in light of the muon g $-$ 2 anomaly}},
  \href{https://dx.doi.org/10.1007/JHEP10(2021)064}{JHEP {\bfseries 10} (2021)
  064} {\ttfamily [\href{https://arxiv.org/abs/2106.04238}{arXiv:2106.04238}]}.

\bibitem{Li:2021pnt}
Z.~Li, G.-L.~Liu, F.~Wang, J.~M.~Yang, and Y.~Zhang, {\em {Gluino-SUGRA
  scenarios in light of FNAL muon g \textendash{} 2 anomaly}},
  \href{https://dx.doi.org/10.1007/JHEP12(2021)219}{JHEP {\bfseries 12} (2021)
  219} {\ttfamily [\href{https://arxiv.org/abs/2106.04466}{arXiv:2106.04466}]}.

\bibitem{Kim:2021suj}
J.~S.~Kim, D.~E.~Lopez-Fogliani, A.~D.~Perez, and R.~R.~de~Austri, {\em {The
  new (g$-$2)$\mu$ and right-handed sneutrino dark matter}},
  \href{https://dx.doi.org/10.1016/j.nuclphysb.2021.115637}{Nucl.\  Phys.\  B
  {\bfseries 974} (2022) 115637} {\ttfamily
  [\href{https://arxiv.org/abs/2107.02285}{arXiv:2107.02285}]}.

\bibitem{Ellis:2021zmg}
J.~Ellis, J.~L.~Evans, N.~Nagata, D.~V.~Nanopoulos, and K.~A.~Olive, {\em
  {Flipped $\mathbf {g_\mu - 2}$}},
  \href{https://dx.doi.org/10.1140/epjc/s10052-021-09829-8}{Eur.\  Phys.\  J.\
  C {\bfseries 81} (2021) 1079} {\ttfamily
  [\href{https://arxiv.org/abs/2107.03025}{arXiv:2107.03025}]}.

\bibitem{Aboubrahim:2021ily}
A.~Aboubrahim, M.~Klasen, P.~Nath, and R.~M.~Syed
\newblock 2021.
\newblock {\ttfamily
  \href{https://arxiv.org/abs/2107.06021}{arXiv:2107.06021}}.

\bibitem{Nakai:2021mha}
Y.~Nakai, M.~Reece, and M.~Suzuki, {\em {Supersymmetric alignment models for (g
  $-$ 2)$_{\mu}$}}, \href{https://dx.doi.org/10.1007/JHEP10(2021)068}{JHEP
  {\bfseries 10} (2021) 068} {\ttfamily
  [\href{https://arxiv.org/abs/2107.10268}{arXiv:2107.10268}]}.

\bibitem{Li:2021cte}
T.~Li, J.~A.~Maxin, and D.~V.~Nanopoulos, {\em {Spinning no-scale ${\mathcal
  {F}}$-SU(5) in the right direction}},
  \href{https://dx.doi.org/10.1140/epjc/s10052-021-09835-w}{Eur.\  Phys.\  J.\
  C {\bfseries 81} (2021) 1059} {\ttfamily
  [\href{https://arxiv.org/abs/2107.12843}{arXiv:2107.12843}]}.

\bibitem{Lamborn:2021snt}
J.~L.~Lamborn, T.~Li, J.~A.~Maxin, and D.~V.~Nanopoulos, {\em {Resolving the (g
  $-$ 2)$_{\mu}$ discrepancy with $ \mathcal{F} $\textendash{}SU(5)
  intersecting D-branes}},
  \href{https://dx.doi.org/10.1007/JHEP11(2021)081}{JHEP {\bfseries 11} (2021)
  081} {\ttfamily [\href{https://arxiv.org/abs/2108.08084}{arXiv:2108.08084}]}.

\bibitem{Ellis:2021vpp}
J.~Ellis, J.~L.~Evans, N.~Nagata, D.~V.~Nanopoulos, and K.~A.~Olive, {\em
  {Flipped SU(5) GUT phenomenology: proton decay and $\mathbf {g_\mu - 2}$}},
  \href{https://dx.doi.org/10.1140/epjc/s10052-021-09896-x}{Eur.\  Phys.\  J.\
  C {\bfseries 81} (2021) 1109} {\ttfamily
  [\href{https://arxiv.org/abs/2110.06833}{arXiv:2110.06833}]}.

\bibitem{Chakraborti:2021mbr}
M.~Chakraborti, S.~Heinemeyer, I.~Saha, and C.~Schappacher, {\em {$(g-2)_\mu$
  and SUSY Dark Matter: Direct Detection and Collider Search Complementarity}},
  {\ttfamily \href{https://arxiv.org/abs/2112.01389}{arXiv:2112.01389}} (2021).

\bibitem{Ali:2021kxa}
M.~I.~Ali, M.~Chakraborti, U.~Chattopadhyay, and S.~Mukherjee, {\em {Muon and
  Electron $(g-2)$ Anomalies with Non-Holomorphic Interactions in MSSM}},
  {\ttfamily \href{https://arxiv.org/abs/2112.09867}{arXiv:2112.09867}} (2021).

\bibitem{Gomez:2022qrb}
M.~E.~Gomez, Q.~Shafi, A.~Tiwari, and C.~S.~Un, {\em {Muon g-2, Neutralino Dark
  Matter and Stau NLSP}}, {\ttfamily
  \href{https://arxiv.org/abs/2202.06419}{arXiv:2202.06419}} (2022).

\bibitem{Chakraborti:2022vds}
M.~Chakraborti, S.~Iwamoto, J.~S.~Kim, R.~Mase\l{}ek, and K.~Sakurai, {\em
  {Supersymmetric explanation of the muon g-2 anomaly with and without stable
  neutralino}}, {\ttfamily
  \href{https://arxiv.org/abs/2202.12928}{arXiv:2202.12928}} (2022).

\bibitem{Agashe:2022uih}
K.~Agashe, M.~Ekhterachian, Z.~Liu, and R.~Sundrum, {\em {Sleptonic SUSY: From
  UV Framework to IR Phenomenology}}, {\ttfamily
  \href{https://arxiv.org/abs/2203.01796}{arXiv:2203.01796}} (2022).

\bibitem{Endo:2022qnm}
M.~Endo, {\em et al.} in {\em {2022 Snowmass Summer Study}}.
\newblock 2022.
\newblock {\ttfamily
  \href{https://arxiv.org/abs/2203.07056}{arXiv:2203.07056}}.

\bibitem{Chigusa:2022xpq}
S.~Chigusa, T.~Moroi, and Y.~Shoji, {\em {Upper bound on the smuon mass from
  vacuum stability in the light of muon $g-2$ anomaly}}, {\ttfamily
  \href{https://arxiv.org/abs/2203.08062}{arXiv:2203.08062}} (2022).

\bibitem{Cao:2022chy}
J.~Cao, J.~Lian, Y.~Pan, Y.~Yue, and D.~Zhang, {\em {Impact of recent (g $-$
  2)$_{\mu}$ measurement on the light CP-even Higgs scenario in general
  Next-to-Minimal Supersymmetric Standard Model}},
  \href{https://dx.doi.org/10.1007/JHEP03(2022)203}{JHEP {\bfseries 03} (2022)
  203} {\ttfamily [\href{https://arxiv.org/abs/2201.11490}{arXiv:2201.11490}]}.

\bibitem{Planck:2018vyg}
{\bfseries Planck} Collaboration, {\em {Planck 2018 results. VI. Cosmological
  parameters}}, \href{https://dx.doi.org/10.1051/0004-6361/201833910}{Astron.\
  Astrophys.\  {\bfseries 641} (2020) A6} {\ttfamily
  [\href{https://arxiv.org/abs/1807.06209}{arXiv:1807.06209}]}. [Erratum:
  Astron.Astrophys. 652, C4 (2021)].

\bibitem{Goldman:1989nd}
I.~Goldman and S.~Nussinov, {\em {Weakly Interacting Massive Particles and
  Neutron Stars}}, \href{https://dx.doi.org/10.1103/PhysRevD.40.3221}{Phys.\
  Rev.\  D {\bfseries 40} (1989) 3221--3230}.

\bibitem{Kouvaris:2007ay}
C.~Kouvaris, {\em {WIMP Annihilation and Cooling of Neutron Stars}},
  \href{https://dx.doi.org/10.1103/PhysRevD.77.023006}{Phys.\  Rev.\  D
  {\bfseries 77} (2008) 023006} {\ttfamily
  [\href{https://arxiv.org/abs/0708.2362}{arXiv:0708.2362}]}.

\bibitem{Bertone:2007ae}
G.~Bertone and M.~Fairbairn, {\em {Compact Stars as Dark Matter Probes}},
  \href{https://dx.doi.org/10.1103/PhysRevD.77.043515}{Phys.\  Rev.\  D
  {\bfseries 77} (2008) 043515} {\ttfamily
  [\href{https://arxiv.org/abs/0709.1485}{arXiv:0709.1485}]}.

\bibitem{Kouvaris:2010vv}
C.~Kouvaris and P.~Tinyakov, {\em {Can Neutron stars constrain Dark Matter?}},
  \href{https://dx.doi.org/10.1103/PhysRevD.82.063531}{Phys.\  Rev.\  D
  {\bfseries 82} (2010) 063531} {\ttfamily
  [\href{https://arxiv.org/abs/1004.0586}{arXiv:1004.0586}]}.

\bibitem{deLavallaz:2010wp}
A.~de~Lavallaz and M.~Fairbairn, {\em {Neutron Stars as Dark Matter Probes}},
  \href{https://dx.doi.org/10.1103/PhysRevD.81.123521}{Phys.\  Rev.\  D
  {\bfseries 81} (2010) 123521} {\ttfamily
  [\href{https://arxiv.org/abs/1004.0629}{arXiv:1004.0629}]}.

\bibitem{Yakovlev:1999sk}
D.~G.~Yakovlev, K.~P.~Levenfish, and Y.~A.~Shibanov, {\em {Cooling neutron
  stars and superfluidity in their interiors}},
  \href{https://dx.doi.org/10.1070/PU1999v042n08ABEH000556}{Phys.\  Usp.\
  {\bfseries 42} (1999) 737--778} {\ttfamily
  [\href{https://arxiv.org/abs/astro-ph/9906456}{astro-ph/9906456}]}.

\bibitem{Yakovlev:2000jp}
D.~G.~Yakovlev, A.~D.~Kaminker, O.~Y.~Gnedin, and P.~Haensel, {\em {Neutrino
  emission from neutron stars}},
  \href{https://dx.doi.org/10.1016/S0370-1573(00)00131-9}{Phys.\  Rept.\
  {\bfseries 354} (2001) 1} {\ttfamily
  [\href{https://arxiv.org/abs/astro-ph/0012122}{astro-ph/0012122}]}.

\bibitem{Yakovlev:2004iq}
D.~G.~Yakovlev and C.~J.~Pethick, {\em {Neutron star cooling}},
  \href{https://dx.doi.org/10.1146/annurev.astro.42.053102.134013}{Ann.\  Rev.\
   Astron.\  Astrophys.\  {\bfseries 42} (2004) 169--210} {\ttfamily
  [\href{https://arxiv.org/abs/astro-ph/0402143}{astro-ph/0402143}]}.

\bibitem{Page:2004fy}
D.~Page, J.~M.~Lattimer, M.~Prakash, and A.~W.~Steiner, {\em {Minimal cooling
  of neutron stars: A New paradigm}},
  \href{https://dx.doi.org/10.1086/424844}{Astrophys.\  J.\  Suppl.\
  {\bfseries 155} (2004) 623--650} {\ttfamily
  [\href{https://arxiv.org/abs/astro-ph/0403657}{astro-ph/0403657}]}.

\bibitem{Page:2009fu}
D.~Page, J.~M.~Lattimer, M.~Prakash, and A.~W.~Steiner, {\em {Neutrino Emission
  from Cooper Pairs and Minimal Cooling of Neutron Stars}},
  \href{https://dx.doi.org/10.1088/0004-637X/707/2/1131}{Astrophys.\  J.\
  {\bfseries 707} (2009) 1131--1140} {\ttfamily
  [\href{https://arxiv.org/abs/0906.1621}{arXiv:0906.1621}]}.

\bibitem{Potekhin:2015qsa}
A.~Y.~Potekhin, J.~A.~Pons, and D.~Page, {\em {Neutron stars - cooling and
  transport}}, \href{https://dx.doi.org/10.1007/s11214-015-0180-9}{Space Sci.\
  Rev.\  {\bfseries 191} (2015) 239--291} {\ttfamily
  [\href{https://arxiv.org/abs/1507.06186}{arXiv:1507.06186}]}.

\bibitem{Baryakhtar:2017dbj}
M.~Baryakhtar, J.~Bramante, S.~W.~Li, T.~Linden, and N.~Raj, {\em {Dark Kinetic
  Heating of Neutron Stars and An Infrared Window On WIMPs, SIMPs, and Pure
  Higgsinos}}, \href{https://dx.doi.org/10.1103/PhysRevLett.119.131801}{Phys.\
  Rev.\  Lett.\  {\bfseries 119} (2017) 131801} {\ttfamily
  [\href{https://arxiv.org/abs/1704.01577}{arXiv:1704.01577}]}.

\bibitem{Gardner:2006ky}
J.~P.~Gardner {\em et~al.}, {\em {The James Webb Space Telescope}},
  \href{https://dx.doi.org/10.1007/s11214-006-8315-7}{Space Sci.\  Rev.\
  {\bfseries 123} (2006) 485} {\ttfamily
  [\href{https://arxiv.org/abs/astro-ph/0606175}{astro-ph/0606175}]}.

\bibitem{Bramante:2017xlb}
J.~Bramante, A.~Delgado, and A.~Martin, {\em {Multiscatter stellar capture of
  dark matter}}, \href{https://dx.doi.org/10.1103/PhysRevD.96.063002}{Phys.\
  Rev.\  D {\bfseries 96} (2017) 063002} {\ttfamily
  [\href{https://arxiv.org/abs/1703.04043}{arXiv:1703.04043}]}.

\bibitem{Raj:2017wrv}
N.~Raj, P.~Tanedo, and H.-B.~Yu, {\em {Neutron stars at the dark matter direct
  detection frontier}},
  \href{https://dx.doi.org/10.1103/PhysRevD.97.043006}{Phys.\  Rev.\  D
  {\bfseries 97} (2018) 043006} {\ttfamily
  [\href{https://arxiv.org/abs/1707.09442}{arXiv:1707.09442}]}.

\bibitem{Chen:2018ohx}
C.-S.~Chen and Y.-H.~Lin, {\em {Reheating neutron stars with the annihilation
  of self-interacting dark matter}},
  \href{https://dx.doi.org/10.1007/JHEP08(2018)069}{JHEP {\bfseries 08} (2018)
  069} {\ttfamily [\href{https://arxiv.org/abs/1804.03409}{arXiv:1804.03409}]}.

\bibitem{Bell:2018pkk}
N.~F.~Bell, G.~Busoni, and S.~Robles, {\em {Heating up Neutron Stars with
  Inelastic Dark Matter}},
  \href{https://dx.doi.org/10.1088/1475-7516/2018/09/018}{JCAP {\bfseries 09}
  (2018) 018} {\ttfamily
  [\href{https://arxiv.org/abs/1807.02840}{arXiv:1807.02840}]}.

\bibitem{Garani:2018kkd}
R.~Garani, Y.~Genolini, and T.~Hambye, {\em {New Analysis of Neutron Star
  Constraints on Asymmetric Dark Matter}},
  \href{https://dx.doi.org/10.1088/1475-7516/2019/05/035}{JCAP {\bfseries 05}
  (2019) 035} {\ttfamily
  [\href{https://arxiv.org/abs/1812.08773}{arXiv:1812.08773}]}.

\bibitem{Camargo:2019wou}
D.~A.~Camargo, F.~S.~Queiroz, and R.~Sturani, {\em {Detecting Dark Matter with
  Neutron Star Spectroscopy}},
  \href{https://dx.doi.org/10.1088/1475-7516/2019/09/051}{JCAP {\bfseries 09}
  (2019) 051} {\ttfamily
  [\href{https://arxiv.org/abs/1901.05474}{arXiv:1901.05474}]}.

\bibitem{Bell:2019pyc}
N.~F.~Bell, G.~Busoni, and S.~Robles, {\em {Capture of Leptophilic Dark Matter
  in Neutron Stars}},
  \href{https://dx.doi.org/10.1088/1475-7516/2019/06/054}{JCAP {\bfseries 06}
  (2019) 054} {\ttfamily
  [\href{https://arxiv.org/abs/1904.09803}{arXiv:1904.09803}]}.

\bibitem{Hamaguchi:2019oev}
K.~Hamaguchi, N.~Nagata, and K.~Yanagi, {\em {Dark Matter Heating vs.
  Rotochemical Heating in Old Neutron Stars}},
  \href{https://dx.doi.org/10.1016/j.physletb.2019.06.060}{Phys.\  Lett.\  B
  {\bfseries 795} (2019) 484--489} {\ttfamily
  [\href{https://arxiv.org/abs/1905.02991}{arXiv:1905.02991}]}.

\bibitem{Garani:2019fpa}
R.~Garani and J.~Heeck, {\em {Dark matter interactions with muons in neutron
  stars}}, \href{https://dx.doi.org/10.1103/PhysRevD.100.035039}{Phys.\  Rev.\
  D {\bfseries 100} (2019) 035039} {\ttfamily
  [\href{https://arxiv.org/abs/1906.10145}{arXiv:1906.10145}]}.

\bibitem{Acevedo:2019agu}
J.~F.~Acevedo, J.~Bramante, R.~K.~Leane, and N.~Raj, {\em {Warming Nuclear
  Pasta with Dark Matter: Kinetic and Annihilation Heating of Neutron Star
  Crusts}}, \href{https://dx.doi.org/10.1088/1475-7516/2020/03/038}{JCAP
  {\bfseries 03} (2020) 038} {\ttfamily
  [\href{https://arxiv.org/abs/1911.06334}{arXiv:1911.06334}]}.

\bibitem{Joglekar:2019vzy}
A.~Joglekar, N.~Raj, P.~Tanedo, and H.-B.~Yu, {\em {Relativistic capture of
  dark matter by electrons in neutron stars}},
  \href{https://dx.doi.org/10.1016/j.physletb.2020.135767}{Phys.\  Lett.\
  {\bfseries B} (2020) 135767} {\ttfamily
  [\href{https://arxiv.org/abs/1911.13293}{arXiv:1911.13293}]}.

\bibitem{Keung:2020teb}
W.-Y.~Keung, D.~Marfatia, and P.-Y.~Tseng, {\em {Heating neutron stars with GeV
  dark matter}}, \href{https://dx.doi.org/10.1007/JHEP07(2020)181}{JHEP
  {\bfseries 07} (2020) 181} {\ttfamily
  [\href{https://arxiv.org/abs/2001.09140}{arXiv:2001.09140}]}.

\bibitem{Yanagi:2020yvg}
K.~Yanagi, {\em {Thermal Evolution of Neutron Stars as a Probe of Physics
  beyond the Standard Model}}, {\ttfamily
  \href{https://arxiv.org/abs/2003.08199}{arXiv:2003.08199}} (2020).

\bibitem{Joglekar:2020liw}
A.~Joglekar, N.~Raj, P.~Tanedo, and H.-B.~Yu, {\em {Dark kinetic heating of
  neutron stars from contact interactions with relativistic targets}},
  \href{https://dx.doi.org/10.1103/PhysRevD.102.123002}{Phys.\  Rev.\  D
  {\bfseries 102} (2020) 123002} {\ttfamily
  [\href{https://arxiv.org/abs/2004.09539}{arXiv:2004.09539}]}.

\bibitem{Bell:2020jou}
N.~F.~Bell, G.~Busoni, S.~Robles, and M.~Virgato, {\em {Improved Treatment of
  Dark Matter Capture in Neutron Stars}},
  \href{https://dx.doi.org/10.1088/1475-7516/2020/09/028}{JCAP {\bfseries 09}
  (2020) 028} {\ttfamily
  [\href{https://arxiv.org/abs/2004.14888}{arXiv:2004.14888}]}.

\bibitem{Bell:2020lmm}
N.~F.~Bell, G.~Busoni, S.~Robles, and M.~Virgato, {\em {Improved Treatment of
  Dark Matter Capture in Neutron Stars II: Leptonic Targets}},
  \href{https://dx.doi.org/10.1088/1475-7516/2021/03/086}{JCAP {\bfseries 03}
  (2021) 086} {\ttfamily
  [\href{https://arxiv.org/abs/2010.13257}{arXiv:2010.13257}]}.

\bibitem{Anzuini:2021lnv}
F.~Anzuini, {\em et al.}, {\em {Improved treatment of dark matter capture in
  neutron stars III: nucleon and exotic targets}},
  \href{https://dx.doi.org/10.1088/1475-7516/2021/11/056}{JCAP {\bfseries 11}
  (2021) 056} {\ttfamily
  [\href{https://arxiv.org/abs/2108.02525}{arXiv:2108.02525}]}.

\bibitem{Zeng:2021moz}
Y.-P.~Zeng, X.~Xiao, and W.~Wang, {\em {Constraints on Pseudo-Nambu-Goldstone
  dark matter from direct detection experiment and neutron star reheating
  temperature}},
  \href{https://dx.doi.org/10.1016/j.physletb.2021.136822}{Phys.\  Lett.\  B
  {\bfseries 824} (2022) 136822} {\ttfamily
  [\href{https://arxiv.org/abs/2108.11381}{arXiv:2108.11381}]}.

\bibitem{Bramante:2021dyx}
J.~Bramante, B.~J.~Kavanagh, and N.~Raj, {\em {Scattering searches for dark
  matter in subhalos: neutron stars, cosmic rays, and old rocks}}, {\ttfamily
  \href{https://arxiv.org/abs/2109.04582}{arXiv:2109.04582}} (2021).

\bibitem{Tinyakov:2021lnt}
P.~Tinyakov, M.~Pshirkov, and S.~Popov, {\em {Astroparticle Physics with
  Compact Objects}}, \href{https://dx.doi.org/10.3390/universe7110401}{Universe
  {\bfseries 7} (2021) 401} {\ttfamily
  [\href{https://arxiv.org/abs/2110.12298}{arXiv:2110.12298}]}.

\bibitem{Maity:2021fxw}
T.~N.~Maity and F.~S.~Queiroz, {\em {Detecting bosonic dark matter with neutron
  stars}}, \href{https://dx.doi.org/10.1103/PhysRevD.104.083019}{Phys.\  Rev.\
  D {\bfseries 104} (2021) 083019} {\ttfamily
  [\href{https://arxiv.org/abs/2104.02700}{arXiv:2104.02700}]}.

\bibitem{Fujiwara:2022uiq}
M.~Fujiwara, K.~Hamaguchi, N.~Nagata, and J.~Zheng, {\em {Capture of
  Electroweak Multiplet Dark Matter in Neutron Stars}}, {\ttfamily
  \href{https://arxiv.org/abs/2204.02238}{arXiv:2204.02238}} (2022).

\bibitem{Ilie:2020vec}
C.~Ilie, J.~Pilawa, and S.~Zhang, {\em {Comment on
  \textquotedblleft{}Multiscatter stellar capture of dark
  matter\textquotedblright{}}},
  \href{https://dx.doi.org/10.1103/PhysRevD.102.048301}{Phys.\  Rev.\  D
  {\bfseries 102} (2020) 048301} {\ttfamily
  [\href{https://arxiv.org/abs/2005.05946}{arXiv:2005.05946}]}.

\bibitem{Hisano:2011cs}
J.~Hisano, K.~Ishiwata, N.~Nagata, and T.~Takesako, {\em {Direct Detection of
  Electroweak-Interacting Dark Matter}},
  \href{https://dx.doi.org/10.1007/JHEP07(2011)005}{JHEP {\bfseries 07} (2011)
  005} {\ttfamily [\href{https://arxiv.org/abs/1104.0228}{arXiv:1104.0228}]}.

\bibitem{Hisano:2015rsa}
J.~Hisano, K.~Ishiwata, and N.~Nagata, {\em {QCD Effects on Direct Detection of
  Wino Dark Matter}}, \href{https://dx.doi.org/10.1007/JHEP06(2015)097}{JHEP
  {\bfseries 06} (2015) 097} {\ttfamily
  [\href{https://arxiv.org/abs/1504.00915}{arXiv:1504.00915}]}.

\bibitem{Shifman:1978zn}
M.~A.~Shifman, A.~I.~Vainshtein, and V.~I.~Zakharov, {\em {Remarks on Higgs
  Boson Interactions with Nucleons}},
  \href{https://dx.doi.org/10.1016/0370-2693(78)90481-1}{Phys.\  Lett.\  B
  {\bfseries 78} (1978) 443--446}.

\bibitem{Ellis:2018dmb}
J.~Ellis, N.~Nagata, and K.~A.~Olive, {\em {Uncertainties in WIMP Dark Matter
  Scattering Revisited}},
  \href{https://dx.doi.org/10.1140/epjc/s10052-018-6047-y}{Eur.\  Phys.\  J.\
  C {\bfseries 78} (2018) 569} {\ttfamily
  [\href{https://arxiv.org/abs/1805.09795}{arXiv:1805.09795}]}.

\bibitem{Pato:2015dua}
M.~Pato, F.~Iocco, and G.~Bertone, {\em {Dynamical constraints on the dark
  matter distribution in the Milky Way}},
  \href{https://dx.doi.org/10.1088/1475-7516/2015/12/001}{JCAP {\bfseries 12}
  (2015) 001} {\ttfamily
  [\href{https://arxiv.org/abs/1504.06324}{arXiv:1504.06324}]}.

\bibitem{Bell:2020obw}
N.~F.~Bell, {\em et al.}, {\em {Nucleon Structure and Strong Interactions in
  Dark Matter Capture in Neutron Stars}},
  \href{https://dx.doi.org/10.1103/PhysRevLett.127.111803}{Phys.\  Rev.\
  Lett.\  {\bfseries 127} (2021) 111803} {\ttfamily
  [\href{https://arxiv.org/abs/2012.08918}{arXiv:2012.08918}]}.

\bibitem{Potekhin:2020ttj}
A.~Y.~Potekhin, D.~A.~Zyuzin, D.~G.~Yakovlev, M.~V.~Beznogov, and
  Y.~A.~Shibanov, {\em {Thermal luminosities of cooling neutron stars}},
  \href{https://dx.doi.org/10.1093/mnras/staa1871}{Mon.\  Not.\  Roy.\
  Astron.\  Soc.\  {\bfseries 496} (2020) 5052--5071} {\ttfamily
  [\href{https://arxiv.org/abs/2006.15004}{arXiv:2006.15004}]}.

\bibitem{tempdata}
{\em {Cooling neutron stars}},
  \url{http://www.ioffe.ru/astro/NSG/thermal/cooldat.html}.

\bibitem{Kargaltsev:2003eb}
O.~Kargaltsev, G.~G.~Pavlov, and R.~W.~Romani, {\em {Ultraviolet emission from
  the millisecond pulsar j0437-4715}},
  \href{https://dx.doi.org/10.1086/380993}{Astrophys.\  J.\  {\bfseries 602}
  (2004) 327--335} {\ttfamily
  [\href{https://arxiv.org/abs/astro-ph/0310854}{astro-ph/0310854}]}.

\bibitem{Mignani:2008jr}
R.~P.~Mignani, G.~G.~Pavlov, and O.~Kargaltsev, {\em {A possible optical
  counterpart to the old nearby pulsar J0108-1431}},
  \href{https://dx.doi.org/10.1051/0004-6361:200810212}{Astron.\  Astrophys.\
  {\bfseries 488} (2008) 1027} {\ttfamily
  [\href{https://arxiv.org/abs/0805.2586}{arXiv:0805.2586}]}.

\bibitem{Durant:2011je}
M.~Durant, {\em et al.}, {\em {The spectrum of the recycled PSR J0437-4715 and
  its white dwarf companion}},
  \href{https://dx.doi.org/10.1088/0004-637X/746/1/6}{Astrophys.\  J.\
  {\bfseries 746} (2012) 6} {\ttfamily
  [\href{https://arxiv.org/abs/1111.2346}{arXiv:1111.2346}]}.

\bibitem{Rangelov:2016syg}
B.~Rangelov, {\em et al.}, {\em {Hubble Space Telescope Detection of the
  Millisecond Pulsar J2124$-$3358 and its Far-ultraviolet Bow Shock Nebula}},
  \href{https://dx.doi.org/10.3847/1538-4357/835/2/264}{Astrophys.\  J.\
  {\bfseries 835} (2017) 264} {\ttfamily
  [\href{https://arxiv.org/abs/1701.00002}{arXiv:1701.00002}]}.

\bibitem{Pavlov:2017eeu}
G.~G.~Pavlov, {\em et al.}, {\em {Old but still warm: Far-UV detection of PSR
  B0950+08}}, \href{https://dx.doi.org/10.3847/1538-4357/aa947c}{Astrophys.\
  J.\  {\bfseries 850} (2017) 79} {\ttfamily
  [\href{https://arxiv.org/abs/1710.06448}{arXiv:1710.06448}]}.

\bibitem{Abramkin:2021fzy}
V.~Abramkin, G.~G.~Pavlov, Y.~Shibanov, and O.~Kargaltsev, {\em {Thermal and
  Nonthermal Emission in the Optical-UV Spectrum of PSR B0950+08*}},
  \href{https://dx.doi.org/10.3847/1538-4357/ac3a6f}{Astrophys.\  J.\
  {\bfseries 924} (2022) 128} {\ttfamily
  [\href{https://arxiv.org/abs/2111.08801}{arXiv:2111.08801}]}.

\bibitem{Yanagi:2019vrr}
K.~Yanagi, N.~Nagata, and K.~Hamaguchi, {\em {Cooling Theory Faced with Old
  Warm Neutron Stars: Role of Non-Equilibrium Processes with Proton and Neutron
  Gaps}}, \href{https://dx.doi.org/10.1093/mnras/staa076}{Mon.\  Not.\  Roy.\
  Astron.\  Soc.\  {\bfseries 492} (2020) 5508--5523} {\ttfamily
  [\href{https://arxiv.org/abs/1904.04667}{arXiv:1904.04667}]}.

\bibitem{Gonzalez:2010ta}
D.~Gonzalez and A.~Reisenegger, {\em {Internal Heating of Old Neutron Stars:
  Contrasting Different Mechanisms}},
  \href{https://dx.doi.org/10.1051/0004-6361/201015084}{Astron.\  Astrophys.\
  {\bfseries 522} (2010) A16} {\ttfamily
  [\href{https://arxiv.org/abs/1005.5699}{arXiv:1005.5699}]}.

\bibitem{Reisenegger:1994be}
A.~Reisenegger, {\em {Deviations from chemical equilibrium due to spindown as
  an internal heat source in neutron stars}},
  \href{https://dx.doi.org/10.1086/175480}{Astrophys.\  J.\  {\bfseries 442}
  (1995) 749} {\ttfamily
  [\href{https://arxiv.org/abs/astro-ph/9410035}{astro-ph/9410035}]}.

\bibitem{1992A&A...262..131H}
P.~{Haensel}, {\em {Non-equilibrium neutrino emissivities and opacities of
  neutron star matter}}, \aap {\bfseries 262} (1992) 131--137.

\bibitem{1993A&A...271..187G}
E.~{Gourgoulhon} and P.~{Haensel}, {\em {Upper bounds on the neutrino burst
  from collapse of a neutron star into a black hole}}, \aap {\bfseries 271}
  (1993) 187.

\bibitem{Fernandez:2005cg}
R.~Fernandez and A.~Reisenegger, {\em {Rotochemical heating in millisecond
  pulsars. Formalism and non-superfluid case}},
  \href{https://dx.doi.org/10.1086/429551}{Astrophys.\  J.\  {\bfseries 625}
  (2005) 291--306} {\ttfamily
  [\href{https://arxiv.org/abs/astro-ph/0502116}{astro-ph/0502116}]}.

\bibitem{Villain:2005ns}
L.~Villain and P.~Haensel, {\em {Non-equilibrium beta processes in superfluid
  neutron star cores}},
  \href{https://dx.doi.org/10.1051/0004-6361:20053313}{Astron.\  Astrophys.\
  {\bfseries 444} (2005) 539} {\ttfamily
  [\href{https://arxiv.org/abs/astro-ph/0504572}{astro-ph/0504572}]}.

\bibitem{Petrovich:2009yh}
C.~Petrovich and A.~Reisenegger, {\em {Rotochemical heating in millisecond
  pulsars: modified Urca reactions with uniform Cooper pairing gaps}},
  \href{https://dx.doi.org/10.1051/0004-6361/200913861}{Astron.\  Astrophys.\
  {\bfseries 521} (2010) A77} {\ttfamily
  [\href{https://arxiv.org/abs/0912.2564}{arXiv:0912.2564}]}.

\bibitem{Pi:2009eq}
C.-M.~Pi, X.-P.~Zheng, and S.-H.~Yang, {\em {Neutrino Emissivity of
  Non-equilibrium beta processes With Nucleon Superfluidity}},
  \href{https://dx.doi.org/10.1103/PhysRevC.81.045802}{Phys.\  Rev.\  C
  {\bfseries 81} (2010) 045802} {\ttfamily
  [\href{https://arxiv.org/abs/0912.2884}{arXiv:0912.2884}]}.

\bibitem{Gonzalez-Jimenez:2014iia}
N.~Gonz\'alez-Jim\'enez, C.~Petrovich, and A.~Reisenegger, {\em {Rotochemical
  heating of millisecond and classical pulsars with anisotropic and
  density-dependent superfluid gap models}},
  \href{https://dx.doi.org/10.1093/mnras/stu2558}{Mon.\  Not.\  Roy.\  Astron.\
   Soc.\  {\bfseries 447} (2015) 2073} {\ttfamily
  [\href{https://arxiv.org/abs/1411.6500}{arXiv:1411.6500}]}.

\bibitem{1984ApJ...276..325A}
M.~A.~{Alpar}, D.~{Pines}, P.~W.~{Anderson}, and J.~{Shaham}, {\em {Vortex
  creep and the internal temperature of neutron stars. I - General theory}},
  \href{https://dx.doi.org/10.1086/161616}{\apj {\bfseries 276} (1984)
  325--334}.

\bibitem{1989ApJ...346..808S}
N.~{Shibazaki} and F.~K.~{Lamb}, {\em {Neutron star evolution with internal
  heating}}, \href{https://dx.doi.org/10.1086/168062}{Astrophys.\  J.\
  {\bfseries 346} (1989) 808--822}.

\bibitem{1991ApJ...381L..47V}
K.~A.~{van Riper}, R.~I.~{Epstein}, and G.~S.~{Miller}, {\em {Soft X-ray pulses
  from neutron star glitches}}, \href{https://dx.doi.org/10.1086/186193}{\apj
  {\bfseries 381} (1991) L47--L50}.

\bibitem{1993ApJ...408..186U}
H.~{Umeda}, N.~{Shibazaki}, K.~{Nomoto}, and S.~{Tsuruta}, {\em {Thermal
  evolution of neutron stars with internal frictional heating}},
  \href{https://dx.doi.org/10.1086/172579}{\apj {\bfseries 408} (1993)
  186--193}.

\bibitem{VanRiper:1994vp}
K.~Van~Riper, B.~Link, and R.~Epstein, {\em {Frictional heating and neutron
  star thermal evolution}},
  \href{https://dx.doi.org/10.1086/175961}{Astrophys.\  J.\  {\bfseries 448}
  (1995) 294}
{\ttfamily [\href{https://arxiv.org/abs/astro-ph/9404060}{astro-ph/9404060}]}.

\bibitem{Larson:1998it}
M.~B.~Larson and B.~Link, {\em {Superfluid friction and late-time thermal
  evolution of neutron stars}},
  \href{https://dx.doi.org/10.1086/307532}{Astrophys.\  J.\  {\bfseries 521}
  (1999) 271}
{\ttfamily [\href{https://arxiv.org/abs/astro-ph/9810441}{astro-ph/9810441}]}.

\bibitem{Gusakov:2015kaa}
M.~E.~Gusakov, E.~M.~Kantor, and A.~Reisenegger, {\em {Rotation-induced deep
  crustal heating of millisecond pulsars}},
  \href{https://dx.doi.org/10.1093/mnrasl/slv095}{Mon.\  Not.\  Roy.\  Astron.\
   Soc.\  {\bfseries 453} (2015) L36--L40} {\ttfamily
  [\href{https://arxiv.org/abs/1507.04586}{arXiv:1507.04586}]}.

\bibitem{Alexandrou:2019brg}
C.~Alexandrou, {\em et al.}, {\em {Nucleon axial, tensor, and scalar charges
  and $\sigma$-terms in lattice QCD}},
  \href{https://dx.doi.org/10.1103/PhysRevD.102.054517}{Phys.\  Rev.\  D
  {\bfseries 102} (2020) 054517} {\ttfamily
  [\href{https://arxiv.org/abs/1909.00485}{arXiv:1909.00485}]}.

\bibitem{ATLAS:2019lff}
{\bfseries ATLAS} Collaboration, {\em {Search for electroweak production of
  charginos and sleptons decaying into final states with two leptons and
  missing transverse momentum in $\sqrt{s}=13$ TeV $pp$ collisions using the
  ATLAS detector}},
  \href{https://dx.doi.org/10.1140/epjc/s10052-019-7594-6}{Eur.\  Phys.\  J.\
  C {\bfseries 80} (2020) 123} {\ttfamily
  [\href{https://arxiv.org/abs/1908.08215}{arXiv:1908.08215}]}.

\bibitem{PandaX-4T:2021bab}
{\bfseries PandaX-4T} Collaboration, {\em {Dark Matter Search Results from the
  PandaX-4T Commissioning Run}},
  \href{https://dx.doi.org/10.1103/PhysRevLett.127.261802}{Phys.\  Rev.\
  Lett.\  {\bfseries 127} (2021) 261802} {\ttfamily
  [\href{https://arxiv.org/abs/2107.13438}{arXiv:2107.13438}]}.

\bibitem{Billard:2021uyg}
J.~Billard {\em et~al.}, {\em {Direct Detection of Dark Matter -- APPEC
  Committee Report}}, {\ttfamily
  \href{https://arxiv.org/abs/2104.07634}{arXiv:2104.07634}} (2021).

\bibitem{XENON:2019rxp}
{\bfseries XENON} Collaboration, {\em {Constraining the spin-dependent
  WIMP-nucleon cross sections with XENON1T}},
  \href{https://dx.doi.org/10.1103/PhysRevLett.122.141301}{Phys.\  Rev.\
  Lett.\  {\bfseries 122} (2019) 141301} {\ttfamily
  [\href{https://arxiv.org/abs/1902.03234}{arXiv:1902.03234}]}.

\bibitem{PICO:2019vsc}
{\bfseries PICO} Collaboration, {\em {Dark Matter Search Results from the
  Complete Exposure of the PICO-60 C$_3$F$_8$ Bubble Chamber}},
  \href{https://dx.doi.org/10.1103/PhysRevD.100.022001}{Phys.\  Rev.\  D
  {\bfseries 100} (2019) 022001} {\ttfamily
  [\href{https://arxiv.org/abs/1902.04031}{arXiv:1902.04031}]}.

\end{thebibliography}\endgroup


\end{document}